\documentclass[aps,prl,reprint,floatfix,amsmath,amssymb,nofootinbib]{revtex4-2}
\usepackage{certified-gap}

\begin{document}

\title{Bridging Gaps with Tensor Networks: From the Haldane Gap and Beyond}
\author{Sirui Lu}
\email{sirui.lu@mpq.mpg.de}
\affiliation{Max-Planck-Institut f\"ur Quantenoptik, Garching, Germany}
\affiliation{Munich Center for Quantum Science and Technology (MCQST), Schellingstrasse 4, D-80799 M\"unchen, Germany}
\date{\today}

\begin{abstract}
    Proofs generated by large language models are often difficult to
understand, generalize, or formalize directly. Here we verify, digest,
and formalize OpenAI's computer-assisted proof of a uniform gap in the
periodic spin-1 Heisenberg chain using tensor-network diagrams and
Lean~4. Reading the partition-function network along space and imaginary
time shows how finite-ring bounds propagate across doubled lengths and
inverse temperatures to all longer even rings, sharpening the spin-1
bound from the same finite inputs by a factor of $19$ to
$\Delta_L>0.0729$ for all even $L\ge36$. We identify reflection
positivity as the general condition behind the proof and certify uniform
gaps in seven additional spin chains and ladders. More broadly,
digesting machine-generated proofs into transparent physical mechanisms
yields sharper bounds, broader theorems, and reusable formal
verifications.
\end{abstract}
\maketitle
\hideMainFromTOC

Haldane mapped one-dimensional Heisenberg antiferromagnets onto the
nonlinear $\sigma$ model to predict a gap for integer spins, in
contrast to half-integer
chains~\cite{Haldane1983Continuum,Haldane1983Nonlinear,Auerbach1994Interacting}
(near $0.4105$ for spin one~\cite{White1993Numerical}). Because the
Heisenberg chain is not frustration free, finite-size gap criteria for
frustration-free models~\cite{Knabe1988Energy,Gosset2016Local,Rai2026Hierarchy,Rozman2026Lower}
and ground-energy relaxations~\cite{Kull2024Lower} do not suffice.

OpenAI recently gave a computer-assisted proof of a unique ground
state and a uniform gap $\Delta_L>\tfrac{\log20}{784}\simeq0.003821$ for all
even periodic lengths $L\ge2304$ (and
$\Delta_L\gtrsim4.79\times10^{-4}$ for $L\ge60$) from certified finite
partition-function bounds~\cite{OpenAI2026Periodic}. After verifying those finite inputs in exact arithmetic, we recast the
proof in tensor-network diagrams, isolating reflection positivity as
the condition behind its spatial inequalities. We then
certify uniform gaps for seven additional spin chains and ladders,
sharpen the periodic spin-1 bound from the same inputs to
$\Delta_L>0.0729$ for all even $L\ge36$, and formalize the general
theorem in Lean~4~\cite{SM}.

\emph{Thermal purity and the gap.---}
Consider $H_L=\sum_{j=1}^{L}\bm S_j\cdot\bm S_{j+1}$ and
$Z_L(\beta)=\Tr e^{-\beta H_L}$ with periodic indices, spin one, even $L\ge4$, and unit coupling.
Ordering the eigenvalues of $H_L$ with multiplicity defines the gap
$\Delta_L=E_1(L)-E_0(L)$~(\levels).
For $\beta>0$, the Gibbs state $\rho_L=e^{-\beta H_L}/Z_L(\beta)$
has ordered weights $w_0\ge w_1\ge\cdots$. Its thermal purity
$\Pt(L,\beta)=\Tr\rho_L^2\le w_0$ gives%
~\leanmark{https://sirui-lu.com/CertifiedGaps/docs/CertifiedGaps/PurityCriterion.html\#CertifiedGaps.exp_neg_mul_sub_le_of_isMin}
\begin{equation}
    e^{-\beta\Delta_L}=\frac{w_1}{w_0}
    \le\frac{1-w_0}{w_0}
    \le\frac{1-\Pt(L,\beta)}{\Pt(L,\beta)}.
    \label{eq:core-purity-gap}
\end{equation}
When $\Pt(L,\beta)$ is close to one (\dominantglyph), $\Pt>1/2$ forces
$w_0>1/2$ and hence a unique ground state, while $0<1-\Pt\le q<1/2$ gives
$\Delta_L\ge\tfrac1\beta\log\tfrac{1-q}{q}$.

\emph{Proof strategy.---}
Certifying a uniform gap from Eq.~\eqref{eq:core-purity-gap} requires
bounding $\Pt(L,\beta)$ for all even $L\ge n_0$ at $\beta\propto L$,
whereas direct calculation reaches only a finite seed $(n_0,\beta_0)$.
Because $H_n$ is Hermitian, the Gibbs state $\rho_n(\beta)$ is a
density operator whose square is proportional to $\rho_n(2\beta)$ and
squares the defect $1-\Pt$. OpenAI's proof pairs $\rho_n(\beta)$
with a spatial density operator $\sigma_n(\beta)$ whose square is
proportional to $\sigma_{2n}(\beta)$ and doubles the ring length
$n\mapsto2n$. From $(n_0,\beta_0)$, the two squarings alternate along
the staircase:
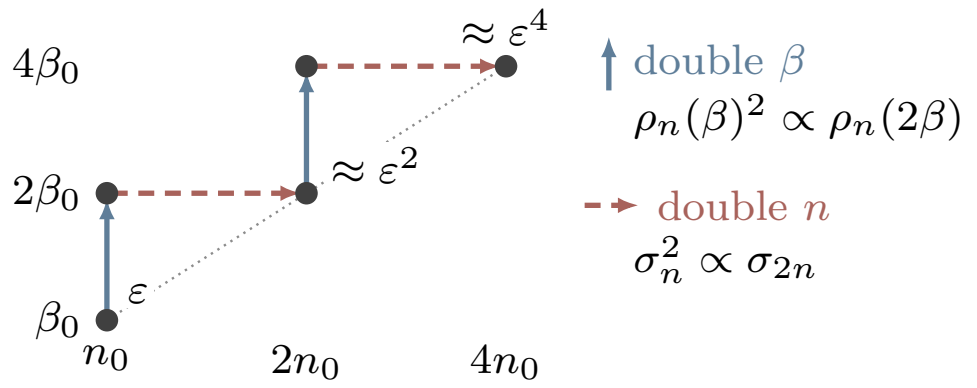
\begin{proofdiagram}
    \strategystairs
\end{proofdiagram}
Each step squares the defect $\varepsilon_j=1-P$ while doubling both
$n_j$ and $\beta_j$, so $\tfrac{\log(1/\varepsilon_j)}{\beta_j}$ stays bounded
below. The proof therefore needs three ingredients: a positive
semidefinite spatial transfer operator to define the density operator
$\sigma_n$; a coupling between the spatial ($n\mapsto2n$) and thermal
($\beta\mapsto2\beta$) squarings; and an interpolation from the dyadic
lengths $n_j=2^jn_0$ to every even $L\in[n_j,2n_j]$.

\emph{Spatial transfer matrix.---}
Set $\tau=\beta/M$ and let $S_L(\tau)$ be the product of
$I-\tau h_{j,j+1}$ over even bonds followed by odd bonds, where
$h=\bm S\cdot\bm S$. Contracting this brickwork network in imaginary
time or space gives~\cite{Suzuki1985Transfermatrix}
\begin{equation}
    \underbrace{\Tr S_L(\beta/M)^M}_{\bricknetwork{1}}
    =Z_L^{(M)}(\beta)
    =\underbrace{\Tr T_{\beta,M}^{L/2}}_{\bricknetwork{2}}.
    \label{eq:core-two-readings}
\end{equation}

In the Cartesian basis $\ket{x},\ket{y},\ket{z}$, the generators
$G_\alpha=iS^\alpha$ are real antisymmetric matrices. Hence
$-h=\sum_{\alpha=x,y,z}G_\alpha\otimes G_\alpha$ and
$I-\tau h=\sum_{\gamma=0,x,y,z}A_\gamma\otimes A_\gamma$,
where $A_0=I$ and $A_\alpha=\sqrt{\tau}\,G_\alpha$; diagrammatically,
$\splitgate=\sum_\gamma\bondends$.
Assign bond-label sequences $a,b\in\{0,x,y,z\}^M$ to the odd and even
bonds adjacent to a site. Tracing that site through the $M$ slices,
with $A_{a_t}$ entering from the odd bond and $A_{b_t}$ leaving on the
even bond, defines a real matrix $\mathcal F$ independent of
$L$%
~\leanmark{https://sirui-lu.com/CertifiedGaps/docs/CertifiedGaps/Ring/Transfer.html\#CertifiedGaps.RPRing.exists_posSemidef_transfer}:
\begin{align}
    \mathcal F_{ba} &=\Tr[(A_{b_1}A_{a_1})\cdots(A_{b_M}A_{a_M})]
    =\sitetraceab,\notag\\
    T_{\beta,M}     &=\mathcal F^{\mathsf T}\mathcal F\ge0,
    \qquad
    \tnchain{\mathcal F^{\mathsf T},\mathcal F}
    =\tnchain{T_{\beta,M}}.
    \label{eq:core-gram}
\end{align}
Because each bond gate $I-\tau h$ splits into matching real matrices
$A_\gamma$, the two sites of an even bond mirror each other across $b$,
contracting $\mathcal F$ with its transpose into the positive
semidefinite Gram matrix $T_{\beta,M}=\mathcal F^{\mathsf T}\mathcal
    F\ge0$—finite-slice reflection
positivity~\cite{Dyson1978Phase,Frohlich1978Phase,Jaffe2016Characterization}.
For even $n\ge4$, $\sigma_n^{(M)}=T_{\beta,M}^{n/2}/Z_n^{(M)}(\beta)$
is a density matrix with $(\sigma_n^{(M)})^2\propto\sigma_{2n}^{(M)}$;
taking $M\to\infty$ at fixed $(L,\beta)$ passes its eigenvalue
inequalities to $Z_L(\beta)$ before doubling begins
(Sec.~\ref{sm:sec:trotter} of Ref.~\cite{SM}).

\emph{Coupling and squaring.---}
With $Z_n(\beta)$ drawn as the $n\times\beta$ torus $\Zt21$, the
spatial purity doubles its width and the thermal purity doubles its
height:
\begin{align*}
    \Ps(n,\beta)
     & =\lim_{M\to\infty}\Tr[(\sigma_n^{(M)})^2]
    =\frac{Z_{2n}(\beta)}{Z_n(\beta)^2}=\frac{\Zt41}{\Ztsq21},\nonumber \\
    \Pt(n,\beta)
     & =\Tr\rho_n(\beta)^2
    =\frac{Z_n(2\beta)}{Z_n(\beta)^2}=\frac{\Zt22}{\Ztsq21}.
\end{align*}
Assembling the $2n\times2\beta$ torus $\Zt42$
from four copies of $\Zt21$ in either order,
\begin{proofdiagram}
    \torusgluing
\end{proofdiagram}
gives the exact coupling identity%
~\leanmark{https://sirui-lu.com/CertifiedGaps/docs/CertifiedGaps/PurityCriterion.html\#CertifiedGaps.spatialPurity_two_mul}
\begin{equation}
    \Ps(n,2\beta)\Pt(n,\beta)^2
    =\Pt(2n,\beta)\Ps(n,\beta)^2.
    \label{eq:core-coupling-s}
\end{equation}

For any density operator with eigenvalues $u_i$ and purity
$P=\sum_i u_i^2$, squaring and renormalizing concentrates the weights
($\specbars[black!55]{.5,.3,.2}\mapsto\specbars[black!55]{.66,.24,.1}$),
giving a new purity $P'$ that obeys%
~\leanmark{https://sirui-lu.com/CertifiedGaps/docs/CertifiedGaps/PurityCriterion.html\#CertifiedGaps.one_sub_squaredPurity_le_squaringDefect}
\begin{equation}
    1-P'=\frac{2\sum_{i<j}(u_i u_j)^2}{P^2}
    \le\frac{(1-P)^2}{2P^2}=f(1-P),
    \label{eq:core-square}
\end{equation}
where $f(z)=z^2/[2(1-z)^2]$.
This squaring map acts identically along space and imaginary time:
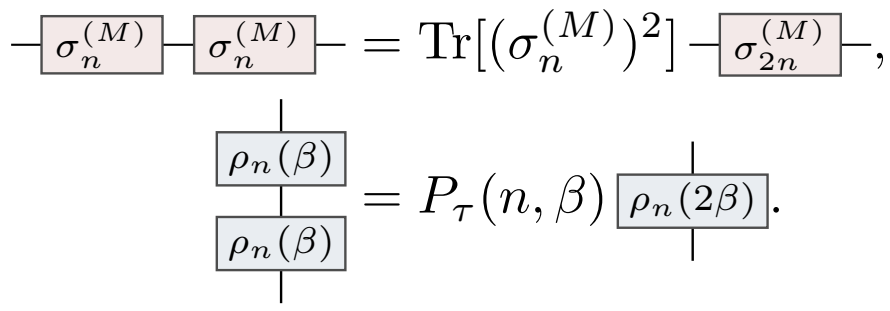
\begin{align*}
    \tnchain{\sigma_n^{(M)},\sigma_n^{(M)}}
     & =\Tr[(\sigma_n^{(M)})^2]\,\tnchain{\sigma_{2n}^{(M)}},\notag \\
    \tncolumn{\rho_n(\beta),\rho_n(\beta)}
     & =\Pt(n,\beta)\,\tncolumn{\rho_n(2\beta)}.
\end{align*}

\begin{figure}[t]
 \centering
 \begin{tikzpicture}
  \begin{axis}[arxiv axis,
   width=.97\columnwidth,height=5.3cm,
   xmin=0,xmax=.23,ymin=0,ymax=.24,
   xlabel={$p\ge1-\Ps(n,\beta)$},
   ylabel={$q\ge1-\Pt(2n,\beta)$},
   xlabel style={text=mpored},
   ylabel style={text=mpoblue},
   xtick={0,.1,.2},ytick={0,.1,.2},
   xticklabels={$0$,$0.1$,$0.2$},
   yticklabels={$0$,$0.1$,$0.2$},
   clip=false]
   \addplot[draw=tncert,line width=.6pt]
    coordinates {
     (0,0) (0,.2006) (.01,.1975) (.02,.194) (.03,.1902)
     (.04,.186) (.05,.1813) (.06,.1762) (.07,.1706)
     (.08,.1645) (.09,.1579) (.10,.1506) (.11,.1427)
     (.12,.1341) (.13,.1247) (.14,.1145) (.15,.1035)
     (.16,.0915) (.17,.0785) (.18,.0644) (.19,.0491)
     (.20,.0326) (.21,.0146) (.2176,0) (0,0)};
   \fill[black!45] (axis cs:0,0) rectangle (axis cs:.01,.01);
   \addplot[black,mark=*,mark size=1.3pt,line width=.7pt]
    coordinates {
     (.0479160,.1815210)
     (.0605003,.1733838)
     (.0686630,.1632461)
     (.0713693,.1447318)
     (.0633146,.1055663)
     (.0376144,.0446518)
     (.0084709,.0055149)};
   \node[anchor=north east,inner sep=2pt,align=right,font=\scriptsize]
    at (axis cs:.0479160,.1815210) {seed\\$n_0=36$};
   \node[anchor=west,inner sep=3pt,font=\scriptsize]
    at (axis cs:.0713693,.1447318) {$288$};
   \node[anchor=south west,inner sep=2pt,font=\scriptsize]
    at (axis cs:.014,.001) {$2304$};
   \node[arxiv cert note,align=center]
    at (axis cs:.142,.071) {numerical\\convergence region};
  \end{axis}
 \end{tikzpicture}
 \caption{Convergence region of the doubling map~\eqref{eq:core-update} in
 the defect bounds $p\ge1-\Ps(n,\beta)$, $q\ge1-\Pt(2n,\beta)$. A seed
 $(p_0,q_0)$ inside the teal boundary gives a uniform gap: doubling
 $(n,\beta)\mapsto(2n,2\beta)$ drives $(p_j,q_j)$ to $(0,0)$, and inside
 the gray square $[0,0.01]^2$ the bound $\varepsilon_{j+1}=5\varepsilon_j^2$
 takes over. Markers: the spin-1 trajectory from $n_0=36$; the small
 $p_0$ pulls $q_j$ down over three steps ($n=288$) and reaches the
 square at step six ($n=2304$).}
 \label{fig:core-region}
\end{figure}
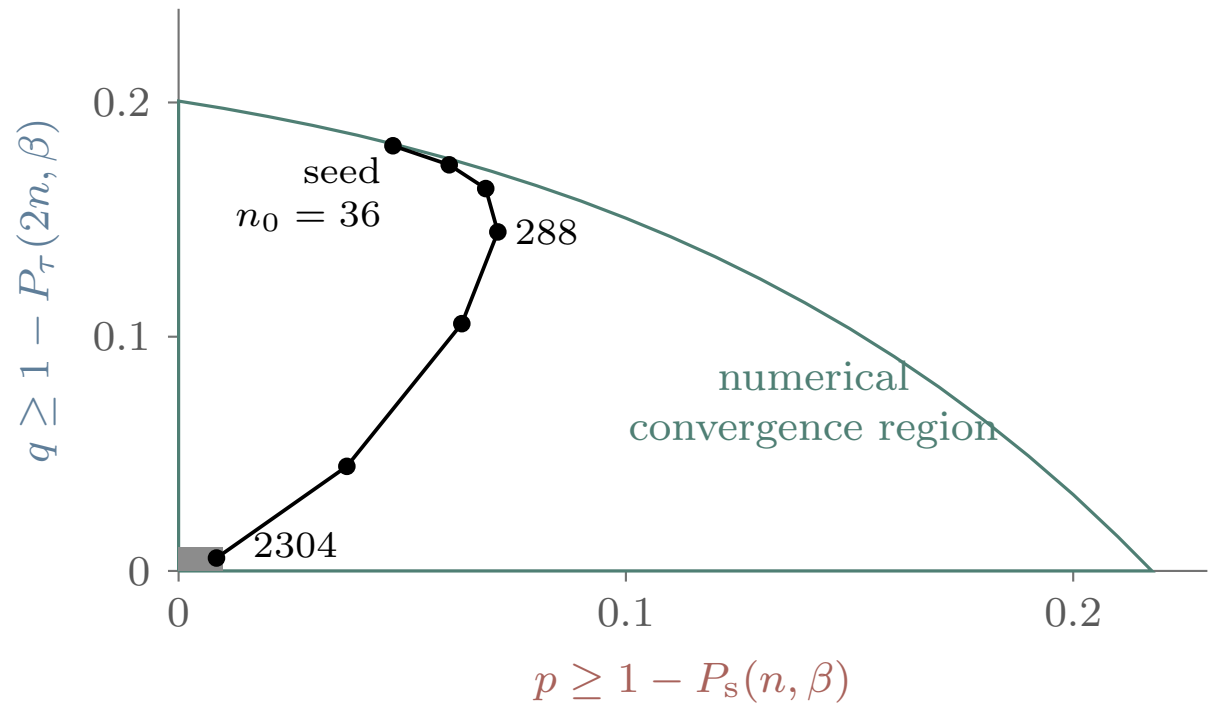

To advance the bounds $1-\Ps(n,\beta)\le p$ and
$1-\Pt(2n,\beta)\le q$ from $(n,\beta)$ to $(2n,2\beta)$, we alternate
torus gluing with squaring along the two directions:
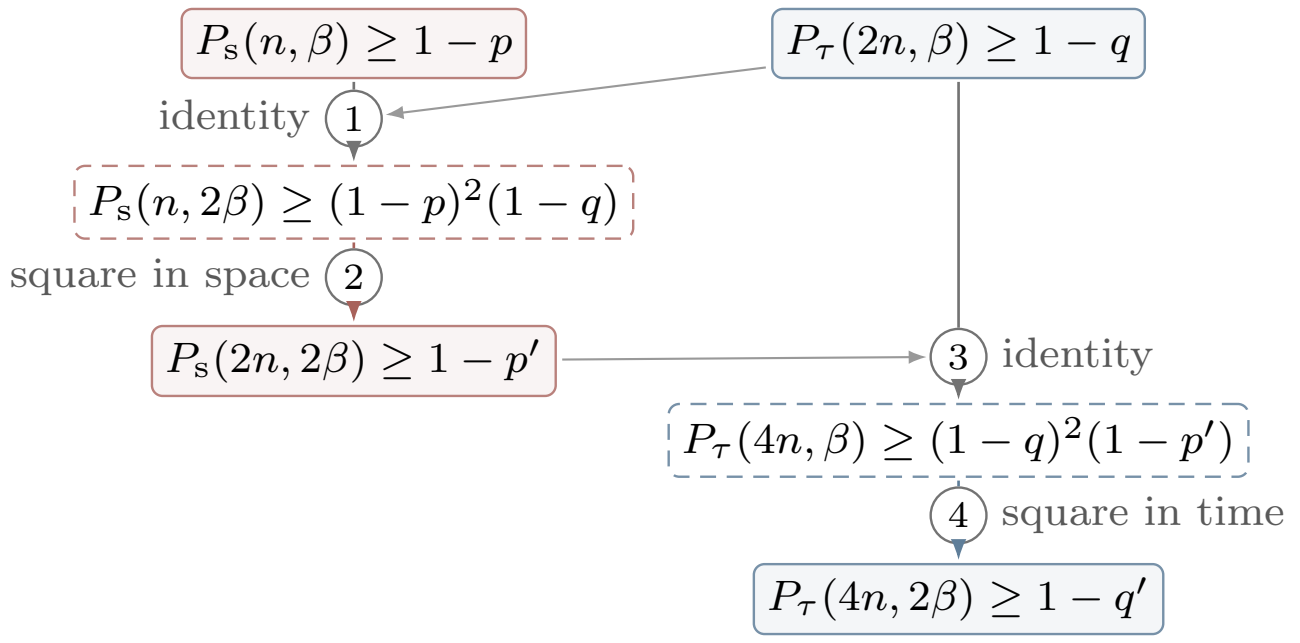
\begin{proofdiagram}
    \pbcsteps
\end{proofdiagram}
Dropping a purity factor $\le1$ in Eq.~\eqref{eq:core-coupling-s}
(moves 1, 3) and applying Eq.~\eqref{eq:core-square} (moves 2, 4) gives%
~\leanmark{https://sirui-lu.com/CertifiedGaps/docs/CertifiedGaps/PurityCriterion.html\#CertifiedGaps.doubling_step}
\begin{align}
    p' &=f(1-(1-p)^2(1-q)),\nonumber\\
    q' &=f(1-(1-q)^2(1-p')),
    \label{eq:core-update}
\end{align}
which maps bounds $(p,q)$ at $(n,\beta)$ to $(p',q')$ at $(2n,2\beta)$
inside the convergent region of Fig.~\ref{fig:core-region}.

\emph{From the finite seed to the uniform gap.---}
Once $(p_j,q_j)$ enters $[0,0.01]^2$ (gray in
Fig.~\ref{fig:core-region}), $f(1-(1-\varepsilon)^3)\le5\varepsilon^2$
drives all further doubling steps analytically (Sec.~\ref{sm:sec:gap}
of Ref.~\cite{SM}). We start from OpenAI's certified seed:

\begin{lemma}[Finite seed~\cite{OpenAI2026Periodic}%
        ~\leanmark{https://sirui-lu.com/CertifiedGaps/docs/CertifiedGaps/Heisenberg/Seed36.html\#CertifiedGaps.SpinOneHeisenberg.seed36_of_table_mps}]
    \label{lem:arxiv-seed36}
    At $\beta_0=49/4$, the periodic spin-1 chain satisfies
    $\Ps(36,\beta_0)\ge1-47916/10^6$ and
    $\Pt(72,\beta_0)\ge1-181521/10^6$.
\end{lemma}

Although $q_0=0.181521$ exceeds the symmetric fixed point
$\varepsilon_*\simeq0.1273$, the smaller $p_0$ pulls $q'$ below $q$,
and six steps of Eq.~\eqref{eq:core-update} reach
$(p_6,q_6)\le(0.0085,0.0056)$ at $(n_6,\beta_6)=(2304,784)$
(Fig.~\ref{fig:core-region}; Sec.~\ref{sm:sec:orbit} of Ref.~\cite{SM}).

Restarting at $(2304,784)$ with $\varepsilon_0=0.01$ and
$\varepsilon_{j+1}=5\varepsilon_j^2$ gives
$\varepsilon_j=20^{-2^j}/5$ at $n_j=2304\,2^j$ and
$\beta_j=784\,2^j$. To cover every even $L\ge2304$, choose $j$ with
$n_j\le L\le2n_j$. Because the eigenvalues of $\sigma_u^{(M)}$ lie in
$[0,1]$ and sum to $1$, raising them to a power $v/u\ge1$ lowers their
sum
($\specbars[black!55]{.6,.3,.1}\mapsto\specbars[black!55]{.36,.09,.01}$),
giving $Z_v(\beta)\le Z_u(\beta)^{v/u}$ for even $v\ge u$ as
$M\to\infty$ and bounding $\Pt(L,\beta_j)$~(\interpolationglyph):%
~\leanmark{https://sirui-lu.com/CertifiedGaps/docs/CertifiedGaps/PurityCriterion.html\#CertifiedGaps.thermalPurity_ge_of_window}
\begin{align}
    \Pt(L,\beta_j)
     &\ge[\Pt(2n_j,\beta_j)\Ps(n_j,\beta_j)^2]^{L/(2n_j)}
    \nonumber\\
     &\ge(1-\varepsilon_j)^3\ge1-3\varepsilon_j.
    \label{eq:core-interpolation}
\end{align}
With $\varepsilon_j\le0.01$, $3\varepsilon_j/(1-3\varepsilon_j)<20^{-2^j}$,
so Eq.~\eqref{eq:core-purity-gap} proves a unique ground state and%
~\leanmark{https://sirui-lu.com/CertifiedGaps/docs/CertifiedGaps/Heisenberg/Main.html\#CertifiedGaps.SpinOneHeisenberg.spin_one_heisenberg_gap}
\begin{equation*}
    \Delta_L>\frac{\log20}{784}\simeq0.003821
    \qquad\text{for all even }L\ge2304.
\end{equation*}

\emph{Bounding the finite seed.---}
Certifying a seed at $(n_0,\beta_0)$ reduces to bounding the
numerators of
\begin{equation}
    \textcolor{mpored}{\Ps(n_0,\beta_0)}=\tfrac{Z_{2n_0}(\beta_0)}{Z_{n_0}(\beta_0)^2},\qquad
    \textcolor{mpoblue}{\Pt(2n_0,\beta_0)}=\tfrac{Z_{2n_0}(2\beta_0)}{Z_{2n_0}(\beta_0)^2}
    \label{seed:ratios}
\end{equation}
from below and their denominators from above. When $2n_0$ is too large
for direct diagonalization ($3^{72}\simeq2.3\times10^{34}$ at
$n_0=36$), Ref.~\cite{OpenAI2026Periodic} bounds these four partition
functions from short-ring traces on $4\le n\le12$ sites
($3^n\le5.3\times10^5$) and a periodic matrix-product trial state at
length $N=2n_0$:
\begin{center}\footnotesize
    \setlength{\tabcolsep}{3pt}
    \begin{tabular}{@{}clll@{}}
        \toprule
                    & Quantity             & Bound & Method                            \\
        \midrule
        $\Zt{4}{1}$ & $Z_{2n_0}(\beta_0)$  & lower & MPS at $2n_0$ + short-ring traces \\
        $\Zt{2}{1}$ & $Z_{n_0}(\beta_0)$   & upper & MPS at $2n_0$ + short-ring traces \\
        $\Zt{4}{2}$ & $Z_{2n_0}(2\beta_0)$ & lower & MPS at $2n_0$                     \\
        $\Zt{4}{1}$ & $Z_{2n_0}(\beta_0)$  & upper & short-ring traces                 \\
        \bottomrule
    \end{tabular}
\end{center}
Five steps turn these inputs into Lemma~\ref{lem:arxiv-seed36} at
$n_0=36$ and $\beta_0=49/4$ (Sec.~\ref{sm:sec:seeds} of
Ref.~\cite{SM}).

\paragraph{Step 1: One-site spatial transfer operator.}
Unlike the two-site Gram matrix $T_{\beta,M}$ (even lengths only), the
continuous-time self-adjoint one-site transfer operator $K_\beta$ of
Ref.~\cite{OpenAI2026Periodic} gives $Z_n(\beta)$ for all $n\ge4$:
\begin{equation*}
    Z_n(\beta)=\Tr K_\beta^n
    =\tnchain[node=xvertex,ring]{,\cdots,}_{\,n},
    \qquad n\ge4.
\end{equation*}
Shifting the energy per site by $\kappa=700741/500000$ (so the trial
ground-state energy in Step~3 is negative) defines
$\widehat K=e^{-\beta_0\kappa}K_{\beta_0}$ with real eigenvalues
$\lambda_i$ of either sign and shifted traces
\begin{equation}
    \widehat Z_n(\beta)=e^{-n\beta\kappa}Z_n(\beta),
    \qquad
    \widehat Z_n(\beta_0)=\sum_i\lambda_i^n,
    \label{seed:shift}
\end{equation}
which leave both purities unchanged and sample the same $\{\lambda_i\}$
at every $n\ge4$.

\paragraph{Step 2: Symmetry sectors from twisted rings.}
To tighten the subleading sum at $n_0$, the commuting $\pi$ spin
rotations split $\widehat K$ into a rotation-invariant sector $J$ and
three degenerate sectors $N$ permuted by axis rotations. Twisting the
closing bond by a $\pi$ rotation $P$ negates two of the three $N$
sectors, giving


\begin{align}
    \seedring{0}\quad &\widehat Z_n=J_n+3N_n,\notag\\
    \seedring{1}\quad &\widehat Z_n(P)=J_n-N_n,
    \label{seed:sectors}
\end{align}
with $J_n=\sum_{i\in J}\lambda_i^n$ and $N_n=\sum_{i\in N}\lambda_i^n$
at $\beta_0$. Integer polynomial approximations to $e^{-\beta_0 H_n/2}$
enclose both traces for $4\le n\le12$; solving Eq.~\eqref{seed:sectors}
at $n=12$ (which divides $n_0$) gives the reference twelfth-moment
bounds $J_{12}\le m_J=1.0657205$ and $N_{12}\le m_N=0.2783155$.

\paragraph{Step 3: Partition-function lower bound from a trial state.}
On a ring of length $N=2n_0$, integer tensors $A$ (bond dimension $26$)
define a periodic MPS $\psi$ whose one-bond energy contracts as an
$N$-step matrix trace,
\begin{equation*}
    \mathcal E=\langle\psi,h_{1,2}\psi\rangle=\seedmps{1},
\end{equation*}
and whose norm $\mathcal N=\langle\psi,\psi\rangle$ omits the $h$ box.
Translation invariance gives
$\langle\psi,H_N\psi\rangle/(N\mathcal N)=\mathcal E/\mathcal N<-\kappa$,
so $E_0(N)<-N\kappa$ and retaining only the ground-state weight gives
$\widehat Z_N(\beta)\ge e^{-\beta(E_0(N)+N\kappa)}>1$ for all $\beta>0$,
supplying both $\sum_i\lambda_i^{2n_0}>1$ at $\beta_0$ and
$\widehat Z_{2n_0}(2\beta_0)>1$.

\paragraph{Step 4: Lower bound on the leading spatial eigenvalue.}
To bound $\widehat Z_{2n_0}(\beta_0)\ge\lambda_0^{2n_0}$ from below, we
show that $J$ contains an eigenvalue with $|\lambda_0|>\ell$
($\ell=0.999$ at $n_0=36$). Write $u_i=\lambda_i^{12}\ge0$, so
$\lambda_i^{2n_0}=u_i^{2n_0/12}$. If every $|\lambda_i|\le\ell$, then
$u_i\le\ell^{12}$; since $u\mapsto u^{2n_0/12}$ is convex and
$\ell^{12}<m_J<2\ell^{12}$, $\sum_{i\in J}u_i^{2n_0/12}$ at mass
$\le m_J$ is maximized by one entry $\ell^{12}$ and remainder
$m_J-\ell^{12}$:
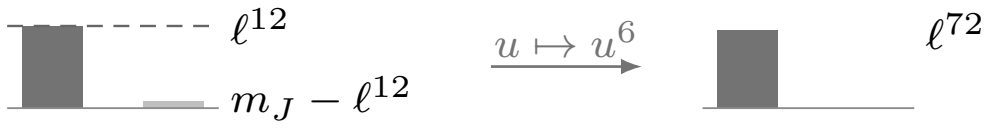
\begin{proofdiagram}
    \seedweightbars
\end{proofdiagram}
Raising to the power $2n_0/12$ suppresses the remainder and bounds
each $N$ sector by $m_N^{2n_0/12}$, giving at $2n_0=72$ ($<0.932$)
\begin{equation*}
    \widehat Z_{2n_0}(\beta_0)
    \le\ell^{2n_0}+(m_J-\ell^{12})^{2n_0/12}+3m_N^{2n_0/12}<1,
\end{equation*}
contradicting $\widehat Z_{2n_0}(\beta_0)>1$ from Step~3; since
$m_N<\ell^{12}$, the eigenvalue $|\lambda_0|>\ell$ lies in $J$.

\begin{table*}[t]
    \caption{Certified uniform gap bounds for reflection-positive spin
        chains and ladders, covering every even length $L$ ($L$ counts sites,
        two-spin dimer cells, or two-spin rungs; $\bm S^2=S(S+1)I$,
        $\sigma^\alpha$ Pauli matrices). ``Gap $\ge$'' gives the certified lower
        bound on $\Delta_L=E_1(L)-E_0(L)$ (the interval bound of the certificate)
        and links to the Lean~4 theorem that proves the rational-chain bound from
        the seed pair as hypothesis, equal in the quoted digits except where the
        chain was rounded more coarsely; the single-ion row's Lean theorem starts
        from the direct seed $(6,9/4)$ and proves $0.6946$;
        ``Numerical'' lists uncertified $12$--$20$-spin estimates (or bulk
        DMRG~\cite{White1993Numerical} for $S=1$ Heisenberg). The mixed-field
        Ising bound is verified unconditionally in the Lean kernel.}
    \label{tab:models}
    \setlength{\tabcolsep}{4pt}
    \renewcommand{\arraystretch}{1.25}
    \begin{tabular}{@{}p{.23\textwidth}p{.39\textwidth}lll@{}}
        \toprule
        Model & Hamiltonian                                                                                              & Even lengths                                                                                                                                                      & Gap $\ge$ & Numerical \\
        \midrule
        Staggered field, $S=\tfrac12$
              & $\sum_j\bm S_j\cdot\bm S_{j+1}-\sum_j(-1)^jS_j^z$
              & $L\ge8$ sites                                                                                            & $1.1489$~\leanmark{https://sirui-lu.com/CertifiedGaps/docs/CertifiedGaps/Models/StaggeredHeisenbergPhysical.html\#CertifiedGaps.StaggeredHeisenberg.physical_gap} & $1.874$               \\
        Dimerized, $S=\tfrac12$
              & $\sum_j[1+(-1)^j/2]\,\bm S_j\cdot\bm S_{j+1}$
              & $L\ge4$ cells                                                                                            & $0.5347$~\leanmark{https://sirui-lu.com/CertifiedGaps/docs/CertifiedGaps/Models/DimerizedHeisenbergPhysical.html\#CertifiedGaps.DimerizedHeisenberg.physical_gap} & $1.189$               \\
        Dimerized, $S=1$
              & $\sum_j[1+9(-1)^j/10]\,\bm S_j\cdot\bm S_{j+1}$
              & $L\ge2$ cells                                                                                            & $0.5288$~\leanmark{https://sirui-lu.com/CertifiedGaps/docs/CertifiedGaps/Models/SpinOneDimerized.html\#CertifiedGaps.SpinOneDimerized.gap}                        & $1.763$               \\
        Single-ion anisotropy, $S=1$
              & $\sum_j\bm S_j\cdot\bm S_{j+1}+4\sum_j(S_j^z)^2$
              & $L\ge12$ sites                                                                                           & $0.8994$~\leanmark{https://sirui-lu.com/CertifiedGaps/docs/CertifiedGaps/Models/SpinOneLargeD.html\#CertifiedGaps.SpinOneLargeD.gap}                              & $2.269$               \\
        Heisenberg ladder, \mbox{$S=\tfrac12$}
              & $\sum_{j,a=1,2}\bm S_{j,a}\cdot\bm S_{j+1,a}
                    +4\sum_j\bm S_{j,1}\cdot\bm S_{j,2}$
              & $L\ge4$ rungs                                                                                            & $1.1024$~\leanmark{https://sirui-lu.com/CertifiedGaps/docs/CertifiedGaps/Models/HeisenbergLadderPhysical.html\#CertifiedGaps.HeisenbergLadder.physical_gap}       & $3.138$               \\
        Ising ladder
              & $-\sum_{j,a}\sigma^z_{j,a}\sigma^z_{j+1,a}-\sum_j\sigma^z_{j,1}\sigma^z_{j,2}-3\sum_{j,a}\sigma^x_{j,a}$
              & $L\ge4$ rungs                                                                                            & $0.6368$~\leanmark{https://sirui-lu.com/CertifiedGaps/docs/CertifiedGaps/Models/TwoLegIsing.html\#CertifiedGaps.TwoLegIsing.gap}                                  & $2.662$               \\
        Mixed-field Ising
              & $-\sum_j\sigma_j^z\sigma_{j+1}^z
                    -\tfrac14\sum_j\sigma_j^x-\tfrac1{20}\sum_j\sigma_j^z$
              & $L\ge6$ sites                                                                                            & $0.3922$~\leanmark{https://sirui-lu.com/CertifiedGaps/docs/Leaderboard/Records/IsingLongitudinal.html\#Leaderboard.Records.IsingLongitudinal.g_1_4_h_1_20}        & $1.190$               \\
        Heisenberg, $S=1$
              & $\sum_j\bm S_j\cdot\bm S_{j+1}$
              & $L\ge36$ sites                                                                                           & $0.0729$~\leanmark{https://sirui-lu.com/CertifiedGaps/docs/Leaderboard/Records/SpinOneHeisenberg.html\#Leaderboard.Records.SpinOneHeisenberg.temperature_lifting} & $0.410$               \\
        \bottomrule
    \end{tabular}
\end{table*}

\paragraph{Step 5: Polynomial upper bounds on the remaining spectrum.}
To bound $\widehat Z_{n_0}(\beta_0)$ and $\widehat Z_{2n_0}(\beta_0)$
from above, we bound the remaining eigenvalues in each sector
$\mathcal S\in\{J,N\}$ using a nonnegative polynomial
$F(x)=\sum_{j=4}^{12}f_jx^j$:
\begin{equation*}
    F(\lambda_i)\le\sum_{a\in\mathcal S}F(\lambda_a)
    =\sum_{j=4}^{12}f_j\mathcal S_j\le B,
    \qquad i\in\mathcal S.
\end{equation*}
No eigenvalue in $\mathcal S$ lies where $F(x)>B$~(\filterglyph).
Two degree-$12$ polynomials rule out $|\lambda|\ge U=1.00139$ in $J$
and $|\lambda|\ge V=0.872$ in $N$ (Sec.~\ref{cert:sec:filters} of
Ref.~\cite{SM}).

Removing $\lambda_0^{12}>\ell^{12}$ leaves twelfth moment at most
$m_J-\ell^{12}$ in $J$, while convexity with upper bound $V^{12}$
($V^{12}<m_N<2V^{12}$) bounds each $N$ sector by one entry at $V^{12}$
and the remainder $m_N-V^{12}$ (\capbars{1,.44}{1}). For
$k\in\{n_0,2n_0\}$, the subleading sum $\sum_{i\ne0}\lambda_i^k$ is at
most
\begin{equation*}
    R_k=(m_J-\ell^{12})^{k/12}+3\bigl[V^k+(m_N-V^{12})^{k/12}\bigr].
\end{equation*}
Bounding each numerator of Eq.~\eqref{seed:ratios} below by its leading
term and each denominator above by its leading term
($|\lambda_0|^{n_0}>\ell^{n_0}$ and $U^{2n_0}$) plus $R_k$ gives%
~\leanmark{https://sirui-lu.com/CertifiedGaps/docs/CertifiedGaps/Seed/Ring.html\#CertifiedGaps.RPRing.seeds_of_certificate}
\begin{align*}
    \Ps(n_0,\beta_0)  & \ge\bigl[1+R_{n_0}/\ell^{n_0}\bigr]^{-2},\notag \\
    \Pt(2n_0,\beta_0) & \ge\bigl[U^{2n_0}+R_{2n_0}\bigr]^{-2};
\end{align*}
exact rational evaluation at $(36,72)$ returns the two bounds of
Lemma~\ref{lem:arxiv-seed36}.

Because Boltzmann weights $z_i>0$ satisfy
$\sum_i z_i^r\le(\sum_i z_i)^r$ for $r>1$, short-ring trace bounds at
$\beta_0$ also bound the moments at $r\beta_0$. Whereas
Ref.~\cite{OpenAI2026Periodic} lifted a separate $n\le10$ table at
$\beta_0'=21/2$ by $r=5/2$ to get $\Delta_L\gtrsim4.79\times10^{-4}$ for
$L\ge60$, lifting the stronger $\beta_0=49/4$ ($n\le12$) table by
$r=17/8$ with a $120$-site trial state and tighter polynomial bounds
$U,V$ certifies $(p_0,q_0)=(0.0306,0.0480)$ at $n_0=60$%
Ten steps of Eq.~\eqref{eq:core-update} and bridging $36\le L\le72$
from $(36,\beta_0)$ then prove $\Delta_L>0.0729$ for all even
$L\ge36$%
~\leanmark{https://sirui-lu.com/CertifiedGaps/docs/Leaderboard/Records/SpinOneHeisenberg.html\#Leaderboard.Records.SpinOneHeisenberg.temperature_lifting}
(Sec.~\ref{cert:sec:lifting} of Ref.~\cite{SM}).

\emph{Reflection-positive models.---}
The spatial Gram factorization~\eqref{eq:core-gram} relies on
locality and bond reflection positivity~\cite{Frohlich1978Phase,Jaffe2016Characterization},
not on $\mathrm{SU}(2)$ symmetry. Consider a periodic ring of $L\ge2$
unit cells with Hamiltonian $H_L=\sum_{j=1}^L h_{j,j+1}+\sum_{j=1}^L o_j$
($j+L\equiv j$), where $h$ couples neighboring cells and $o$ acts within a cell.
Suppose in some local spin basis there is a real symmetric involution
$F$ ($F^2=I$, such as dimer spin exchange) for which $o$ is real
symmetric with $FoF=o$ and
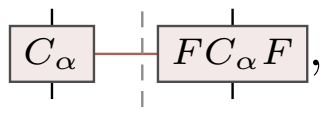
\begin{equation*}
 -h=\sum_\alpha C_\alpha\otimes FC_\alpha F,\qquad\scopebond,
\end{equation*}
with each $C_\alpha$ real symmetric or antisymmetric and the mirror
(dashed) between cells. Conjugating odd cells by $F$ gives both
sublattices the same trace matrix $\mathcal F$, so the Gram
form~\eqref{eq:core-gram} and the doubling recurrence apply unchanged
(Sec.~\ref{sm:sec:transfer} of Ref.~\cite{SM}).

A convergent seed at even $n_0$ then certifies a unique ground state
and a uniform gap for all even $L\ge n_0$ by applying
Eq.~\eqref{eq:core-interpolation} on each dyadic interval $[2^j n_0,2^{j+1}n_0]$%
~\leanmark{https://sirui-lu.com/CertifiedGaps/docs/CertifiedGaps/Ring/Gap.html\#CertifiedGaps.RPRing.gapGT_of_chainCertificate}.
Table~\ref{tab:models} applies this theorem to staggered-field
chains~\cite{Oshikawa1997Field}, bond-alternating
chains~\cite{Barnes1999Alternating,Yamamoto1995Elementary}, chains
with single-ion anisotropy~\cite{Chen2003Ground}, Heisenberg
ladders~\cite{Dagotto1996Surprises}, and Ising models, with finite
seeds certified from verified short-ring spectra or exact matrix
arithmetic (Sec.~\ref{sm:sec:models} of Ref.~\cite{SM}).

\emph{Formal verification.---}
References~\cite{OpenAI2026Periodic,OpenAI2026Boundaryfield} were not
accompanied by a Lean proof, and formalizing their continuous-time
kernel directly would require path-space trace identities not yet in
Mathlib. Our finite Gram matrix~\eqref{eq:core-gram} and
fixed-$(L,\beta)$ limit~\eqref{eq:core-two-readings} instead reduce
the spatial inequalities to finite-dimensional matrix algebra%
~\leanmark{https://sirui-lu.com/CertifiedGaps/docs/CertifiedGaps/Ring/Gap.html\#CertifiedGaps.RPRing.spatialMomentBounds}
and yield a general reflection-positive gap theorem in
\href{https://github.com/LionSR/CertifiedGaps}{Lean~4}~\cite{Moura2021Lean,Lu2026CertifiedGaps,Lu2026Blueprint}
(Sec.~\ref{sm:sec:blueprint} of Ref.~\cite{SM}; each
$\leanmark{https://sirui-lu.com/CertifiedGaps/docs/CertifiedGaps/Ring/Gap.html}$
links to a Lean theorem).

A computer-assisted proof also relies on finite seed computations that
are too large for direct diagonalization inside a proof-assistant
kernel. Separating the universal doubling theorem from the finite seed
lets the formalization work at two levels: taking certified seed bounds
as hypotheses instantiates all eight models in Table~\ref{tab:models},
whereas reducing the seed to exact integer identities lets the Lean
kernel verify the finite arithmetic itself—proving the mixed-field
Ising gap unconditionally%
~\leanmark{https://sirui-lu.com/CertifiedGaps/docs/SeedKernel/Ising/Seed.html\#SeedKernel.Ising.ising_gap}
(Sec.~\ref{models:sec:ising} of Ref.~\cite{SM}) and checking the
spin-1 trial states and shortest ring traces so that only the
remaining short-ring trace bounds enter as hypotheses%
~\leanmark{https://sirui-lu.com/CertifiedGaps/docs/SeedKernel/Heisenberg/Integers.html\#CertifiedGaps.SpinOneHeisenberg.gap_of_hornerTotalsLarge}.

\emph{Discussion.---}
Beyond periodic rings, the same diagrammatic method extends to open
spin-1 chains with endpoint magnetic
fields~\cite{OpenAI2026Boundaryfield}: replacing the spatial trace by
boundary vectors and tracking their overlap (\overlapglyph) with the
leading spatial eigenvector gives a diagrammatic proof and Lean
formalization that sharpen the gap bound to $\Delta_n>0.01556$ for all
odd $n\ge1921$%
~\leanmark{https://sirui-lu.com/CertifiedGaps/docs/CertifiedGaps/Heisenberg/OpenChain.html\#CertifiedGaps.SpinOneHeisenberg.gapGT_of_boundaryFieldInputs_sharp}
(Sec.~\ref{sm:sec:obc} of Ref.~\cite{SM}). More broadly, as
machine-assisted proofs become more common, raw proofs are often hard
to understand, generalize, or formalize directly; verifying the finite
inputs first, digesting the argument into a transparent mechanism, and
only then formalizing the general theorem turns a raw proof into
sharper bounds, new certified models, and a reusable formal library.

In two dimensions, a spacetime cube identity couples two spatial
purities with the thermal purity when both lattice directions are
reflection positive~\cite{Lu2026CertifiedGaps}; for the honeycomb AKLT
model~\cite{Affleck1987Rigorous,Lemm2020Existence,Pomata2020Demonstrating},
the spin-$3/2$ bond projector violates standard reflection positivity,
leaving spatial gluing open.

Finally, uniform gap bounds supply the hypothesis of correlation-decay
and entanglement area-law theorems in one and two
dimensions~\cite{Hastings2006Spectral,Hastings2007Area,Arad2013Area,OpenAI2026Twodimensional,OpenAI2026Polynomial},
whose proofs have also been formalized in
Lean~\cite{Lu2026Multiagent,Lu2026TNLean}. Checking the remaining
spin-1 short-ring trace recursions in the kernel motivates verified
numerical linear algebra in the proof assistant.

\begin{acknowledgments}
    We sincerely thank Xiao-Liang Qi for insightful discussions and hospitality at Stanford
    University, and thank Ignacio Cirac, Shengqi Sang, Yimu Bao, and Junkai Dong for helpful
    discussions. Part of this work was also carried out during a visit to the Kavli Institute for
    Theoretical Physics (KITP), supported in part by grant NSF PHY-2309135; we thank KITP for its
    hospitality. This research was funded by the Deutsche Forschungsgemeinschaft (DFG, German
    Research Foundation) under Germany's Excellence Strategy---EXC-2111---390814868. Following the
    Agentic Publication Protocol~\cite{Lu2026APP}, the code, gap certificates, Lean~4
    formalization~\cite{Lu2026CertifiedGaps}, and proof blueprint~\cite{Lu2026Blueprint} are
    available in the companion repository and interactive project website~\cite{Lu2026Contribute}
    (Sec.~\ref{sm:sec:blueprint} of Ref.~\cite{SM}). We welcome open-science contributions of new
    or improved gap certificates and Lean~4 formalizations~\cite{Lu2026Contribute}.
\end{acknowledgments}

\emph{Note added.---}%
An initial version of this work was circulated at Stanford on
October~7, 2026, and presented at the KITP \emph{Intermediate-Scale AI
    for Physics} conference~\cite{Lu2026Bridging}. We note the independent
expository account by Tasaki~\cite{Tasaki2026Haldane} and the
independent preprint by Grover~\cite{Grover2026Extensions}, which
appeared after this work was completed.

\emph{AI-use declaration.---}%
S.\,L. prompted GPT-6-Astra (OpenAI) to recast the proofs of
Refs.~\cite{OpenAI2026Periodic,OpenAI2026Boundaryfield} in tensor-network diagrams, and
used TeXRA~\cite{TeXRA2025}, a multi-agent autoformalization
workflow~\cite{Lu2026Multiagent,Lu2026TNLean}, and OpenAI Codex with GPT-6-Astra and
GPT-6-Sol for subsequent calculations, formalization, and manuscript preparation. All
mathematical arguments, numerical certificates, and formal proofs were verified by the
author.

\par\label{main:text-end}

\bibliographystyle{apsrev4-2}
\bibliography{references}

\clearpage
\appendix
\renewcommand{\appendixname}{}
\setcounter{secnumdepth}{3}
\def\theHequation{supp.\arabic{equation}}
\def\theHtable{supp.\arabic{table}}
\def\theHfigure{supp.\arabic{figure}}
\def\theHtheorem{supp.\arabic{theorem}}
\def\theHlemma{supp.\arabic{lemma}}
\def\theHproposition{supp.\arabic{proposition}}
\renewcommand{\thesection}{S\Roman{section}}
\renewcommand{\thesubsection}{\thesection.\Alph{subsection}}
\makeatletter
\@removefromreset{equation}{section}
\def\p@subsection{}
\def\p@subsubsection{\thesubsection.}
\makeatother
\setcounter{figure}{0}
\renewcommand{\thefigure}{S\arabic{figure}}
\setcounter{theorem}{0}
\renewcommand{\thetheorem}{S\arabic{theorem}}
\setcounter{lemma}{0}
\renewcommand{\thelemma}{S\arabic{lemma}}
\setcounter{proposition}{0}
\renewcommand{\theproposition}{S\arabic{proposition}}
\onecolumngrid
\setcounter{equation}{0}
\renewcommand{\theequation}{S\arabic{equation}}
\setcounter{table}{0}
\renewcommand{\thetable}{S\arabic{table}}
\begin{center}\large Supplemental Material\end{center}
\medskip
\noindent
This Supplemental Material contains the full proofs, seed
certificates, and Lean~4 formalization details.
Section~\ref{sm:sec:trotter} constructs the finite-dimensional spatial
transfer matrix $T_{\beta,M}$ for any reflection-positive ring
Hamiltonian, takes the continuum limit $M\to\infty$, and proves the
periodic scale-doubling theorem: a finite seed $(n_0,\beta_0)$ with both
defects at most $1/100$ gives a uniform gap on all even lengths
$L\ge n_0$, applied to the spin-1 chain from $(2304,784)$.
Section~\ref{sm:sec:seeds} proves the periodic spin-1 Heisenberg seeds
from the one-site Poisson transfer operator $K_\beta$, symmetry-sector
trace recursions, polynomial spectral filters, and matrix-product trial
states, first at $(n_0,\beta_0)=(36,12)$ and then at $\beta_0=49/4$ to
obtain $\Delta_L>0.07290$ for all even $L\ge36$.
Section~\ref{sm:sec:models} gives the reflection-positive
factorizations, finite seeds, and gap bounds for the seven other spin
chains and ladders in Table~\ref{tab:models}, including the
unconditional Lean-kernel proof for the mixed-field Ising chain.
Section~\ref{sm:sec:blueprint} describes the Lean~4 formalization of the
general criterion, the model theorems, and the kernel-checked seed
arithmetic. Section~\ref{sm:sec:obc} extends the diagrammatic doubling
proof and Lean~4 formalization to open spin-1 Heisenberg chains with
endpoint fields, proving $\Delta_n>0.01556$ for all odd $n\ge1921$.

\tableofcontents

\subsection*{Graphical notation}
\label{sm:sec:notation}

Every diagram below is a tensor network built from the elements in the
following legend. Red marks space, blue marks imaginary time, teal marks
certified conclusions, grays mark spectra, weights, and boundary vectors, and ochre
marks regions ruled out.
\begin{center}
    \renewcommand{\arraystretch}{1.8}
    \begin{tabular}{@{}cl@{\qquad}cl@{}}
        \toprule
        \rule[-2.4ex]{0pt}{6.6ex}\physleg{}                       & physical leg: one site, through imaginary time                 &
        \tncolumn[ring]{\rho_n(\beta)}                            & closed physical loop: the trace $\Tr$                            \\
        \rule[-1.8ex]{0pt}{4.8ex}\virtleg{}                       & virtual leg: the bond between two sites                        &
        \tnchain[node=xvertex,ring]{,\cdots,}                     & closed virtual loop: the trace $\trv$                            \\
        \rule[-1.8ex]{0pt}{4.8ex}\tnchain[node=xvertex]{,}        & joined legs are contracted                                     &
        \tnchain[capl=\bar v,capr=v]{X_\beta}                     & boundary vectors on an open chain                                \\
        \rule[-2.4ex]{0pt}{6.0ex}\tnchain[node=mpo,legs]{W_\beta} & local MPO tensor (thermal, blue)                               &
        \tnchain[node=xvertex]{{}}                                & transfer vertex: $W_\beta$ with physical legs traced             \\
        \rule[-2.4ex]{0pt}{6.0ex}\tncolumn{\rho_n(\beta)}         & Gibbs state on $n$ sites (thermal, blue)                       &
        \tnchain{X_\beta}                                         & spatial transfer operator ($T_{\beta,M},K_\beta,X_\beta$, red)   \\
        \rule[-2.0ex]{0pt}{5.0ex}\Zt21                            & $Z_n(\beta)$: a torus of width $n$ and height $\beta$          &
        \Zt41,\ \Zt22                                             & length doubled, inverse temperature doubled                      \\
        \bottomrule
    \end{tabular}
\end{center}

\section{Finite Trotter slices and the periodic doubling proof}
\label{sm:sec:trotter}

\subsection{Finite Trotter slices and the two-site transfer matrix}
\label{sm:sec:transfer}

OpenAI's proof uses a continuous-time spatial integral
kernel~\cite{OpenAI2026Periodic}, whose trace identities require
infinite-dimensional path-space integration. To formalize the general
gap theorem in Lean with finite-dimensional linear algebra only, we
discretize imaginary time into $M$ brickwork slices. The spatial
transfer operator is then a finite Gram matrix
$\smT=\smF^{\mathsf T}\smF\succeq0$, and a fixed-$(L,\beta)$ matrix
product limit passes its moment inequalities to the exact partition
functions $Z_L(\beta)$.

Consider the periodic spin-1 Heisenberg Hamiltonian
\begin{equation}
    H_L=\sum_{j=0}^{L-1}h_{j,j+1},
    \qquad h=\sum_{\alpha=x,y,z}S^\alpha\otimes S^\alpha,
    \qquad Z_L(\beta)=\Tr e^{-\beta H_L},
    \label{sm:eq:model}
\end{equation}
with indices taken modulo $L$, exchange coupling equal to one, and
$\beta>0$. In the Cartesian basis $s,s'\in\{x,y,z\}$, set
\begin{equation}
    (S^\alpha)_{ss'}=-i\epsilon_{\alpha ss'},
    \qquad G_\alpha=iS^\alpha,
    \qquad -h=\sum_\alpha G_\alpha\otimes G_\alpha.
    \label{sm:eq:generators}
\end{equation}
Here $\epsilon_{\alpha ss'}$ is the Levi-Civita symbol, $\Tr$ denotes
the matrix trace, and $\smI$ denotes the identity on the relevant space.
Each $G_\alpha$ is real antisymmetric and has operator norm one.

For $L=2N$, $N\ge1$, and a positive integer $M$, let
$\tau=\beta/M$. Define the linearized brickwork slice and its trace by
\begin{align}
    S_{2N}(\tau)
                    &=\prod_{k=0}^{N-1}(\smI-\tau h_{2k,2k+1})
    \prod_{k=0}^{N-1}(\smI-\tau h_{2k+1,2k+2}), \notag\\
    \smZ{2N}(\beta) &=\Tr S_{2N}(\beta/M)^M.
    \label{sm:eq:slice}
\end{align}
Products run from left to right in increasing index, and the trace is
real because every factor is.
Equation~\eqref{sm:eq:slice} multiplies $M$ slices; the same network read
along space multiplies two-site columns,
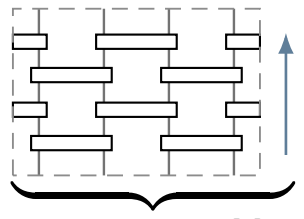
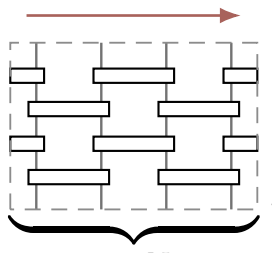
\begin{equation}
    \underbrace{\bricknetwork{1}}_{\Tr S_{2N}(\tau)^M}
    \;=\;\smZ{2N}(\beta)\;=\;
    \underbrace{\bricknetwork{2}}_{\Tr\smT^N}.
    \label{sm:eq:two-readings}
\end{equation}

The moment inequalities of Sec.~\ref{sm:sec:moments} need
$\smZ{2N}(\beta)=\Tr\smT^N$ with a positive semidefinite column. A
one-site column, like the continuous-time thermal-MPO kernel of
Ref.~\cite{Tang2021Continuous}, is real symmetric but need not be
positive; two sites mirrored across a bond give a Gram matrix instead.
Let $\Gamma=\{0,x,y,z\}$, with $A_0=\smI$ and
$A_\alpha=\sqrt{\tau}\,G_\alpha$. Each bond gate $\smI-\tau h$ splits
into matching matrices $A_\gamma$ on its two sites,
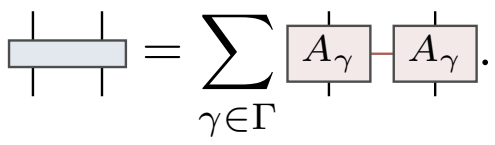
\begin{equation}
    \smI-\tau h=\sum_{\gamma\in\Gamma}A_\gamma\otimes A_\gamma,
    \qquad
    \splitgate=\sum_{\gamma\in\Gamma}\bondendsA.
    \label{sm:eq:bond}
\end{equation}
An imaginary-time bond sequence $p\in\smD=\Gamma^M$ lists one label per
slice. Define the single-site trace matrix, independent of $N$, by
\begin{equation}
    \smF_{pq}
    =\Tr\!\left[\prod_{t=0}^{M-1}(A_{p_t}A_{q_t})\right]
    =\sitetrace,
    \qquad p,q\in\smD,
    \label{sm:eq:F}
\end{equation}
drawn for $M=3$ with $A_{q_t}$ entering from the odd bond on the left
and $A_{p_t}$ leaving on the even bond to the right.

To contract the network in space, assign sequences $b_k$ to the even
bonds and $c_k$ to the odd bonds. Across the mirror of an even bond the
left site is the reflected right site, so the column is $\smF$
contracted with its own transpose:
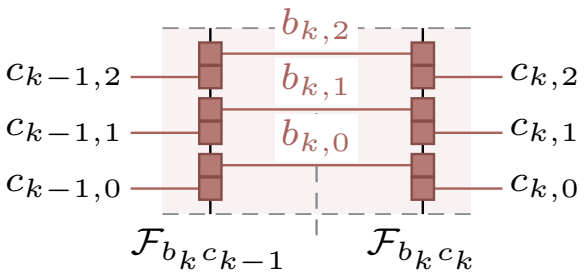
\begin{equation}
    \twositecolumn
    \qquad
    \sum_{b_k\in\smD}\smF_{b_kc_{k-1}}\smF_{b_kc_k}=(\smF^{\mathsf T}\smF)_{c_{k-1}c_k}.
    \label{sm:eq:column}
\end{equation}
Summing all bond labels and tracing every site gives%
~\leanmark{https://sirui-lu.com/CertifiedGaps/docs/CertifiedGaps/Ring/Transfer.html\#CertifiedGaps.RPRing.trace_trotterSlice_pow_eq_trace_transfer}
\begin{align}
    \smZ{2N}(\beta)
     &=\sum_{\substack{b_0,\ldots,b_{N-1}\in\smD\\
               c_0,\ldots,c_{N-1}\in\smD}}
    \prod_{k=0}^{N-1}
    \smF_{b_kc_{k-1}}\smF_{b_kc_k} \notag\\
     &=\sum_{c_0,\ldots,c_{N-1}\in\smD}
    \prod_{k=0}^{N-1}(\smF^{\mathsf T}\smF)_{c_{k-1}c_k}
    =\Tr(\smF^{\mathsf T}\smF)^N
    =\columnring,
    \label{sm:eq:factorization}
\end{align}
where $c_{-1}=c_{N-1}$: the middle equality is Eq.~\eqref{sm:eq:column}
at each even bond, and the last closes the $N$ columns into a ring.

Thus the spatial transfer matrix%
~\leanmark{https://sirui-lu.com/CertifiedGaps/docs/CertifiedGaps/Ring/Transfer.html\#CertifiedGaps.RPRing.exists_posSemidef_transfer}
\begin{equation}
    \smT=\smF^{\mathsf T}\smF=\smF^\dagger\smF\succeq0
    \label{sm:eq:positive}
\end{equation}
advances \emph{two sites}. Because $\smF$ is real,
$v^\dagger\smT v=\smnorm{\smF v}^2\ge0$ without $\smF$ itself being
symmetric or positive. The mirror across the even bond in
Eq.~\eqref{sm:eq:column} is the matching of real factors in
Eq.~\eqref{sm:eq:bond}, that is, reflection
positivity~\cite{Dyson1978Phase,Frohlich1978Phase,Jaffe2016Characterization}.

The models of Sec.~\ref{sm:sec:models} carry on-site terms and
multi-site cells, which Eq.~\eqref{sm:eq:F} omits. The same contraction
covers any real reflection-positive cell model, with lengths counting
cells. Take a finite-dimensional cell space with
real symmetric or antisymmetric bond matrices $C_\alpha$, a real
symmetric on-site matrix $o$, and a real symmetric involution $F$ in
the bond mirror $\scopebond$, such that%
~\leanmark{https://sirui-lu.com/CertifiedGaps/docs/CertifiedGaps/Ring/Model.html\#CertifiedGaps.RPRing}
\begin{equation}
    F^2=\smI,\qquad FoF=o,\qquad
    h=-\sum_\alpha C_\alpha\otimes FC_\alpha F.
    \label{sm:eq:cell}
\end{equation}
The transpose signs cancel between the two factors of each bond term,
so $h^{\mathsf T}=h$ and $H_L=\sum_j h_{j,j+1}+\sum_j o_j$ is
self-adjoint. Insert the on-site layer $\prod_i(\smI-\tau o_i)$ between
the even and odd bond layers and expand with
\[
    A_0=\smI,\qquad A_\alpha=\sqrt{\tau}\,C_\alpha,\qquad
    O=\smI-\tau o,\qquad
    \smI-\tau h=\sum_\gamma A_\gamma\otimes FA_\gamma F.
\]
Each even cell now carries the on-site factor $O$ (blue box) and the
two reflections $F$ (gray ticks) on its worldline, so
Eq.~\eqref{sm:eq:F} becomes
\begin{equation}
    \smF_{pq}=\Tr\prod_{t=0}^{M-1}(A_{p_t}OFA_{q_t}F)=\cellsitetrace=\oddcellsitetrace,
    \label{sm:eq:cell-F}
\end{equation}
with sequences over the enlarged bond-label alphabet. The last
equality is the odd cell, $FA_{p_t}FOA_{q_t}$ ($FO=OF$, $F^2=\smI$),
so both sublattices give the same $\smF$ and
Eqs.~\eqref{sm:eq:factorization} and~\eqref{sm:eq:positive} follow;
everything below applies with $L$ counting cells.

\subsection{Spatial and thermal purity inequalities at finite $M$}
\label{sm:sec:moments}

The recurrence of Sec.~\ref{sm:sec:asymmetric} needs three facts: the
spatial purity is at most one, squaring a distribution shrinks its
defect quadratically, and power sums interpolate between lengths.
Positivity of $\smT$ expresses $\smZ{L}(\beta)$ as an even moment sum of
nonnegative weights, which supplies all three at finite $M$.
At fixed $M,\beta$, let $\lambda_i\ge0$ be the eigenvalues of the
transfer matrix (\specbars[black!55]{1,.45,.3,.1}) and put
$\mu_i=\sqrt{\lambda_i}$.
For every even $L\ge2$, Eq.~\eqref{sm:eq:factorization} gives
\begin{equation}
    \smZ{L}(\beta)=\sum_i\mu_i^L.
    \label{sm:eq:moments}
\end{equation}

Doubling $n$ replaces the weights $\mu_i^n$ by their squares, so we
need the defect of a squared family in terms of the original defect.
For any finite nonnegative family $a_i$, let $S_k=\sum_i a_i^k$.
Then%
~\leanmark{https://sirui-lu.com/CertifiedGaps/docs/CertifiedGaps/PurityCriterion.html\#CertifiedGaps.sq_sub_sum_sq_finset}
\begin{equation}
    S_2\le S_1^2,\qquad
    2(S_2^2-S_4)\le(S_1^2-S_2)^2.
    \label{sm:eq:powers}
\end{equation}
Cut the square of side $S_1$ by the weights (\pairsquare): the
dark diagonal blocks have area $S_2$ and each symmetric pair of light
blocks has area $2a_ia_j$, so
\[
    S_1^2-S_2=\sum_{i<j}2a_ia_j,\qquad
    2(S_2^2-S_4)=\sum_{i<j}(2a_ia_j)^2\le\Big(\sum_{i<j}2a_ia_j\Big)^2,
\]
a sum of squares being at most the square of the sum.

If $\sum_i a_i=1$, the purity is $P=\sum_i a_i^2>0$.
Squaring and normalizing concentrates the distribution
($\specbars[black!55]{.5,.3,.2}\mapsto\specbars[black!55]{.66,.24,.1}$)
onto $a_i^2/P$, of purity $P'=S_4/P^2$. Thus%
~\leanmark{https://sirui-lu.com/CertifiedGaps/docs/CertifiedGaps/PurityCriterion.html\#CertifiedGaps.one_sub_squaredPurity_le_squaringDefect}
\begin{equation}
    1-P'\le f(1-P),\qquad
    f(z)=\frac{z^2}{2(1-z)^2}.
    \label{sm:eq:squaring}
\end{equation}
The map $f$ is nondecreasing on $0\le z<1$, since
$z/(1-z)$ is nonnegative and increasing there. For $0<z<1/2$,
$f(z)<z$ (\fbelowdiag), and squaring makes a small $1-P$ quadratically
smaller.

Section~\ref{sm:sec:gap} must also bound even lengths between dyadic
scales, which it does by comparing power sums of different orders.
For $0<u\le v$, nonnegative weights also satisfy%
~\leanmark{https://sirui-lu.com/CertifiedGaps/docs/CertifiedGaps/PurityCriterion.html\#CertifiedGaps.hasSum_pow_le_rpow}
\begin{equation}
    \sum_i\mu_i^v\le\left(\sum_i\mu_i^u\right)^{v/u}.
    \label{sm:eq:lp}
\end{equation}
If the sum on the right is zero, the claim is immediate.
Otherwise normalize $\mu_i^u$ to probabilities $a_i$; the power
$v/u\ge1$ lowers every bar
($\specbars[black!55]{.6,.3,.1}\mapsto\specbars[black!55]{.36,.09,.01}$),
so $\sum_i a_i^{v/u}\le\sum_i a_i=1$.

Substituting Eq.~\eqref{sm:eq:moments} gives, at fixed $\beta$, the
closed scalar inequalities%
~\leanmark{https://sirui-lu.com/CertifiedGaps/docs/CertifiedGaps/Ring/Gap.html\#CertifiedGaps.RPRing.hasMomentForm_trotterTrace}
\begin{align}
    \smZ{2n} &\le(\smZ{n})^2,
             &                             &\text{Eq.~\eqref{sm:eq:powers} with }a_i=\mu_i^n, \notag\\
    2((\smZ{2n})^2-\smZ{4n})
             &\le((\smZ{n})^2-\smZ{2n})^2,
             &                             &\text{Eq.~\eqref{sm:eq:powers} with }a_i=\mu_i^n, \notag\\
    \smZ{v}  &\le(\smZ{u})^{v/u},
             &                             &\text{Eq.~\eqref{sm:eq:lp}},
    \label{sm:eq:closed}
\end{align}
for even $n\ge2$ and even $v\ge u\ge2$ (temperature argument suppressed
only here). These three inequalities are all the recurrence uses;
Sec.~\ref{sm:sec:limit} passes them to the exact $Z_L(\beta)$.

\subsection{Taking the Trotter limit $M\to\infty$}
\label{sm:sec:limit}

To pass the finite-$M$ inequalities in Eq.~\eqref{sm:eq:closed} to the
exact partition functions, we need the matrix product limit
$S_L(\beta/M)^M\to e^{-\beta H_L}$. Each inequality involves
finitely many fixed pairs $(L,\beta)$, so the error may grow with $L$
and a telescoping estimate at fixed $L$ and $\beta$ suffices.
Use the operator norm, which is submultiplicative and satisfies
$\smnorm{\smI}=1$.
For $\smnorm{X},\smnorm{Y}\le K$, $K\ge1$, the difference of $k$-fold
products telescopes into $k$ chains that differ in one link, each of
norm at most $K^{k-1}\smnorm{X-Y}\le K^k\smnorm{X-Y}$:
\begin{equation}
    X^k-Y^k=\sum_{r=0}^{k-1}\;\telescopechain\,,
    \qquad
    \smnorm{X^k-Y^k}\le kK^k\smnorm{X-Y}.
    \label{sm:eq:telescoping}
\end{equation}

Suppose $M(X_M-\smI)\to A$, and set $Y_M=e^{A/M}$.
Then $M(Y_M-\smI)\to A$ and $Y_M^M=e^A$.
With $C=\smnorm{A}+1$, convergence implies
$\smnorm{X_M},\smnorm{Y_M}\le1+C/M$ for sufficiently large $M$, and
Eq.~\eqref{sm:eq:telescoping} gives the quantitative comparison
\begin{align}
    \smnorm{X_M^M-e^A}
     &\le (1+C/M)^M\smnorm{M(X_M-Y_M)} \notag\\
     &\le e^C\smnorm{M(X_M-Y_M)}\longrightarrow0.
    \label{sm:eq:error}
\end{align}
In particular, if $g(0)=\smI$ and $g'(0)=A$, then
$g(1/M)^M\to e^A$.

For fixed $L,\beta$, list the negated bond operators in the order
of Eq.~\eqref{sm:eq:slice}, including the negated on-site operators
when present, as $B_1,\ldots,B_k$.
Their sum is $-H_L$, so the polynomial
$g(t)=\prod_{i=1}^k(\smI+\beta tB_i)$ has $g'(0)=-\beta H_L$. Hence%
~\leanmark{https://sirui-lu.com/CertifiedGaps/docs/CertifiedGaps/Ring/Gap.html\#CertifiedGaps.RPRing.tendsto_trotterTrace}
\begin{equation}
    S_L(\beta/M)^M\longrightarrow e^{-\beta H_L},
    \qquad \smZ{L}(\beta)\longrightarrow Z_L(\beta).
    \label{sm:eq:limit}
\end{equation}
More explicitly, Taylor expansion gives
$g(1/M)-e^{-\beta H_L/M}=O(M^{-2})$.
Equation~\eqref{sm:eq:error} therefore bounds the operator error by
$e^C\kappa/M$ for a finite $\kappa$ depending on $L$ and $\beta$, with
$C=\smnorm{\beta H_L}+1$, once $M$ is large enough.
Trace convergence follows from
$\lvert\Tr A\rvert\le3^L\smnorm{A}$, with $3^L$ the dimension of the
spin-1 ring.

Taking $M\to\infty$ at each fixed $(L,\beta)$ occurring in
Eq.~\eqref{sm:eq:closed} passes all three inequalities to the exact
partition functions $Z_L(\beta)$%
~\leanmark{https://sirui-lu.com/CertifiedGaps/docs/CertifiedGaps/Ring/Gap.html\#CertifiedGaps.RPRing.spatialMomentBounds}:
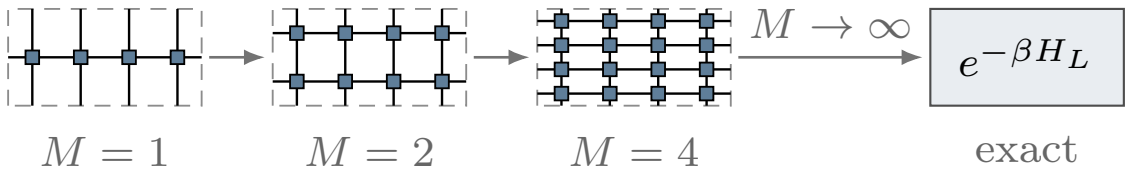
\begin{proofdiagram}
    \limitfirst
\end{proofdiagram}
The recurrence therefore starts from exact partition functions, and no
Trotter error enters any later step.

\subsection{Torus coupling identities and the asymmetric recurrence}
\label{sm:sec:asymmetric}

Once the spatial moment inequalities hold for the exact partition
functions $Z_L(\beta)$, we couple the spatial and thermal purities on a
torus. A torus of width $n$ and height $\beta$ (\Zt21) represents
$Z_n(\beta)$; the spatial purity joins two copies side by side and the
thermal purity stacks them in imaginary time:
\begin{equation}
    \Ps(n,\beta)=\frac{Z_{2n}(\beta)}{Z_n(\beta)^2}=\frac{\Zt41}{\Ztsq21},
    \qquad
    \Pt(L,\beta)=\frac{Z_L(2\beta)}{Z_L(\beta)^2}=\frac{\Zt22}{\Ztsq21}.
    \label{sm:eq:purities}
\end{equation}
All partition functions are strictly positive, and the thermal purity
is the purity of the Gibbs state $e^{-\beta H_L}/Z_L(\beta)$.
The limiting spatial inequalities and the Boltzmann power sums give
$0<\Ps,\Pt\le1$ at even lengths and
\begin{align}
    1-\Ps(2n,\beta) &\le f(1-\Ps(n,\beta)),
                    &                       &\text{second line of Eq.~\eqref{sm:eq:closed}, divided by }
    2Z_{2n}(\beta)^2, \notag\\
    1-\Pt(L,2\beta) &\le f(1-\Pt(L,\beta)),
                    &                       &\text{Eq.~\eqref{sm:eq:squaring} on the Boltzmann weights}.
    \label{sm:eq:two-squares}
\end{align}

Equation~\eqref{sm:eq:two-squares} doubles $n$ at fixed $\beta$ or
$\beta$ at fixed $L$, but one recurrence step must double both; an
exact identity between tori supplies the missing move. Gluing four
copies into the doubled torus in either order gives the exact
cancellation
\begin{equation}
    \Ps(n,2\beta)\Pt(n,\beta)^2
    =\frac{Z_{2n}(2\beta)}{Z_n(\beta)^4}
    =\frac{\Zt42}{\Ztsq[4]21}
    =\Ps(n,\beta)^2\Pt(2n,\beta).
    \label{sm:eq:torus}
\end{equation}
Solving for one factor and repeating at length $2n$ (substitute
Eq.~\eqref{sm:eq:purities}) yields%
~\leanmark{https://sirui-lu.com/CertifiedGaps/docs/CertifiedGaps/PurityCriterion.html\#CertifiedGaps.spatialPurity_two_mul}~\leanmark{https://sirui-lu.com/CertifiedGaps/docs/CertifiedGaps/PurityCriterion.html\#CertifiedGaps.thermalPurity_four_mul}
\begin{align}
    \Ps(n,2\beta)
     &=\frac{\Ps(n,\beta)^2\Pt(2n,\beta)}{\Pt(n,\beta)^2}, \notag\\
    \Pt(4n,\beta)
     &=\frac{\Pt(2n,\beta)^2\Ps(2n,2\beta)}{\Ps(2n,\beta)^2}.
    \label{sm:eq:coupling}
\end{align}
Dropping each denominator, a purity at most one, turns the identity
into a lower bound on the purity the recurrence lacks.

Let $n\ge2$ be even, and let $0\le p,q<1$ bound the seed purities:
\begin{equation}
    \Ps(n,\beta)\ge1-p,\qquad
    \Pt(2n,\beta)\ge1-q.
    \label{sm:eq:seed-pair}
\end{equation}
To double both $n$ and $\beta$ (\torusseq), write
$z(a,b)=1-(1-a)^2(1-b)$. The asymmetric update is%
~\leanmark{https://sirui-lu.com/CertifiedGaps/docs/CertifiedGaps/PurityCriterion.html\#CertifiedGaps.doubling_step}
\begin{equation}
    p'=f(z(p,q)),\qquad q'=f(z(q,p')).
    \label{sm:eq:update}
\end{equation}
Assuming $p'<1$, one doubling step advances $(p,q)$ to $(p',q')$ in four
moves across space and imaginary time (thin arrows mark the second
input at each torus step):
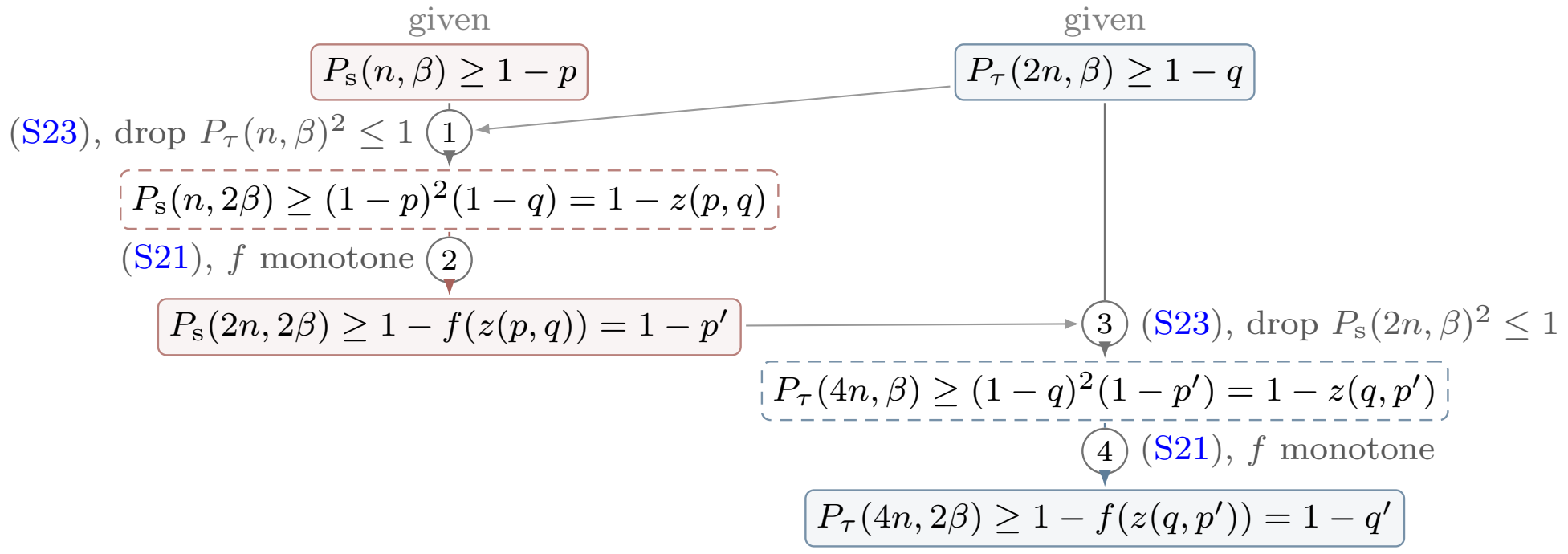
\begin{proofdiagram}
    \asymsteps
\end{proofdiagram}
Move~3 substitutes the updated spatial bound $p'$, not $p$, into the
thermal step, so a small $p'$ pulls $q'$ down even when $q$ starts above
the symmetric fixed point $\varepsilon_*$ of Sec.~\ref{sm:sec:orbit}.
Any upper bound below one may be substituted for $p'$ in Move~3.

\subsection{Six rational steps from the seed to $[0,1/100]^2$}
\label{sm:sec:orbit}

The analytic tail estimate in Sec.~\ref{sm:sec:gap} requires both
$p$ and $q$ to be at most $1/100$ (there one step costs at most
$5\varepsilon^2$). The seed bounds of Sec.~\ref{sm:sec:seeds} at
$(n_0,\beta_0)=(36,49/4)$,
\begin{equation}
    \Ps(36,49/4)\ge1-\frac{47916}{10^6},
    \qquad
    \Pt(72,49/4)\ge1-\frac{181521}{10^6},
    \label{sm:eq:initial}
\end{equation}
have larger $(p_0,q_0)$. Moreover $q_0=0.1815$ exceeds the fixed point
$\varepsilon_*\simeq0.1273$ of the symmetric map
$\varepsilon\mapsto f(1-(1-\varepsilon)^3)$, so a symmetric recurrence
cannot start from this seed. We therefore iterate the asymmetric
update~\eqref{sm:eq:update} six times, to
$(n_6,\beta_6)=(36\cdot2^6,\tfrac{49}{4}2^6)=(2304,784)$.
Iterating $f$ exactly squares the denominators at every step, so the
exact orbit is not a usable certificate. Following
Ref.~\cite{OpenAI2026Periodic}, each update is rounded upward to seven
decimal places: with $D=10^7$, the stored integer $a_{j+1}=Dp_{j+1}$
is the ceiling of the exact value $Df$, \ceilcheck, and likewise for
$q$. Rounding up keeps every bound valid, and the kernel checks the
twelve rational inequalities, each exact update at most its stored
entry.
Table~\ref{sm:tab:orbit} lists the resulting sequence, drawn here on a
logarithmic scale:
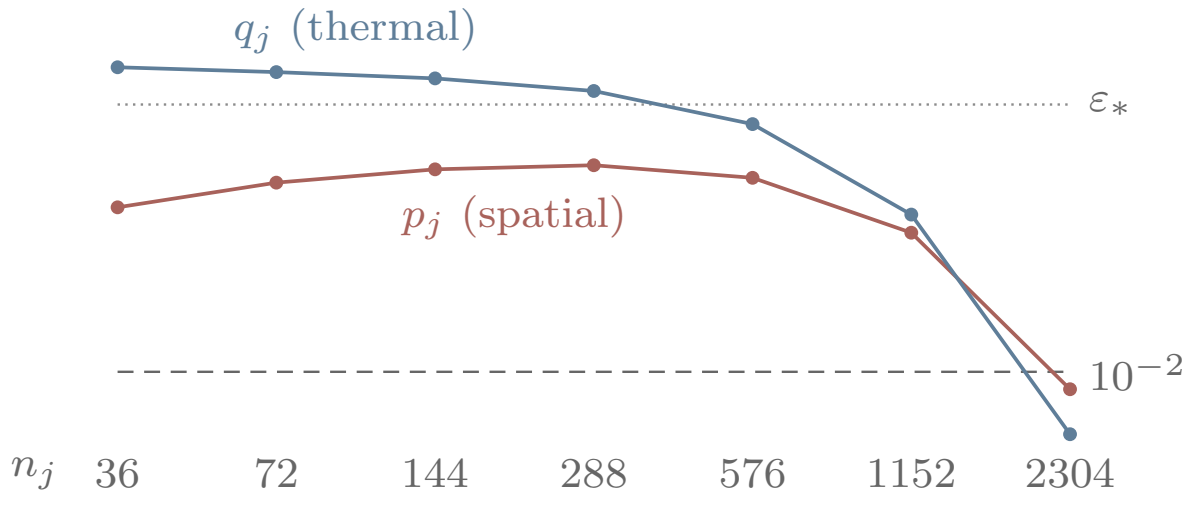
\begin{proofdiagram}
    \defectladder
\end{proofdiagram}
The spatial bound $p_j$ first rises, to $0.0714$ at $j=3$, because its
input $z(p,q)\ge q$ stays large while $q$ does; once $q$ drops below
$0.15$, both sequences fall rapidly. The thermal bound $q_j$ falls from
the first step, although it starts above $\varepsilon_*$ (dotted): the
updated $p'$ in Move~3 keeps $f(z(q,p'))<q$ at every step.

\begin{table}[ht]
    \centering
    \caption{Six steps of the asymmetric recurrence~\eqref{sm:eq:update}
        along the dyadic scales $n_j=36\,2^j$ and $\beta_j=(49/4)2^j$, starting
        from the seed bounds~\eqref{sm:eq:initial} at $j=0$. Column $p_j$
        bounds $1-\Ps(n_j,\beta_j)$ and column $q_j$ bounds
        $1-\Pt(2n_j,\beta_j)$, each rounded upward to seven decimal places.}
    \label{sm:tab:orbit}
    \begin{tabular}{rrrrr}
        \toprule
        $j$ & $n_j$ & $\beta_j$ & $p_j$     & $q_j$     \\
        \midrule
        0   & 36    & $49/4$    & 0.0479160 & 0.1815210 \\
        1   & 72    & $49/2$    & 0.0605003 & 0.1733838 \\
        2   & 144   & 49        & 0.0686630 & 0.1632461 \\
        3   & 288   & 98        & 0.0713693 & 0.1447318 \\
        4   & 576   & 196       & 0.0633146 & 0.1055663 \\
        5   & 1152  & 392       & 0.0376144 & 0.0446518 \\
        6   & 2304  & 784       & 0.0084709 & 0.0055149 \\
        \bottomrule
    \end{tabular}
\end{table}

At $j=6$, Table~\ref{sm:tab:orbit} gives%
~\leanmark{https://sirui-lu.com/CertifiedGaps/docs/CertifiedGaps/Heisenberg/Main.html\#CertifiedGaps.SpinOneHeisenberg.seeds2304_of_seeds36}
\begin{align}
    \Ps(2304,784) &\ge1-0.0084709>0.99,\notag\\
    \Pt(4608,784) &\ge1-0.0055149>0.99,
    \label{sm:eq:terminal}
\end{align}
with both $p_6$ and $q_6$ below $1/100$, the entry condition of
Sec.~\ref{sm:sec:gap}.

\subsection{Quadratic tail, spatial interpolation, and the uniform gap}
\label{sm:sec:gap}

Rounded rational steps cannot continue to arbitrarily large $j$, so the
remaining steps need a closed-form bound. With $p,q\le\varepsilon$ the
input $z(p,q)$ is at most $1-(1-\varepsilon)^3$ and $f$ is quadratic
near zero, so one scalar recurrence bounds every further step.
For $0\le\varepsilon\le1/100$%
~\leanmark{https://sirui-lu.com/CertifiedGaps/docs/CertifiedGaps/PurityCriterion.html\#CertifiedGaps.squaringDefect_le},
\begin{equation}
    f(1-(1-\varepsilon)^3)
    \le\frac{9\varepsilon^2}{2(1-3\varepsilon)^2}
    \le5\varepsilon^2\le\varepsilon.
    \label{sm:eq:scalar}
\end{equation}
The first inequality uses $1-(1-\varepsilon)^3\le3\varepsilon$ and the
monotonicity of $f$, the second uses $1-3\varepsilon\ge97/100$ so that
$9/[2(97/100)^2]<5$, and the third uses $5\varepsilon\le1/20\le1$.
With $p,q\le\varepsilon$, the spatial update is
therefore at most $5\varepsilon^2\le\varepsilon$; the thermal coupling
input is then again at most $1-(1-\varepsilon)^3$, so the thermal
update is at most $5\varepsilon^2$ as well.

Starting from Eq.~\eqref{sm:eq:terminal}, reset the scale index and put
\begin{equation}
    n_j=2304\,2^j,\qquad
    \beta_j=784\,2^j,\qquad
    \varepsilon_j=\frac{20^{-2^j}}5,\qquad j\ge0.
    \label{sm:eq:tail-scales}
\end{equation}
The step $\varepsilon\mapsto5\varepsilon^2$ squares $5\varepsilon$, and
$5\varepsilon_0=1/20$ at the threshold: hence the base $20$, and
$\log_{10}\varepsilon_j$ falls doubly exponentially from $-2$ at $j=0$
to $-21.5$ at $j=4$ (\trendline{-22}{0}{-2,-3.3,-5.9,-11.1,-21.5}).
Induction using Eq.~\eqref{sm:eq:scalar} proves%
~\leanmark{https://sirui-lu.com/CertifiedGaps/docs/CertifiedGaps/PurityCriterion.html\#CertifiedGaps.purities_ge_defect}
\begin{equation}
    \Ps(n_j,\beta_j)\ge1-\varepsilon_j,\qquad
    \Pt(2n_j,\beta_j)\ge1-\varepsilon_j.
    \label{sm:eq:tail}
\end{equation}

Equation~\eqref{sm:eq:tail} holds only at the dyadic scales, and an
even $L$ between them has no torus bound of its own. To cover every
even length $L\ge2304$, choose $j$ with $n_j\le L\le2n_j$. The two
known tori supply the missing bound, from above at $\beta_j$ and
from below at $2\beta_j$:
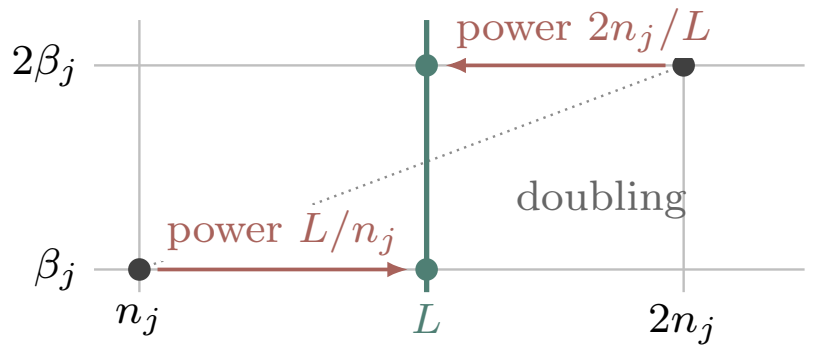
\begin{proofdiagram}
    \dyadicinterval
\end{proofdiagram}
The power-sum inequality in Eq.~\eqref{sm:eq:closed} applies twice,
with the powers $v/u$ drawn above. At $\beta_j$ take $(u,v)=(n_j,L)$; at
$2\beta_j$ take $(u,v)=(L,2n_j)$ and solve
for the unknown torus:
\begin{align}
    Z_L(\beta_j)^2 &\le Z_{n_j}(\beta_j)^{2L/n_j}, \notag\\
    Z_L(2\beta_j)  &\ge Z_{2n_j}(2\beta_j)^{L/(2n_j)}.
    \label{sm:eq:interpolation-input}
\end{align}
Set $r=L/(2n_j)\in[1/2,1]$. Let $x$ be the four-copy ratio of
Eq.~\eqref{sm:eq:torus} at scale $j$,
\[
    x=\frac{Z_{2n_j}(2\beta_j)}{Z_{n_j}(\beta_j)^4}
    =\frac{\Zt42}{\Ztsq[4]21}
    =\Pt(2n_j,\beta_j)\Ps(n_j,\beta_j)^2.
\]
By Eq.~\eqref{sm:eq:tail}, $(1-\varepsilon_j)^3\le x\le1$, so dividing
the second inequality in Eq.~\eqref{sm:eq:interpolation-input} by the
first yields
\begin{equation}
    \Pt(L,\beta_j)\ge x^r\ge x
    \ge(1-\varepsilon_j)^3\ge1-3\varepsilon_j>\frac12.
    \label{sm:eq:interpolation}
\end{equation}
Every even $L\ge2304$ thus has thermal purity above $1/2$ at a
temperature $\beta_j$ of its own scale.

Finally, we convert this thermal purity bound into a lower bound on
the spectral gap. A purity above $1/2$ at one finite temperature bounds
the largest Gibbs weight, and the ratio of the two largest weights is
$e^{-\beta_j\Delta_L}$. Bounding that ratio strictly below $20^{-2^j}$
gives a strict gap independent of $L$. List the energies with
multiplicity,
$E_0(L)\le E_1(L)\le\cdots$. Each level $m$ carries the Gibbs weight
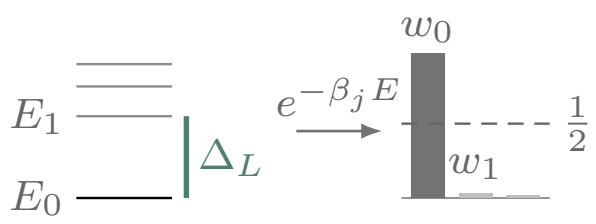
\[
    w_m=\frac{e^{-\beta_jE_m(L)}}{Z_L(\beta_j)},\qquad\levelweights
\]
The purity is at most the largest weight, $\sum_m w_m^2\le w_0$, so
Eq.~\eqref{sm:eq:interpolation} gives $w_0>1/2$. The ground state is
therefore unique, with $w_1\le1-w_0$. With $\Delta_L$ the gap drawn
above, we obtain
\begin{align}
    e^{-\beta_j\Delta_L}
     &=\frac{w_1}{w_0}
    \le\frac{3\varepsilon_j}{1-3\varepsilon_j} \notag\\
     &\le\frac{60}{97}\,20^{-2^j}
    <20^{-2^j},
    \label{sm:eq:strict-ratio}
\end{align}
where $3\varepsilon_j=(3/5)20^{-2^j}$ and
$1-3\varepsilon_j\ge97/100$ give the prefactor $(3/5)(100/97)=60/97$.
Taking logarithms and dividing by $\beta_j=784\,2^j$ at this finite,
length-dependent temperature gives%
~\leanmark{https://sirui-lu.com/CertifiedGaps/docs/CertifiedGaps/Heisenberg/Main.html\#CertifiedGaps.SpinOneHeisenberg.spin_one_heisenberg_gap}
\begin{equation}
    \boxed{
        \Delta_L>\frac{\log20}{784}
        \quad\text{for every even }L\ge2304
    }.
    \label{sm:eq:gap}
\end{equation}
The strict inequality comes from the finite prefactor $60/97<1$,
without taking a fixed-length limit $\beta\to\infty$. On the $j$th dyadic
interval, Eq.~\eqref{sm:eq:strict-ratio} gives the larger bound
$\tfrac{1}{\beta_j}\log\tfrac{1-3\varepsilon_j}{3\varepsilon_j}$, which decreases
from $4.43\times10^{-3}$ at $j=0$ to $3.86\times10^{-3}$ at $j=4$
toward the uniform bound $\tfrac{\log20}{784}\approx3.82\times10^{-3}$
(\trendline[3.821]{3.7}{4.5}{4.434,4.146,3.984,3.903,3.862}, dashed).

\section{Seed bounds and gap estimates for the spin-1 chain}
\label{sm:sec:seeds}

The unformalized computer-assisted proof of Ref.~\cite{OpenAI2026Periodic}
establishes two periodic spin-1 gap bounds from two sets of finite
inputs (in this section we write $b_0=\beta_0$ and $b=rb_0$ for the seed
inverse temperatures). First, short-ring trace enclosures on
$4\le n\le12$ sites at $b_0=49/4$ together with a $72$-site
matrix-product trial state certify the seed bounds at
$(n_0,b_0)=(36,49/4)$ (Lemma~\ref{lem:arxiv-seed36})%
~\leanmark{https://sirui-lu.com/CertifiedGaps/docs/CertifiedGaps/Heisenberg/Seed36.html\#CertifiedGaps.SpinOneHeisenberg.seed36_of_table_mps},
\begin{align}
    1-\Ps(36,49/4) &\le\frac{47916}{10^6},\notag\\
    1-\Pt(72,49/4) &\le\frac{181521}{10^6},
    \label{cert:eq:base}
\end{align}
from which six asymmetric doubling steps reach $(2304,784)$ and prove
$\Delta_L>\tfrac{\log20}{784}\simeq0.003821$ for every even $L\ge2304$
(Sec.~\ref{cert:sec:base}). Second (Sec.~2.2 of
Ref.~\cite{OpenAI2026Periodic}), a separate table of nine traces on
$n\in\{6,8,10\}$ sites at $b_0'=21/2$ combined with a $120$-site trial
state and a temperature lift $r=5/2$ to $b=105/4$ certifies
$1-\Ps(60,105/4),\,1-\Pt(120,105/4)<1/8$, proving
$\Delta_L>\frac{4}{105}\log\frac{80}{79}\simeq0.000479$ for every even
$L\ge60$.

We first independently reproduced and verified all of OpenAI's finite
trace and matrix-product trial contractions in exact arithmetic.
Section~\ref{cert:sec:base} reviews the derivation of
Eq.~\eqref{cert:eq:base} at $(36,49/4)$. Section~\ref{cert:sec:lifting}
then adapts the temperature-lifting and threshold-refinement mechanism
from Sec.~2.2 of Ref.~\cite{OpenAI2026Periodic} to the stronger
$b_0=49/4$ ($4\le n\le12$) trace table together with the $120$-site
trial state at $r=17/8$ ($b=833/32$), tighter twelve-decimal polynomial
caps $U,V$, and per-interval gap evaluation across ten doubling steps
plus $[36,72]$ from $(36,b_0)$. From the same verified finite inputs,
this proves a unique ground state and%
~\leanmark{https://sirui-lu.com/CertifiedGaps/docs/Leaderboard/Records/SpinOneHeisenberg.html\#Leaderboard.Records.SpinOneHeisenberg.temperature_lifting}
\begin{equation}
    \Delta_L>0.072901771
    \quad\text{for every even }L\ge36.
    \label{cert:eq:improved}
\end{equation}
Following Ref.~\cite{OpenAI2026Periodic}, intermediate bounds are
rounded outward to rationals so that every step reduces to exact
rational comparisons. In Lean~4, we formalize the deduction from the short-ring trace table
to Eq.~\eqref{cert:eq:base} and the doubling chains behind
Eqs.~\eqref{cert:eq:base} and~\eqref{cert:eq:improved}; the lifted seed
bounds of Table~\ref{cert:tab:lifted} and the twelve-decimal bridge
seeds enter the Lean theorem for Eq.~\eqref{cert:eq:improved} as
hypotheses, verified in exact rational arithmetic outside Lean. We further
verify the $72$- and $120$-site integer matrix-product trial contractions
and the $n=4$ trace totals directly in the Lean kernel, reducing the
remaining hypotheses to the printed integer trace totals for
$5\le n\le12$%
~\leanmark{https://sirui-lu.com/CertifiedGaps/docs/SeedKernel/Heisenberg/Integers.html\#CertifiedGaps.SpinOneHeisenberg.gap_of_hornerTotalsLarge}.

\subsection{Seed purities at length $36$ and inverse temperature $49/4$}
\label{cert:sec:base}

The seed inverse temperature $b_0=49/4$ is the one chosen in
Ref.~\cite{OpenAI2026Periodic}. At this scale the two ratios
\begin{equation}
    \textcolor{mpored}{\Ps(36,b_0)}=\frac{Z_{72}(b_0)}{Z_{36}(b_0)^2}
    =\frac{\Zt41}{\Ztsq21},\qquad
    \textcolor{mpoblue}{\Pt(72,b_0)}=\frac{Z_{72}(2b_0)}{Z_{72}(b_0)^2}
    =\frac{\Zt42}{\Ztsq41}
    \label{cert:eq:seed-ratios}
\end{equation}
need lower bounds on their numerators and upper bounds on their
denominators. A $72$-site trial state and exact traces on rings of $4$
to $12$ sites supply them:
\begin{center}
    \setlength{\tabcolsep}{6pt}
    \begin{tabular}{@{}clll@{}}
        \toprule
                & Partition function & Bound & Supplied by                                                                \\
        \midrule
        $\Zt41$ & $Z_{72}(b_0)$      & lower & trial state, through $|\lambda_0|>\ell$ (Sec.~\ref{cert:sec:base-defects}) \\
        $\Zt21$ & $Z_{36}(b_0)$      & upper & traces, through the bounds $U,V$ (Sec.~\ref{cert:sec:filters})             \\
        $\Zt42$ & $Z_{72}(2b_0)$     & lower & trial state (Sec.~\ref{cert:sec:inputs})                                   \\
        $\Zt41$ & $Z_{72}(b_0)$      & upper & traces, through the bounds $U,V$                                           \\
        \bottomrule
    \end{tabular}
\end{center}

\subsubsection{Shifted Hamiltonian, twisted rings, and the spatial transfer operator}
\label{cert:sec:transfer}

The seed ratios~\eqref{cert:eq:seed-ratios} need partition functions at
$36$ and $72$ sites, while certified traces stop at $12$ sites. Three
objects connect them: a Hamiltonian shift that turns the trial bound into
$Z>1$, twisted rings that label spin-rotation sectors, and a transfer
operator whose eigenvalues govern every ring length.
We consider the periodic spin-1 Heisenberg Hamiltonian
\begin{equation}
    H_n=\sum_{j=1}^n\bm S_j\cdot\bm S_{j+1},
    \qquad \bm S_j^2=2I,\qquad \bm S_{n+1}=\bm S_1,
    \label{cert:eq:hamiltonian}
\end{equation}
with unit exchange coupling. Lengths count sites, and
$\Delta_L=E_1(L)-E_0(L)$, with eigenvalues ordered with multiplicity.
We shift the Hamiltonian by $\kappa$ per site, a rational with $-\kappa$
just above the trial energy per site of Sec.~\ref{cert:sec:inputs}. The
shifted ground energy is then negative and the leading transfer
eigenvalue sits near one:
\begin{equation}
    \kappa=\frac{700741}{500000}\simeq1.401482,\qquad
    \widehat H_n=H_n+\kappa nI,\qquad
    Z_n(\beta)=\Tr e^{-\beta\widehat H_n}.
    \label{cert:eq:shift}
\end{equation}
We drop the hats of Eq.~\eqref{seed:shift}: in this section $Z_n$ and
$K_\beta$ denote the shifted quantities.
This extensive shift cancels in both purities%
~\leanmark{https://sirui-lu.com/CertifiedGaps/docs/CertifiedGaps/Seed/Limits.html\#CertifiedGaps.purities_of_shifted}:
\begin{equation}
    \Ps(n,\beta)=\frac{Z_{2n}(\beta)}{Z_n(\beta)^2},
    \qquad
    \Pt(n,\beta)=\frac{Z_n(2\beta)}{Z_n(\beta)^2}.
    \label{cert:eq:purities}
\end{equation}

To separate the spatial transfer spectrum by spin-rotation symmetry, we
also use rings whose closing bond is twisted by a spin rotation,

\seedring{1}, alongside the plain ring \seedring{0}. For an on-site spin
rotation $u_g$ representing $g$, we define
\begin{align}
    H_n(g)       &=\sum_{j=1}^{n-1}h_{j,j+1}+\sum_{\alpha=x,y,z}
    S_n^\alpha(u_gS^\alpha u_g^\dagger)_1,\notag\\
    Z_n(\beta,g) &=\Tr e^{-\beta(H_n(g)+\kappa nI)},
    \label{cert:eq:twisted-hamiltonian}
\end{align}
where $h_{j,j+1}=\bm S_j\cdot\bm S_{j+1}$.

The continuous-time construction of Ref.~\cite{OpenAI2026Periodic}
provides a self-adjoint one-site spatial transfer operator $K_\beta$ for
the shifted Hamiltonian and a virtual symmetry operator $U_g$ for the
twist. The twisted trace is then a ring of $n$ transfer vertices closed
through $U_g$:
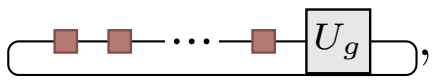
\begin{equation}
    Z_n(\beta,g)=\Tr(K_\beta^nU_g)
    =\tnchain[node=xvertex,last=sym,ring]{,,\cdots,,U_g},\qquad n\ge4.
    \label{cert:eq:transfer-trace}
\end{equation}
At $g=1$ the trace is the $n$th power sum of the signed spectrum

\signedspectrum{} of $K_\beta$. Its two signs nearly cancel at odd $n$,
so the odd traces in Table~\ref{cert:tab:traces} are small (\tracebars{}
on a logarithmic scale) and both parities are needed. Because the same
eigenvalues govern every length $n\ge4$, trace bounds on $4\le n\le12$
constrain the power sums at $n_0$ and $2n_0$.

\subsubsection{Short-ring traces and the matrix-product trial state}
\label{cert:sec:inputs}

Two numerical inputs from Ref.~\cite{OpenAI2026Periodic} supply the
short-ring moments and ground-state energy bound at $b_0=49/4$. The
first input is the table of shifted traces $Z_n(b_0)$ and $Z_n(b_0,P)$
for $4\le n\le12$, where $P$ is a $\pi$ rotation about a Cartesian axis. Table~\ref{cert:tab:traces} lists the
rational centers, each certified within the absolute error
\begin{equation}
    \epsilon_Z=3\times10^{-5},
    \label{cert:eq:trace-error}
\end{equation}
the error bound of the integer recursion described next.

\begin{table}[t]
    \caption{Shifted short-ring traces $Z_n(b_0)=\Tr e^{-b_0(H_n+\kappa nI)}$
        and twisted traces $Z_n(b_0,P)=\Tr e^{-b_0(H_n(P)+\kappa nI)}$ at
        $b_0=49/4$ and $\kappa=700741/500000$ from Ref.~\cite{OpenAI2026Periodic},
        where $P$ is a $\pi$ spin rotation on the closing bond. Each tabulated
        rational center is certified within absolute error
        $\epsilon_Z=3\times10^{-5}$ and supplies the sector moments $J_n,N_n$ in
        Eq.~\eqref{cert:eq:sector-inversion}.}
    \label{cert:tab:traces}
    \begin{tabular}{@{}r rr r rr@{}}
        \toprule
        $n$ & $Z_n(b_0)$ center & $Z_n(b_0,P)$ center &
        $n$ & $Z_n(b_0)$ center & $Z_n(b_0,P)$ center                            \\
        \midrule
        4   & 124.885379        & 0.065757            & 9  & 0.195455 & 1.514846 \\
        5   & 0.002925          & 2.960742            & 10 & 2.654823 & 0.675419 \\
        6   & 12.870614         & 0.286599            & 11 & 0.377629 & 1.314544 \\
        7   & 0.054069          & 1.920414            & 12 & 1.900547 & 0.787405 \\
        8   & 4.639283          & 0.510140            &    &          &          \\
        \bottomrule
    \end{tabular}
\end{table}

The exact integer recursion of Ref.~\cite{OpenAI2026Periodic} writes
each trace as $\Tr A^2$ with $A$ real symmetric and folds it into
squared columns,
\begin{equation}
    Z_n(b_0,g)=\halfgibbstrace,\qquad A=e^{-b_0(H_n(g)+\kappa nI)/2},
    \label{cert:eq:half-gibbs-fold}
\end{equation}
so the recursion runs at $b_0/2$, one truncated Poisson polynomial of a
rescaled Hamiltonian of norm at most one per column, in scaled integer
arithmetic; the rounding errors add column by column over symmetry
representatives with their multiplicities, within total error
$\epsilon_Z$.

The numerators need lower bounds at $72$ sites, which no certified trace
reaches; a variational state supplies them once its energy per site lies
below $-\kappa$.
The second input is a periodic matrix-product trial state $\psi_N$ of
bond dimension $26$ with integer tensors $A$, used at $N=72$ and at
$N=120$. Its norm and its energy, with $h$ on one of the $N$ bonds, are
exact integer contractions,
\begin{align}
    \|\psi_N\|^2 &=\seedmps0,\notag\\
    \langle\psi_N,H_N\psi_N\rangle
                 &=N\,\seedmps1.
    \label{cert:eq:mps-contractions}
\end{align}
Exact integer arithmetic (verified in the Lean kernel for both
$N=72$%
~\leanmark{https://sirui-lu.com/CertifiedGaps/docs/SeedKernel/Heisenberg/Seed36.html\#CertifiedGaps.SpinOneHeisenberg.one_lt_shiftedZ_seventytwo}
and $N=120$%
~\leanmark{https://sirui-lu.com/CertifiedGaps/docs/SeedKernel/Heisenberg/Seed60.html\#CertifiedGaps.SpinOneHeisenberg.exists_spectrum_lt_onetwenty})
verifies that for $N\in\{72,120\}$,
\begin{equation}
    \frac{\langle\psi_N,H_N\psi_N\rangle}{N\,\|\psi_N\|^2}<-\kappa,
    \label{cert:eq:trial-input}
\end{equation}
so the shifted ground energy $E_0(N)+\kappa N$ is negative by the
variational principle. Retaining only the ground-state
Boltzmann weight gives the lower bounds%
~\leanmark{https://sirui-lu.com/CertifiedGaps/docs/CertifiedGaps/Ring/Variational.html\#CertifiedGaps.exp_lt_gibbsTrace_of_re_dotProduct_lt}
\begin{equation}
    Z_{72}(\beta)>1,\qquad Z_{120}(\beta)>1
    \quad(\beta>0).
    \label{cert:eq:trial-consequence}
\end{equation}
In the base seed, $Z_{72}(2b_0)>1$ bounds the thermal numerator directly,
whereas $Z_{72}(b_0)>1$ forces a large leading spatial eigenvalue.

\subsubsection{Spin-rotation sectors and polynomial eigenvalue bounds}
\label{cert:sec:filters}

Bounding all eigenvalues of $K_{b_0}$ from the untwisted traces alone
would give the subleading eigenvalues the same bound $U>1$ as the
leading one, leaving the $36$-site remainder too large. Instead, we split the spectrum of $K_{b_0}$ into spin-rotation
sectors so that the subleading sectors carry a strictly smaller bound
$V<1$. The four virtual $\pi$ rotations $D_2=\{1,P_x,P_y,P_z\}$ commute
with $K_{b_0}$,
\begin{equation}
    \tnchain[last=sym]{K_{b_0},U_g}
    =\tnchain[node=sym,last=xop]{U_g,K_{b_0}},\qquad g\in D_2,
    \label{cert:eq:twist-diagram}
\end{equation}
and define the character projectors
\begin{equation}
    \Pi_\chi=\frac14\sum_{g\in D_2}\chi(g)U_g,\qquad
    \Tr(K_{b_0}^n\Pi_\chi)
    =\tnchain[node=xvertex,last=sym,ring]{,,\cdots,,\Pi_\chi}.
    \label{cert:eq:projectors}
\end{equation}
Cyclic permutation of the coordinate axes conjugates the three nontrivial
characters into one another, so their three sectors share a common
eigenvalue list $N$, while the $D_2$-invariant sector has eigenvalue
list $J$. Writing $J_n=\sum_{\lambda\in J}\lambda^n$ and
$N_n=\sum_{\lambda\in N}\lambda^n$ for their $n$th power sums at $b_0$,
the character table of the plain and $P_z$-twisted rings,


\begin{proofdiagram}
    \sectortable
\end{proofdiagram}
inverts to the sector moments:
\begin{align}
    J_n &=\frac{Z_n+3Z_n(P)}4, &
    N_n &=\frac{Z_n-Z_n(P)}4.
    \label{cert:eq:sector-inversion}
\end{align}
The inversion maps the trace error~\eqref{cert:eq:trace-error} to an
error at most $\epsilon_Z$ on every sector moment for $4\le n\le12$.
Adding $\epsilon_Z$ to both twelfth-moment centers of
Table~\ref{cert:tab:traces} gives the upper bounds
\begin{equation}
    J_{12}\le m_J=\frac{2131441}{2000000},\qquad
    N_{12}\le m_N=\frac{556631}{2000000}.
    \label{cert:eq:masses}
\end{equation}

The twelfth moment alone bounds a $J$ eigenvalue only by
$m_J^{1/12}\simeq1.0053$, whose $72$nd power exceeds $1.46$; the thermal
denominator below needs a far tighter upper bound. A polynomial that is
nonnegative on $\mathbb R$ makes its sector sum an upper bound for each
eigenvalue. One that nearly vanishes on the subleading spectrum draws
its total from the leading eigenvalue alone, tightening the bound.
Ref.~\cite{OpenAI2026Periodic} takes degree at most $12$, a fourth power
times a square:
\begin{align}
    F_J(x) &=x^4(7+17x-280x^2-185x^3+1000x^4)^2,\notag\\
    F_N(x) &=x^4(12-394x^2+1000x^4)^2.
    \label{cert:eq:filters}
\end{align}
Expanded, each polynomial is a combination $\sum_jf_jx^j$ of
$x^4,\ldots,x^{12}$ ($F_N$ of the even powers only), so its sum over a
sector is the same combination of the sector moments. Taking each moment
at the end of its $\epsilon_Z$ interval that bounds its term (upper for
$f_j>0$, lower for $f_j<0$) gives the certified totals
\begin{equation}
    \sum_{\lambda\in J}F_J(\lambda)\le B_J=318604.54848175,\qquad
    \sum_{\lambda\in N}F_N(\lambda)\le B_N=48169.880699.
    \label{cert:eq:filter-bounds}
\end{equation}
Each polynomial rises above its total outside a bounded interval:
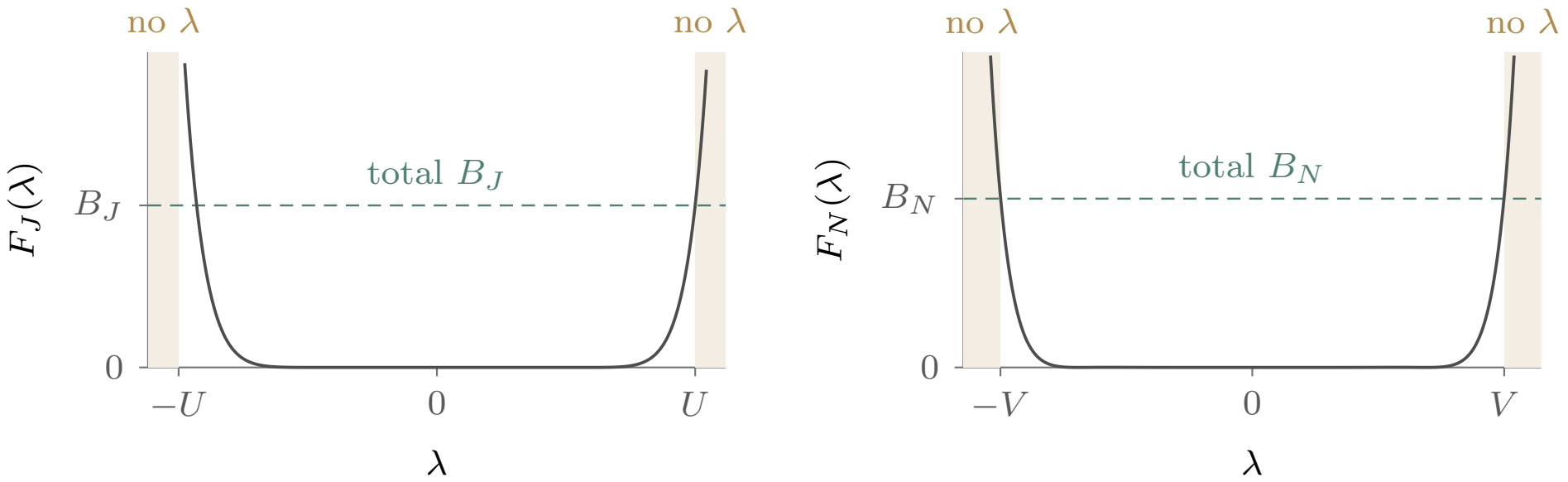
\begin{proofdiagram}
    \filterpair
\end{proofdiagram}
Each equation $F(x)=B$ has one positive root, $x_J\simeq1.0013506440113$
for $J$ and $x_N\simeq0.8712557046526$ for $N$; the negative root for $J$
has the smaller modulus, near $0.9315$. For the base seed,
Ref.~\cite{OpenAI2026Periodic} rounds the roots up to the short rationals
$U=1.00139$ and $V=0.872$. Shifting $x=\pm(U+y)$ with $y\ge0$ gives
polynomials with positive coefficients, so Lean checks the bounds (with its own,
slightly tighter rational caps) without root extraction, ruling out both
exterior rays%
~\leanmark{https://sirui-lu.com/CertifiedGaps/docs/CertifiedGaps/Heisenberg/Seed36.html\#CertifiedGaps.SpinOneHeisenberg.seed36}:
\begin{equation}
    |\lambda|<U\quad(\lambda\in J),\qquad
    |\lambda|<V\quad(\lambda\in N).
    \label{cert:eq:caps}
\end{equation}
Section~\ref{cert:sec:base-defects} uses $U$ in the thermal denominator
and $V$ for the three $N$ sectors of both residuals.

\subsubsection{Purity bounds at lengths $36$ and $72$ from twelfth moments}
\label{cert:sec:base-defects}

The seed ratios need power sums at $n=36$ and $n=72$, far beyond the
certified rings. The twelfth moments~\eqref{cert:eq:masses} and the
eigenvalue bounds~\eqref{cert:eq:caps} still bound every higher moment through one
extremal function. For $m\ge0$, $a>0$, and $r\ge1$,
set%
~\leanmark{https://sirui-lu.com/CertifiedGaps/docs/CertifiedGaps/Seed/CappedMass.html\#CertifiedGaps.sum_pow_le_capped}
\begin{equation}
    \mathcal C(m,a,r)=qa^r+(m-qa)^r,\qquad
    q=\lfloor m/a\rfloor.
    \label{cert:eq:capped}
\end{equation}
Whenever $0\le t_i\le a$ and $\sum_i t_i\le m$, convexity of $t^r$ lets
mass move from a smaller entry to a larger one without lowering
$\sum_i t_i^r$ (\massmoves). The maximum therefore fills $q$ entries to
the bound $a$ and leaves $m-qa$ in one entry, so $\sum_i t_i^r\le\mathcal C(m,a,r)$.
Applied to $t_i=\lambda_i^{12}\ge0$, this bounds higher moments of the
twelfth-power weights.

The spatial numerator is bounded by the leading eigenvalue alone, so that
eigenvalue must also dominate the $36$-site denominator; the trial bound
$Z_{72}(b_0)>1$ constrains only a sum. We first show that $J$ contains a dominant eigenvalue with
$|\lambda_0|>\ell=0.999$, the short rational of
Ref.~\cite{OpenAI2026Periodic}; any $\ell$ that keeps the
sum~\eqref{cert:eq:dominance-check} under one would serve. If every $\lambda\in J$
satisfied $|\lambda|\le\ell$, Eq.~\eqref{cert:eq:capped} with $r=6$ would
place the twelfth-power masses as follows,
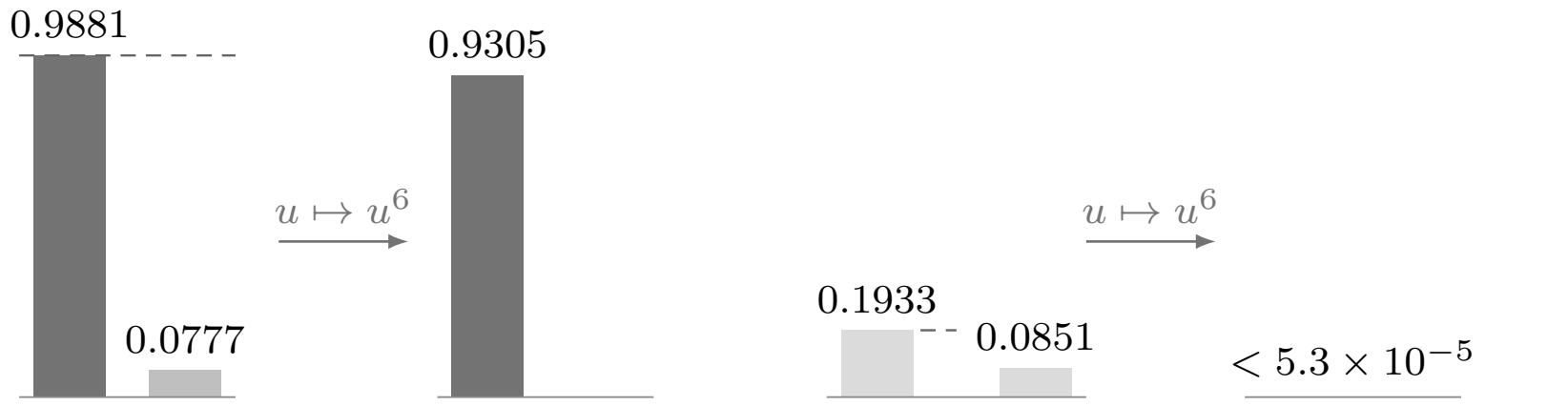
\begin{proofdiagram}
    \cappedmassfigure
\end{proofdiagram}
and give
\begin{equation}
    Z_{72}(b_0)
    \le\mathcal C(m_J,\ell^{12},6)
    +3\mathcal C(m_N,V^{12},6)
    <0.931<1,
    \label{cert:eq:dominance-check}
\end{equation}
contradicting the $72$-site trial bound $Z_{72}(b_0)>1$ in
Eq.~\eqref{cert:eq:trial-consequence}. Because $m_N<\ell^{12}$, no $N$
eigenvalue reaches $\ell$ in modulus, so $J$ holds an eigenvalue
$\lambda_0$ with $|\lambda_0|>\ell$ (\sectorspectrum). This is where the sector split
is needed: the plain traces alone could not exclude three copies of
$\lambda_0$ in $N$.

Removing $\lambda_0^{12}>\ell^{12}$ leaves residual twelfth moment at
most $m_J-\ell^{12}$ in $J$, below $U^{12}$, so it counts as a
single entry. For even $k\ge12$, the subleading moment
$\sum_{i\ne0}\lambda_i^k$ is therefore at most
\begin{equation}
    R_k=(m_J-\ell^{12})^{k/12}
    +3\mathcal C(m_N,V^{12},k/12).
    \label{cert:eq:base-residual}
\end{equation}
At $k=36$ and $k=72$, keeping only the leading eigenvalue in the spatial
numerator, the trial bound~\eqref{cert:eq:trial-consequence} in the
thermal numerator, and $R_k$ in both denominators gives the two base
purity bounds:
\begin{align}
    \Ps(36,b_0) &=\frac{Z_{72}(b_0)}{Z_{36}(b_0)^2}
    \ge\frac{\lambda_0^{72}}{(|\lambda_0|^{36}+R_{36})^2}
    \ge\left(1+\frac{R_{36}}{\ell^{36}}\right)^{-2}=1-p,\notag\\
    \Pt(72,b_0) &=\frac{Z_{72}(2b_0)}{Z_{72}(b_0)^2}
    \ge\frac{1}{(U^{72}+R_{72})^2}=1-q.
    \label{cert:eq:base-purities}
\end{align}
The sizes entering these bounds are
\begin{center}
    \footnotesize
    \setlength{\tabcolsep}{4pt}
    \begin{tabular}{@{}llllll@{}}
        \toprule
        $k$ & $(m_J-\ell^{12})^{k/12}$  & $3\,\mathcal C(m_N,V^{12},k/12)$        & $R_k$
            & leading term              & bound on $1-P$                                                      \\
        \midrule
        36  & $0.0777^3<0.00047$        & $3(0.1933^3+0.0851^3)<0.02352$          & $<0.024$
            & $\ell^{36}>0.9646$        & $p<1-(0.9646/0.9886)^2<0.048$                                       \\
        72  & $0.0777^6<3\times10^{-7}$ & $3(0.1933^6+0.0851^6)<1.6\times10^{-4}$
            & $<1.6\times10^{-4}$       & $U^{72}<1.1052$                         & $q<1-1.10536^{-2}<0.1816$ \\
        \bottomrule
    \end{tabular}
\end{center}

Evaluating Eq.~\eqref{cert:eq:base-purities} gives%
~\leanmark{https://sirui-lu.com/CertifiedGaps/docs/CertifiedGaps/Seed/Ring.html\#CertifiedGaps.RPRing.seeds_of_certificate}
\begin{equation}
    p\simeq0.0479158\le\frac{47916}{10^6},\qquad
    q\simeq0.1815202\le\frac{181521}{10^6},
    \label{cert:eq:scalar-checks}
\end{equation}
which completes the proof of Eq.~\eqref{cert:eq:base}.
Section~\ref{cert:sec:coverage} reuses these formulas with sharper
constants as the bridge on $36\le L\le72$.

\subsection{Seed purities at length $60$ and the bound $\Delta_L>0.07290$}
\label{cert:sec:lifting}

The gap bound of the doubling argument is inversely proportional to the
seed inverse temperature, so a seed at $b=rb_0$ with $r>1$ gives a larger constant if
$1-\Ps$ and $1-\Pt$ stay small. Ref.~\cite{OpenAI2026Periodic}, Sec.~2.2,
introduced this temperature lifting on a separate trace table at
$b_0'=21/2$ (rings of $6$, $8$, and $10$ sites, twists $1,P,C$), with $r=5/2$
and the $120$-site trial state. That seed at $n_0=60$ proves
$\Delta_L>\frac{4}{105}\log\frac{80}{79}\simeq0.000479$ for even
$L\ge60$. Here we apply the same lift to the two-sector trace table at
$b_0=49/4$ (Table~\ref{cert:tab:traces}), raised to $b=rb_0$ by a
convexity inequality and combined with the $120$-site trial bound of
Eq.~\eqref{cert:eq:trial-consequence}, twelve-decimal polynomial caps,
and per-interval gap bounds over ten doubling steps. The length $n_0=60$
is half the trial ring, and the $(36,b_0)$ bound covers $36\le L\le72$,
so every even length $L\ge36$ is reached with $\Delta_L>0.072901771$.

\subsubsection{Spatial purity at length $60$ and inverse temperature $b=rb_0$}
\label{cert:sec:lifted-moments}

At $120$ sites and inverse temperature $b=rb_0$ ($r\simeq2$), the
eigenvalue bound $U$ enters raised to the power $120r\ge240$. At that
exponent the five-decimal rounding $U=1.00139$ of Eq.~\eqref{cert:eq:caps}
inflates $U^{120r}$ by nearly $1\%$ relative to the root $x_J$. Rounding
both roots up to twelve decimal places instead gives
\begin{equation}
    U=1.001350644012,\qquad V=0.871255704653.
    \label{cert:eq:tight-caps}
\end{equation}
The step below needs trace bounds at $b_0$ on a ring of $k>12$ sites
(Table~\ref{cert:tab:lifted}); the extremal moment bound supplies them.
With $m_J,m_N$ from Eq.~\eqref{cert:eq:masses}, since
$U^{12}<m_J<2U^{12}$ and $V^{12}<m_N<2V^{12}$, Eq.~\eqref{cert:eq:capped}
bounds the sector moments at any even length $k\ge12$ by
\begin{equation}
    \mathcal J_k=U^k+(m_J-U^{12})^{k/12},\qquad
    \mathcal N_k=V^k+(m_N-V^{12})^{k/12},\qquad
    M_k=\mathcal J_k+3\mathcal N_k.
    \label{cert:eq:moment-upper}
\end{equation}
Because $N_k\ge0$ at even $k$, these bound both traces,
$Z_k(b_0)\le M_k$ and $Z_k(b_0,P)\le\mathcal J_k$.

Trace bounds exist only at $b_0$, so the step to $b=rb_0$ must act on
the Boltzmann weights themselves. Fix a temperature ratio $r\ge1$ and an even
length $12\le k\le n_0$ (here $n_0=60$ and $2n_0=120$). Both
$H_k$ and $H_k(P)$ are self-adjoint, so their Boltzmann weights $z_i>0$
satisfy $\sum_i z_i^r\le(\sum_i z_i)^r$. This raises the trace bounds
from $b_0$ to $b$, and with them the $J$ share of
Eq.~\eqref{cert:eq:sector-inversion}:
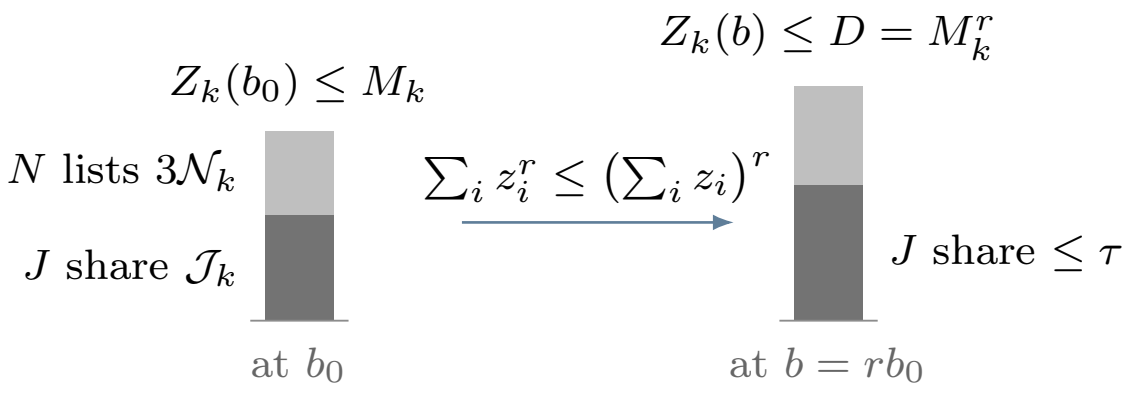
\begin{proofdiagram}
    \liftbars
\end{proofdiagram}
\begin{equation}
    Z_k(b)\le D:=M_k^r,\qquad
    Z_k(b,P)\le E:=\mathcal J_k^r.
    \label{cert:eq:temperature-lift}
\end{equation}

The totals $D$ and $E$ are all the lifted argument knows at $b$; as at
$b_0$, the trial bound must force one weight that dominates the
$n_0$-site denominator.
At $b$, let $w_i=\lambda_i(b)^k\ge0$ be the $k$th powers of the
eigenvalues of the one-site transfer operator ($k$ is even). With
$h=n_0/k$, so that $w_i^h$ and $w_i^{2h}$ are the $n_0$- and
$2n_0$-site weights, Eq.~\eqref{cert:eq:temperature-lift} and the
trial bound $Z_{2n_0}(b)>1$ of Eq.~\eqref{cert:eq:trial-consequence}
give
\begin{align}
    \sum_i w_i     &=Z_k(b)\le D,\qquad
    \sum_{\lambda_i\in J}w_i=\frac{Z_k(b)+3Z_k(b,P)}4\le\tau:=\frac{D+3E}4,
    \label{cert:eq:spatial-weights}\\
    \sum_iw_i^{2h} &=Z_{2n_0}(b)>1.
    \label{cert:eq:lifted-trial}
\end{align}
If a threshold $t>0$ satisfies $Dt^{2h-1}<1$ and $3t>D$, then some
weight $w_0$ exceeds $t$ and lies in $J$:
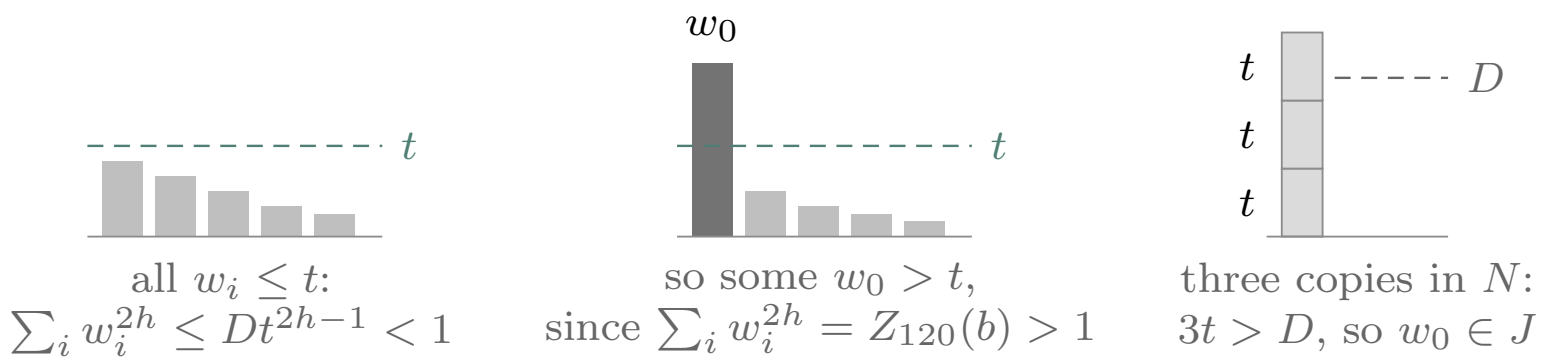
\begin{proofdiagram}
    \weightforcing
\end{proofdiagram}

The $n_0$-site denominator is $w_0^h$ plus the $h$th moment of the other
weights, whose split between $J$ and the $N$ lists is unknown.
Removing $w_0$ leaves residual $J$ mass $x\in[0,\tau-t]$ and at most
$(D-t-x)/3$ in each of the three identical $N$ sectors. For $s\ge1$ the
residual moment $\sum_{i\ne0}w_i^s\le x^s+3[(D-t-x)/3]^s$ is convex in
$x$, so its maximum sits at an endpoint,
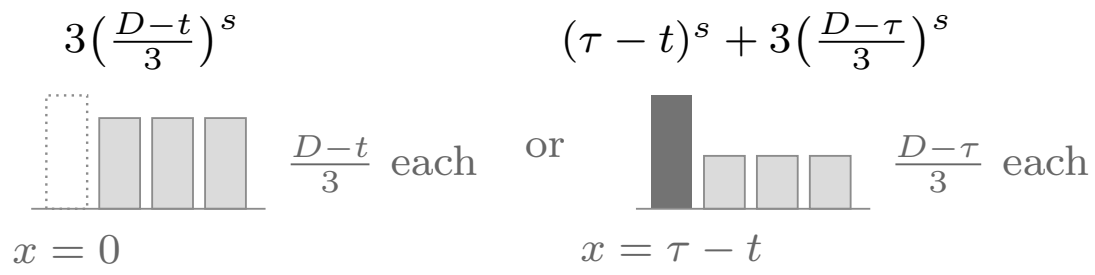
\begin{proofdiagram}
    \residualends
\end{proofdiagram}
\begin{equation}
    \mathcal R_s(t)=
    \max\left\{3\left(\frac{D-t}{3}\right)^s,
    (\tau-t)^s+3\left(\frac{D-\tau}{3}\right)^s\right\}.
    \label{cert:eq:lifted-residual}
\end{equation}
All choices below satisfy $0<t\le\tau\le D$.

Once $w_0>t$ holds, any larger rational $t'$ with
$(t')^{2h}+\mathcal R_{2h}(t)<1$ also satisfies $w_0>t'$, since
otherwise the $2n_0$-site sum would fall below one. Testing rationals
avoids a $2h$th root, and iterating raises the threshold
(\thresholdup). At $n_0=60$, the leading $n_0$-site weight $w_0^h$
exceeds $t^h$ and the rest sum to at most $\mathcal R_h(t)$, so the
spatial purity satisfies%
\begin{equation}
    \Ps(n_0,b)\ge
    \left(1+\frac{\mathcal R_h(t)}{t^h}\right)^{-2}.
    \label{cert:eq:lifted-spatial}
\end{equation}
Its complement is the $p_0$ column of Table~\ref{cert:tab:lifted}.

\subsubsection{Thermal purity at length $120$ and inverse temperature $b=rb_0$}
\label{cert:sec:thermal-residual}

Neither $120$-site partition function in the thermal purity at $b$ is
certified. Both the trial bound and the moment bound~\eqref{cert:eq:moment-upper}
concern the Boltzmann weights $z_i$ of the shifted $120$-site Hamiltonian
at $b_0$; the lift to $b$ raises each to the $r$th power. We therefore bound $\Pt(2n_0,b)$ from the ordered
weights at $b_0$ (\boltzmannbars): the trial state gives $z_0>1$ and
Eq.~\eqref{cert:eq:moment-upper} gives $\sum_i z_i\le M_{2n_0}$, hence
\begin{equation}
    \sum_{i>0}\frac{z_i}{z_0}\le M_{2n_0}-1,\qquad
    \sum_{i>0}\left(\frac{z_i}{z_0}\right)^r
    \le(M_{2n_0}-1)^r.
    \label{cert:eq:boltzmann-residual}
\end{equation}
At $b=rb_0$ the weights become $z_i^r$, which concentrates them on the
ground state ($\specbars{.6,.25,.15}\mapsto\specbars{.83,.13,.04}$ at
$r=17/8$). Keeping only the ground term $z_0^{2r}$ in the numerator
$Z_{2n_0}(2b)$ and dividing by it gives the thermal purity bound at
$2n_0=120$%
\begin{equation}
    \Pt(2n_0,rb_0)\ge
    \left[1+(M_{2n_0}-1)^r\right]^{-2}.
    \label{cert:eq:lifted-thermal}
\end{equation}
Its complement is the $q_0$ column of Table~\ref{cert:tab:lifted}; with
$p_0$ it seeds the recurrence of Sec.~\ref{cert:sec:coverage}.

\subsubsection{Combining the dyadic intervals for all even $L\ge36$}
\label{cert:sec:coverage}

The quadratic tail of Sec.~\ref{sm:sec:gap} needs both $p$ and $q$ below
$1/100$, reached only after $K$ doublings, so it covers only $L\ge n_K$;
the $K$ earlier doubling intervals need their own bounds.
Write $(p_j,q_j)$ for the bounds on $1-\Ps(n_j,b_j)$ and
$1-\Pt(2n_j,b_j)$ after $j$ of $K$ steps of the
asymmetric recurrence~\eqref{sm:eq:update} from the seed, with the last
step below the threshold $1/100$ of Sec.~\ref{sm:sec:gap}:
\[
    (n_j,b_j)=(2^jn_0,2^jb),\qquad A_j=(1-p_j)^2(1-q_j),\qquad
    u=\max(p_K,q_K)\le\frac1{100}.
\]
On the interval between two consecutive scales, the spatial
interpolation inequality~\eqref{sm:eq:interpolation} bounds the largest
normalized Boltzmann weight $a_0$ by $A_j$, which exceeds $1/2$ at every
step below. The ground state is therefore unique, the next weight is
$a_1\le1-a_0$, and
\begin{align}
    a_0      &\ge\Pt(L,b_j)\ge A_j^{L/(2n_j)}\ge A_j,\qquad
    e^{-b_j\Delta_L}=\frac{a_1}{a_0}\le\frac{1-A_j}{A_j},\notag\\
    \Delta_L &\ge\frac{\log(A_j/(1-A_j))}{b_j}
    \quad(n_j\le L\le2n_j,\ j<K).
    \label{cert:eq:interval-gap}
\end{align}
For even $L\ge n_K$, the quadratic tail of Sec.~\ref{sm:sec:gap} takes
over with its exact coefficient $G(u)=f(1-(1-u)^3)/u^2$, which rises from
$G(0)=9/2$ to $4.69$ at $u=1/100$ (Sec.~\ref{sm:sec:gap} rounds it to
$5$). With a rational $C\ge G(u)$, the bounds $(p_m,q_m)$ at scale $m\ge K$ are at
most $C^{-1}(Cu)^{2^{m-K}}$, and $\Delta_L>\tfrac{\log(1/(Cu))}{b_K}$ (using
$3u/(1-3u)\le Cu$, which holds because $C\ge9/2$ and $u\le1/100$). The
minimum of the $K$ interval bounds~\eqref{cert:eq:interval-gap} and this
tail bound therefore bounds $\Delta_L$ for every even $L\ge n_0$.

For the seeds at $(60,rb_0)$, we run $K=10$ doubling steps, enough for
the per-interval bounds to converge.

\begin{table}[t]
    \caption{Certified seed bounds $p_0\ge1-\Ps(60,rb_0)$ and
        $q_0\ge1-\Pt(120,rb_0)$%
        and resulting uniform gap bounds at length
        $n_0=60$ and inverse temperature $b=rb_0$ ($b_0=49/4$). Here $k$ is the
        even ring length at which the trace bounds~\eqref{cert:eq:moment-upper}
        are raised to the power $r$ in Eq.~\eqref{cert:eq:temperature-lift}.
        Running $K=10$ doubling steps from $(60,rb_0)$ certifies the stated gap
        bound for all even $L\ge60$%
        ~\leanmark{https://sirui-lu.com/CertifiedGaps/docs/Leaderboard/Chains/SpinOneHeisenberg.html\#Leaderboard.Chains.SpinOneHeisenberg.temperature_lifting_valid},
        and combining with the $(36,b_0)$ bound on
        $36\le L\le72$ extends it to all even $L\ge36$%
        ~\leanmark{https://sirui-lu.com/CertifiedGaps/docs/Leaderboard/Records/SpinOneHeisenberg.html\#Leaderboard.Records.SpinOneHeisenberg.temperature_lifting}.}
    \label{cert:tab:lifted}
    \begin{tabular}{@{}ccccc@{}}
        \toprule
        $k$ & $r$    & $p_0\ge1-\Ps$ & $q_0\ge1-\Pt$ & $\Delta_L>$ \\
        \midrule
        30  & $2$    & 0.0307608679  & 0.0590740709  & 0.070450002 \\
        28  & $17/8$ & 0.0305302052  & 0.0479556776  & 0.072901771 \\
        \bottomrule
    \end{tabular}
\end{table}

Choosing $r$ balances two effects. A larger $r$ shrinks the thermal
residual $(M_{120}-1)^r$ in Eq.~\eqref{cert:eq:lifted-thermal} and
increases $b=rb_0$. Too large an $r$, however, makes $D=M_k^r$ too large
for a threshold $t$ with $Dt^{2h-1}<1$ and $3t>D$ to isolate a leading spatial
eigenvalue.

Table~\ref{cert:tab:lifted} lists two choices near the optimum
$r\approx2$. The integer choice $(k,r)=(30,2)$ has $h=60/k=2$, so
Eq.~\eqref{cert:eq:lifted-residual} uses only second and fourth powers;
$(k,r)=(28,17/8)$ maximizes the gap bound.

To cover the remaining even lengths $36\le L<60$, we evaluate the
$(36,b_0)$ formulas~\eqref{cert:eq:base-residual}--\eqref{cert:eq:base-purities}
with the twelve-decimal bounds~\eqref{cert:eq:tight-caps} and
$\ell=0.999997937629$. This $\ell$ is the largest twelve-decimal value
satisfying Eq.~\eqref{cert:eq:dominance-check}, and its $A_0$ in
Eq.~\eqref{cert:eq:interval-gap} gives $\Delta_L>0.09$ on $36\le L\le72$.
The intervals and their bounds for $(k,r)=(28,17/8)$ are
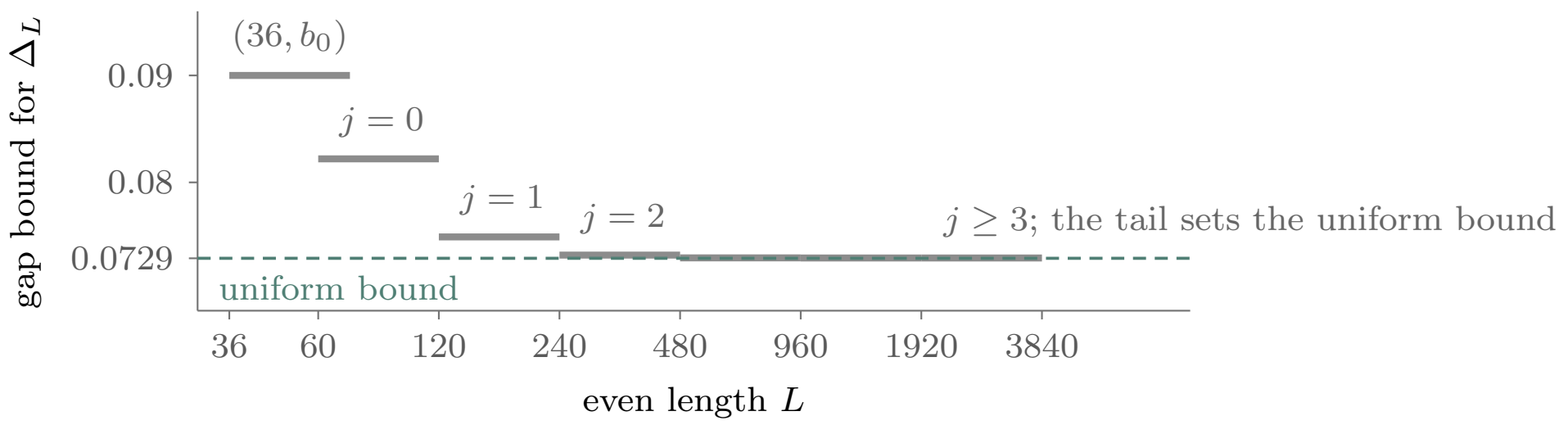
\begin{proofdiagram}
    \coveragefigure
\end{proofdiagram}
(per-interval values are floating-point illustrations; the certified
minimum is the dashed line). Because $[36,72]$ and the dyadic intervals
$[60\,2^j,120\,2^j]$ overlap on $[60,72]$, together they prove
Eq.~\eqref{cert:eq:improved} for every even $L\ge36$.

\section{Reflection-positive spin chains and ladders}
\label{sm:sec:models}
Applying the periodic doubling argument of Section~\ref{sm:sec:trotter} to a
new Hamiltonian requires two inputs:
\begin{enumerate}
    \item write $H_L=\sum_j h_{j,j+1}+\sum_j o_j$ in the real
          bond-reflection-positive form $-h=\sum_\alpha C_\alpha\otimes FC_\alpha F$
          with $FoF=o$ on a chosen unit cell, so that the two-cell Trotter
          transfer matrix $\smT=\smF^{\mathsf T}\smF\succeq0$ of
          Eq.~\eqref{sm:eq:positive} is positive semidefinite at every even cell
          count $L$;
    \item bound $1-\Ps(n_0,\beta_0)$ and $1-\Pt(2n_0,\beta_0)$ at one
          finite scale by rationals $(p_0,q_0)$ inside the convergence region of
          the doubling recurrence.
\end{enumerate}

Reflection positivity carries no gap information (gapless and
spontaneously dimerized chains are reflection positive too;
Proposition~\ref{models:cor:obstruct}), so the seed computation must be
matched to the correlation length $\xi$ and gap $\Delta$ of the model:
\begin{center}
    \renewcommand{\arraystretch}{1.3}
    \setlength{\tabcolsep}{4pt}
    \begin{tabular}{@{}llp{4.1cm}p{5.3cm}@{}}
        \toprule
        Regime                           & Finite inputs          & Seed strategy                           & Models in Table~\ref{tab:models}                                     \\
        \midrule
        Short $\xi$, small $d^{2n_0}$    & $2n_0\le16$ spins      & Symmetry blocks + Weyl bound            & Staggered, dimerized, ladders, Ising (Sec.~\ref{models:sec:spectra}) \\
        Moderate $\xi$, large $d^{4n_0}$ & $6,12$ spins           & Block spectra + Eq.~\eqref{sm:eq:torus} & Single-ion anisotropy, $D=4$ (Sec.~\ref{models:sec:spectra})         \\
        Weak off-diagonal term           & $6,12$ spins           & Trial residuals + diagonal count        & Mixed-field Ising, kernel seed (Sec.~\ref{models:sec:ising})         \\
        Long $\xi$ ($n_0\ge36$)          & $4$--$12$ traces + MPS & Twisted-ring moments + integer MPS      & Heisenberg, $S=1$ (Sec.~\ref{sm:sec:seeds})                          \\
        \bottomrule
    \end{tabular}
\end{center}

\subsection{Reflection-positive spin chains, unit cells, and obstructions}
\label{models:sec:rp}

To apply the finite-Trotter Gram factorization of
Section~\ref{sm:sec:transfer} to a periodic ring of $L\ge2$ unit cells,
\begin{equation}
    H_L=\sum_{j=1}^L h_{j,j+1}+\sum_{j=1}^L o_j,
    \qquad j+L\equiv j,
    \label{models:eq:ring}
\end{equation}
we require a real local cell basis, a real symmetric involution $F$
($F^2=\smI$), a real symmetric on-site matrix $o$ satisfying $FoF=o$,
and real symmetric or antisymmetric bond matrices $C_\alpha$ such that
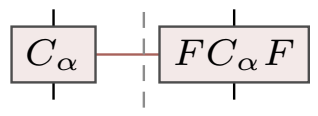
\begin{equation}
    -h=\sum_\alpha C_\alpha\otimes FC_\alpha F
    =\sum_\alpha\;\scopebond\,.
    \label{models:eq:rp}
\end{equation}
Hermiticity of $h$ alone does not supply this form: the factors must be
real and $o$ even under $F$, so basis and cell are chosen model by model.

\subsubsection{Bilinear, staggered, and biquadratic spin chains}
\label{models:sec:catalogue}

For a bilinear bond, a factor $v\cdot\bm S$ and its mirror image must
multiply back to $-\bm S^{\mathsf T}J\bm S$, so the mirror has to act on
the spin components by a $3\times3$ matrix $O$. Only maps realized by an
antiunitary symmetry qualify, which restricts the sign pattern of $J$.
Consider first a single-site spin-$S$ chain
($F=\smI$) with a real symmetric exchange matrix $J$ and a uniform
field $\bm h$,
\begin{equation}
    H_L=\sum_{j=1}^L\bm S_j^{\mathsf T}J\bm S_{j+1}
    +\sum_{j=1}^L\bm h\cdot\bm S_j.
    \label{models:eq:bilinear}
\end{equation}
In the standard $S^z$ eigenbasis, $S^x$ and $S^z$ are real and $S^y$ is
purely imaginary, so complex conjugation $K$ acts on the spin vector by
$\operatorname{diag}(1,-1,1)$. Composing $K$ with a $\pi$ rotation $U_R$
about a principal axis gives an antiunitary reflection $\Theta=U_RK$ that
acts on the spin vector by a symmetric orthogonal matrix
$O=R\operatorname{diag}(1,-1,1)$ with $\det O=-1$; the rotation about $y$
gives time reversal $T=e^{i\pi S^y}K$ with $O=-I$.

\begin{proposition}[Bilinear spin chains]
    \label{models:prop:xyz}
    If there exists a symmetric orthogonal matrix $O\in\mathrm O(3)$ with
    $\det O=-1$ such that $-JO\succeq0$ and $O\bm h=\bm h$, then
    Eq.~\eqref{models:eq:bilinear} admits a reflection-positive bond
    factorization for every spin $S$. For diagonal exchange
    $J=\operatorname{diag}(J_x,J_y,J_z)$ with nonzero entries and $\bm h=0$,
    such an $O$ exists if and only if an even number of $J_x,J_y,J_z$ are
    negative.
\end{proposition}

\begin{proof}
    Factor the positive semidefinite matrix $-JO=\sum_\alpha v_\alpha v_\alpha^{\mathsf T}$
    with real vectors $v_\alpha\in\mathbb R^3$ and set
    $C_\alpha=v_\alpha\cdot\bm S$. Under $\Theta=U_RT$,
    $\Theta(C_\alpha)=(Ov_\alpha)\cdot\bm S$, so
    $\sum_\alpha C_\alpha\otimes\Theta(C_\alpha)
        =\bm S^{\mathsf T}(-JO)O\bm S=-\bm S^{\mathsf T}J\bm S$, while
    $\Theta(\bm h\cdot\bm S)=(O\bm h)\cdot\bm S=\bm h\cdot\bm S$. When $J$
    is diagonal with nonzero entries, $-JO\succeq0$ forces
    $O=\operatorname{sign}(-J)$, whose determinant $(-1)^{n_-}$ is $-1$ if
    and only if the number $n_-$ of ferromagnetic ($J_\alpha<0$) components
    is even.
\end{proof}

Proposition~\ref{models:prop:xyz} sorts the standard single-site spin
chains by the sign parity $n_-$ of their exchange couplings; a uniform
field fails because $O=-\smI$ reverses it:
\begin{center}
    \setlength{\tabcolsep}{5pt}
    \begin{tabular}{@{}llll@{}}
        \toprule
        Chain                                         & $(J_x,J_y,J_z)$, $\bm h$ & Mirror $O$                                  & Reflection positive \\
        \midrule
        Heisenberg antiferromagnet, any $S$           & $(1,1,1)$, $0$           & $-\smI$ ($n_-=0$)                           & yes                 \\
        XXZ, $\Delta\ge0$                             & $(1,1,\Delta)$, $0$      & $-\smI$                                     & yes                 \\
        XY, either sign                               & $(J_x,J_y,0)$, $0$       & $J_z=0$ frees $O_{zz}$                      & yes                 \\
        Transverse- or mixed-field Ising              & $(0,0,-1)$, $(g,0,h)$    & $\operatorname{diag}(1,-1,1)$ fixes $\bm h$ & yes                 \\
        Heisenberg ferromagnet                        & $(-1,-1,-1)$, $0$        & $n_-=3$                                     & no                  \\
        XXZ, $\Delta<0$                               & $(1,1,\Delta)$, $0$      & $n_-=1$                                     & no                  \\
        Heisenberg antiferromagnet in a uniform field & $(1,1,1)$, $\bm h\ne0$   & $-\smI$ reverses $\bm h$                    & no                  \\
        \bottomrule
    \end{tabular}
\end{center}

Although a uniform field breaks reflection positivity of the Heisenberg
antiferromagnet, a \emph{staggered} field restores it on every even
ring: a $\pi$ rotation about the $y$ axis on every odd site makes the
field uniform,
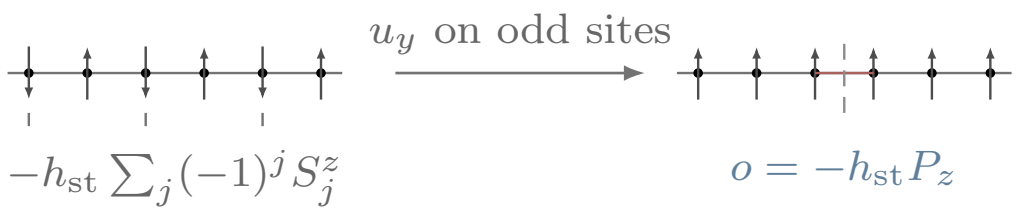
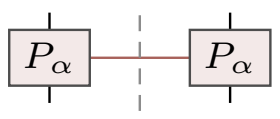
\begin{equation}
    \staggerrotation\qquad
    -h_{j,j+1}=\sum_\alpha\;\rpbondpic{P_\alpha}{P_\alpha}\,,
    \label{models:eq:stagger-rotation}
\end{equation}
and the bond sign parity stays even, $n_-=2$, because the remaining sign
sits in the real antisymmetric factor $P_y=iS^y$. The same rotation
underlies the staggered-field chain in Table~\ref{tab:models} and, on
open chains, relates opposite endpoint fields \fieldopp\ to equal
endpoint fields \fieldsame\ (Section~\ref{sm:sec:obc}).

For $S=1$, adding a biquadratic interaction gives the family
\begin{equation}
    H_L(\vartheta)=\sum_{j=1}^L\Bigl[
        \cos\vartheta\,\bm S_j\cdot\bm S_{j+1}
        +\sin\vartheta\,(\bm S_j\cdot\bm S_{j+1})^2\Bigr].
    \label{models:eq:blbq}
\end{equation}
Proposition~\ref{models:prop:xyz} does not reach the biquadratic term,
which mixes products $S^\alpha S^\beta$ across the bond; factoring it
needs one-site real operators without cross terms across the mirror. In
the Cartesian basis $\ket{x},\ket{y},\ket{z}$ of
Eq.~\eqref{sm:eq:generators}, the spin generators $G_\alpha=iS^\alpha$
are real antisymmetric and the products $Q_{\alpha\beta}=S^\alpha S^\beta$
are real matrices whose symmetric part
$Q_{\alpha\beta}^+=\tfrac12(Q_{\alpha\beta}+Q_{\beta\alpha})$ and
antisymmetric part $\tfrac12\epsilon_{\alpha\beta\gamma}G_\gamma$ are
orthogonal under the trace. Expanding $(\bm S_1\cdot\bm S_2)^2$ into
$G_\alpha$ and $Q_{\alpha\beta}^+$ diagonalizes the bond across the
mirror:
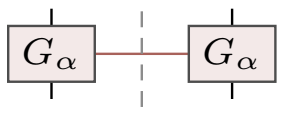
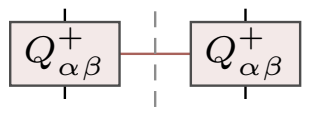
\begin{equation}
    -h(\vartheta)=\bigl(\cos\vartheta-\tfrac12\sin\vartheta\bigr)
    \sum_\alpha\;\rpbondpic{G_\alpha}{G_\alpha}
    \;-\sin\vartheta\sum_{\alpha,\beta}\;
    \rpbondpic{Q_{\alpha\beta}^+}{Q_{\alpha\beta}^+}\,.
    \label{models:eq:blbq-decomp}
\end{equation}

\begin{proposition}[Reflection positivity of the bilinear--biquadratic chain]
    \label{models:prop:blbq}
    In the Cartesian basis, the bond~\eqref{models:eq:blbq-decomp} is
    reflection positive if and only if both coefficients are nonnegative,
    that is,
    \begin{equation}
        -\pi+\arctan2\le\vartheta\le0.
        \label{models:eq:blbq-window}
    \end{equation}
\end{proposition}

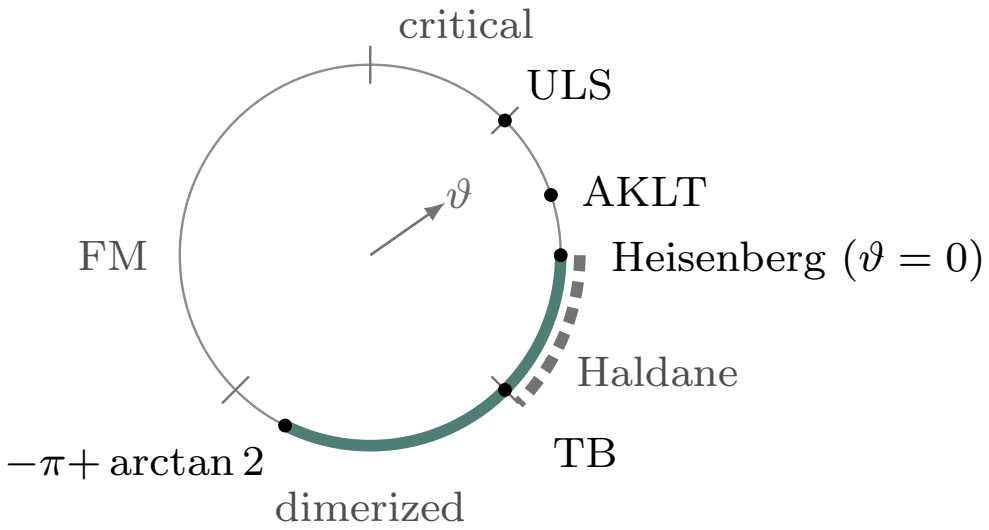
\begin{proofdiagram}
    \blbqcircle
\end{proofdiagram}
\noindent Reflection positivity of the whole family thus reduces to the
signs of two coefficients; the frustration-free AKLT point
($\tan\vartheta=1/3>0$) lies outside the arc because its positive
biquadratic term has the wrong sign in Eq.~\eqref{models:eq:blbq-decomp}.
The doubling argument applies only on the Haldane sub-arc $(-\pi/4,0]$: the
spontaneously dimerized phase $\vartheta\in[-\pi+\arctan2,-\pi/4]$ and
the gapless spin-$1/2$ Heisenberg chain are reflection positive but
have no seed in the convergence region, for reasons no larger ring or lower
temperature removes:

\begin{proposition}[Obstructions to a convergent seed]
    \label{models:cor:obstruct}
    For any reflection-positive ring Hamiltonian~\eqref{models:eq:ring}, no
    scale $(n_0,\beta_0)$ has purities $\Ps(n_0,\beta_0)$ and
    $\Pt(2n_0,\beta_0)$ inside the convergence region of the doubling
    recurrence if either
    \begin{enumerate}
        \item[(i)] $H_{2n_0}$ has a $k$-fold degenerate ground level with
              $k\ge2$: the ordered Gibbs weights satisfy $w_0=w_1\le1/k\le1/2$
              (\specbars{.49,.49,.02}), so $\Pt(2n_0,\beta_0)=\sum_m w_m^2\le w_0\le1/2$;
              or
        \item[(ii)] $\liminf_{L\to\infty,\;L\text{ even}}\Delta_L=0$: a
              seed in the convergence region would give a uniform positive lower bound on $\Delta_L$
              for all even $L\ge n_0$.
    \end{enumerate}
\end{proposition}
\noindent Reflection positivity thus supplies the squaring map, and only
a finite seed inside the convergence region separates gapped chains with
a unique ground state from critical or symmetry-broken ones.

\subsubsection{Multi-spin unit cells, dimer reflections, and explicit decompositions}
\label{models:sec:explicit-rp}

When interactions alternate from bond to bond or couple two parallel
legs, we block sites into a multi-spin unit cell. A single-site mirror
then fails, because the inter-cell bond couples the last spin of one
cell to the first spin of the next. The cell reflection $F$ repairs
this: across an inter-cell mirror cut, a reflection of the ring
reverses the left-to-right order of the sites inside each cell:
\begin{equation}
    -h_{j,j+1}=\sum_\alpha\;\cellbondform\,.
    \label{models:eq:cell-mirror}
\end{equation}
In a two-spin dimer cell $(a,b)$, $F$ is the two-spin swap
$F(x\otimes y)=y\otimes x$%
~\leanmark{https://sirui-lu.com/CertifiedGaps/docs/CertifiedGaps/Local/CellSwap.html\#CertifiedGaps.cellSwap_mul_cellSwap},
which maps $C_\alpha=\smI\otimes G_\alpha$ on spin $b$ to
$FC_\alpha F=G_\alpha\otimes\smI$ on spin $a$ of the next cell as drawn, and the intra-cell bond is a
swap-invariant on-site matrix $FoF=o$. In a two-leg ladder the cell is
a rung and the mirror keeps each leg on its own side,
\begin{equation}
    -h_{j,j+1}=\sum_\alpha\;\rungcellform\,,
    \label{models:eq:rung-mirror}
\end{equation}
so the rung interaction is even under the reflection and belongs to
$o$. These two reflections, the dimer swap and the identity on a rung,
are all that Table~\ref{models:tab:rp} needs.

The two-cell Gram form makes $Z_n(\beta)$ a sum of nonnegative powers
only at even $n$, so an odd seed length needs positivity of the one-cell
transfer operator itself. A dimer cell whose intra-cell bond is
reflection positive supplies it: cutting the cell across its internal
bond factors the \emph{one-cell} spatial transfer operator into a
half-cell product with its transpose,
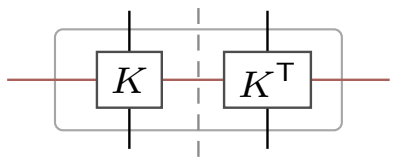
\begin{equation}
    X_\beta=\oddcellpos=KK^{\mathsf T}\succeq0.
    \label{models:eq:odd-cell}
\end{equation}
Every eigenvalue of $KK^{\mathsf T}$ is nonnegative, so
$Z_n(\beta)=\Tr X_\beta^n=\sum_i\lambda_i(\beta)^n$ is a sum of
nonnegative powers at \emph{odd} cell counts $n_0$ too.
In the dimerized spin-$1$ chain at $\delta=7/10$,
Eq.~\eqref{models:eq:odd-cell} allows an odd seed of $n_0=3$ dimer cells
($6$ and $12$ spins), avoiding the $16$-spin ($3^{16}\simeq4.3\times10^7$)
ring that an even seed $n_0=4$ would require.

Proposition~\ref{models:prop:xyz} guarantees a factorization, but the
exact checkers behind Table~\ref{models:tab:rp} need its factors as
explicit real matrices, and $S^y$ is imaginary in the $S^z$ basis.
For $S=1/2$, we write $P_x=S^x=\tfrac12\sigma^x$,
$P_y=iS^y=\tfrac12(i\sigma^y)$, and $P_z=S^z=\tfrac12\sigma^z$, which
are real matrices satisfying $P_x^{\mathsf T}=P_x$,
$P_y^{\mathsf T}=-P_y$, and $P_z^{\mathsf T}=P_z$. Rotating one site of
an antiferromagnetic bond by $u_y=e^{i\pi S^y}=2P_y$ maps
$(S^x,S^y,S^z)\mapsto(-S^x,S^y,-S^z)$ and converts $-\bm S_1\cdot\bm S_2$
into the real reflection-positive form%
~\leanmark{https://sirui-lu.com/CertifiedGaps/docs/CertifiedGaps/Local/SpinHalfRotation.html\#CertifiedGaps.SpinHalf.rot_conj_spinHalf}
\begin{equation}
    -u_{y,2}(\bm S_1\cdot\bm S_2)u_{y,2}^\dagger
    =\sum_{\alpha=x,y,z}P_\alpha\otimes P_\alpha
    =\sum_{\alpha=x,y,z}\;\rpbondpic{P_\alpha}{P_\alpha}\,.
    \label{models:eq:spinhalf-bond}
\end{equation}
Applying $u_y$ on every odd site of an even spin-$1/2$ chain (or on the
second spin $b$ of every dimer cell, or on the checkerboard sublattice
of a two-leg ladder for even $L$) is a unitary transformation that
preserves the spectrum%
~\leanmark{https://sirui-lu.com/CertifiedGaps/docs/CertifiedGaps/Ring/Conjugation.html\#CertifiedGaps.RPRing.eigenvalues_eq_spectrum}
and puts every antiferromagnetic bond into the
form~\eqref{models:eq:spinhalf-bond}. For $S=1$, we work directly in the
Cartesian basis $\ket{x},\ket{y},\ket{z}$ of
Eq.~\eqref{sm:eq:generators}, where $G_\alpha=iS^\alpha$ is real
antisymmetric, $-\bm S_1\cdot\bm S_2=\sum_\alpha G_\alpha\otimes G_\alpha$,
and $(S^z)^2=\operatorname{diag}(1,1,0)$ is real diagonal without any
sublattice rotation. Table~\ref{models:tab:rp} lists the resulting cell,
reflection $F$, bond operators $C_\alpha$, and on-site matrix $o$ for all
seven additional models in Table~\ref{tab:models}.

\begin{table*}[t]
    \caption{Explicit reflection-positive cell decompositions
        $-h=\sum_\alpha C_\alpha\otimes FC_\alpha F$ and $FoF=o$ for the seven
        additional models in Table~\ref{tab:models}, with links to their
        Lean~4 reflection-positivity proofs. Here $F$ is the cell
        reflection involution, $u_y=e^{i\pi S^y}$ is a $\pi$ spin rotation about
        $y$, $P_\alpha\in\{S^x,iS^y,S^z\}$ are the real spin-$1/2$ operators of
        Eq.~\eqref{models:eq:spinhalf-bond}, $G_\alpha=iS^\alpha$ are the real
        antisymmetric spin-$1$ generators in the Cartesian basis, and
        $S_{ab}(x\otimes y)=y\otimes x$ swaps the two spins of a dimer cell.}
    \label{models:tab:rp}
    \setlength{\tabcolsep}{5pt}
    \footnotesize
    \begin{tabular}{@{}lllll@{}}
        \toprule
        Model                                                                                                                                                                   & Unit cell ($F$)           & Local basis / rotation & Bond operators $C_\alpha$                    & On-site cell matrix $o$                                               \\
        \midrule
        Staggered field, $S=\tfrac12$~\leanmark{https://sirui-lu.com/CertifiedGaps/docs/CertifiedGaps/Models/StaggeredHeisenberg.html\#CertifiedGaps.StaggeredHeisenberg.model} & $1$ spin ($\smI$)         & $u_y$ on odd sites     & $P_\alpha$ ($\alpha=x,y,z$)                  & $-h_{\mathrm{st}}P_z$                                                 \\
        Dimerized, $S=\tfrac12$~\leanmark{https://sirui-lu.com/CertifiedGaps/docs/CertifiedGaps/Models/DimerizedHeisenberg.html\#CertifiedGaps.DimerizedHeisenberg.model}       & $2$-spin dimer ($S_{ab}$) & $u_y$ on spin $b$      & $\sqrt{1-\delta}\,(\smI\otimes P_\alpha)$    & $-(1+\delta)\sum_\alpha P_\alpha\otimes P_\alpha$                     \\
        Dimerized, $S=1$~\leanmark{https://sirui-lu.com/CertifiedGaps/docs/CertifiedGaps/Models/SpinOneDimerized.html\#CertifiedGaps.SpinOneDimerized.model}                    & $2$-spin dimer ($S_{ab}$) & Cartesian basis        & $\sqrt{1+\delta}\,(\smI\otimes G_\alpha)$    & $-(1-\delta)\sum_\alpha G_\alpha\otimes G_\alpha$                     \\
        Single-ion anisotropy, $S=1$~\leanmark{https://sirui-lu.com/CertifiedGaps/docs/CertifiedGaps/Models/SpinOneLargeD.html\#CertifiedGaps.SpinOneLargeD.model}              & $1$ spin ($\smI$)         & Cartesian basis        & $G_\alpha$ ($\alpha=x,y,z$)                  & $D\operatorname{diag}(1,1,0)$                                         \\
        Heisenberg ladder, $S=\tfrac12$~\leanmark{https://sirui-lu.com/CertifiedGaps/docs/CertifiedGaps/Models/HeisenbergLadder.html\#CertifiedGaps.HeisenbergLadder.model}     & $2$-spin rung ($\smI$)    & checkerboard $u_y$     & $P_\alpha\otimes\smI,\;\smI\otimes P_\alpha$ & $-J_\perp\sum_\alpha P_\alpha\otimes P_\alpha$                        \\
        Ising ladder~\leanmark{https://sirui-lu.com/CertifiedGaps/docs/CertifiedGaps/Models/TwoLegIsing.html\#CertifiedGaps.TwoLegIsing.model}                                  & $2$-spin rung ($\smI$)    & $\sigma^z$ basis       & $\sigma^z\otimes\smI,\;\smI\otimes\sigma^z$  & $-\sigma^z\otimes\sigma^z-g(\sigma^x\otimes\smI+\smI\otimes\sigma^x)$ \\
        Mixed-field Ising~\leanmark{https://sirui-lu.com/CertifiedGaps/docs/CertifiedGaps/Models/IsingLongitudinal.html\#CertifiedGaps.IsingLongitudinal.model}                 & $1$ spin ($\smI$)         & $\sigma^z$ basis       & $\sigma^z$                                   & $-g\sigma^x-h\sigma^z$                                                \\
        \bottomrule
    \end{tabular}
\end{table*}

\subsection{Seed bounds from verified finite-ring block spectra}
\label{models:sec:spectra}

Once a model is in the reflection-positive form~\eqref{models:eq:rp},
the doubling recurrence of Section~\ref{sm:sec:trotter} applies. For
the first six models in
Table~\ref{tab:models}, the correlation length is short enough that the
seed scale $n_0$ lies between $2$ and $12$ unit cells. Even there, the
upper bounds $(p_0,q_0)$ need two-sided enclosures of three partition
functions, and a dense eigensolve returns only floating-point numbers
whose accuracy nothing certifies. The remedy is a posteriori: take the
floating-point eigenbasis as a candidate and certify it by residual
norms whose own rounding error is bounded.

To enclose the complete spectrum of $H_L$ at $L\in\{n_0,2n_0\}$ (up to
$16$ spin-$1/2$ or $12$ spin-$1$ sites), we block-diagonalize $H_L$ by
translation, lattice reflection, and conserved spin or parity quantum
numbers into self-adjoint blocks $B_q$ whose exact integer multiplicities
$m_q$ sum to $d^L$, with every entry enclosed in rational or interval
arithmetic. From a floating-point eigenbasis $V$ and candidate
eigenvalues $\tilde\lambda_i$ of $B_q$, we bound the Frobenius norms of
the residual $R=B_qV-V\operatorname{diag}(\tilde\lambda)$ and orthogonality
error $G=V^{\mathsf T}V-\smI$ (including floating-point rounding
error~\cite{Rump2010Verification}); Ostrowski's and Weyl's theorems then
enclose the sorted exact eigenvalues $\lambda_i$ of $B_q$ within
\begin{equation}
    |\lambda_i-\tilde\lambda_i|\le
    \frac{2\|G\|_F\max_j|\tilde\lambda_j|+(1+\|G\|_F)\|R\|_F}{1-\|G\|_F}
    \qquad(\|G\|_F<1).
    \label{models:eq:weyl}
\end{equation}
Summing $m_qe^{-\beta\lambda_i}$ over all blocks in interval arithmetic
encloses $\log Z_{n_0}(\beta_0)$, $\log Z_{2n_0}(\beta_0)$, and
$\log Z_{2n_0}(2\beta_0)$, yielding the certified rational upper bounds
$(p_0,q_0)$. The staggered-field, dimerized, ladder, and mixed-field
Ising models in Table~\ref{models:tab:summary} come from this pipeline
directly; the single-ion anisotropy model at $D=4$ uses the half-length
lift below.

When $2n_0$ reaches the limit of block diagonalization, a direct seed at
length $2n_0$ is out of reach, since both of its purities need the
partition function of a ring of $4n_0$ cells. The spectra already in
hand at lengths $n_0$ and $2n_0$ bound $1-\Ps(n_0,\beta_0)\le s$,
$1-\Ps(n_0,2\beta_0)\le s_2$, and $1-\Pt(2n_0,\beta_0)\le t$. The
obstacle is the thermal purity at $4n_0$, which no computed spectrum
reaches; the torus identity~\eqref{sm:eq:torus} at length $2n_0$
supplies it:
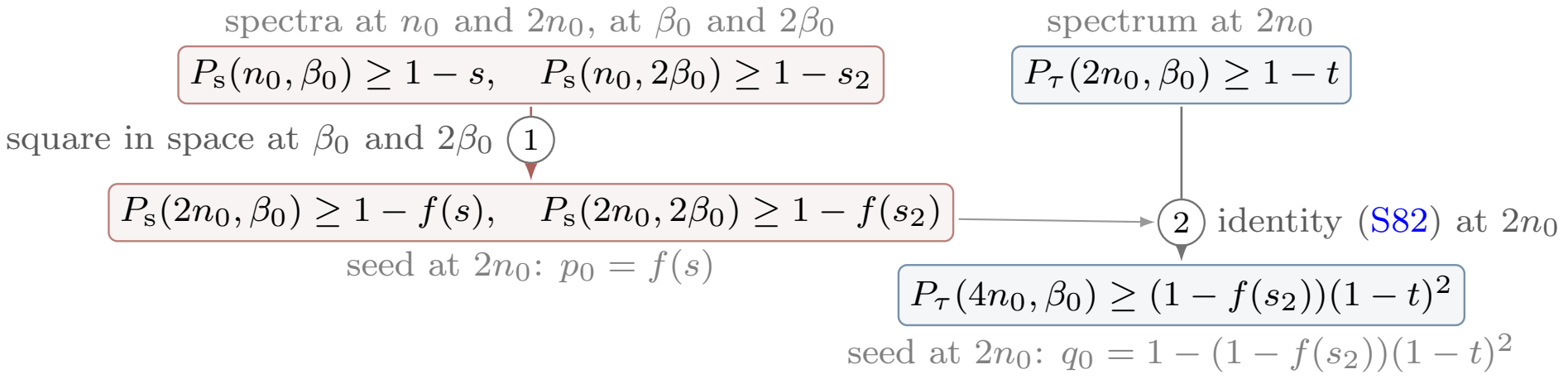
\begin{proofdiagram}
    \halflengthlift
\end{proofdiagram}
Move 2 is this identity with the denominator $\Ps(2n_0,\beta_0)^2\le1$
dropped%
~\leanmark{https://sirui-lu.com/CertifiedGaps/docs/CertifiedGaps/PurityCriterion.html\#CertifiedGaps.thermalPurity_four_mul}:
\begin{equation}
    \Pt(4n_0,\beta_0)
    =\frac{\Ps(2n_0,2\beta_0)\,\Pt(2n_0,\beta_0)^2}{\Ps(2n_0,\beta_0)^2}
    \ge\bigl(1-f(s_2)\bigr)(1-t)^2.
    \label{models:eq:half-length}
\end{equation}
For the spin-$1$ chain with single-ion anisotropy $D=4$ at
$\beta_0=29/10$, the $L=6$ and $L=12$ spectra ($3^{12}=531{,}441$
states) supply $s$, $s_2$, and $t$, and the lift gives the seed at
$n_0=12$ without the $24$-site ring ($3^{24}\simeq2.8\times10^{11}$
states). It sharpens the uniform gap bound at $D=4$ from $0.6946$ (the
Lean chain from the direct seed $(6,9/4)$%
~\leanmark{https://sirui-lu.com/CertifiedGaps/docs/Leaderboard/Records/SpinOneLargeD.html\#Leaderboard.Records.SpinOneLargeD.D_4})
to $0.8994$ for all even
$L\ge12$; the short rings $L=6,8,10$ are covered directly by their
verified spectral gaps.

\begin{table*}[t]
    \caption{Seed scales $(n_0,\beta_0)$, certified starting bounds
        $p_0\ge1-\Ps(n_0,\beta_0)$ and $q_0\ge1-\Pt(2n_0,\beta_0)$, and
        resulting uniform gap bounds for all eight models in
        Table~\ref{tab:models} ($n_0$ counts sites, two-spin dimer cells, or
        two-spin rungs as in Table~\ref{tab:models}). The pairs $(p_0,q_0)$ are
        four-digit upper bounds on the enclosures of the model certificates and
        enter the Lean theorems as hypotheses; the last row is the lifted seed
        $(k,r)=(28,17/8)$ of Table~\ref{cert:tab:lifted}. The printed gap is the
        interval bound of the certificate; each linked Lean theorem proves the
        rational-chain bound~\eqref{models:eq:chain-cert} from its seed pair,
        which agrees in the quoted digits except for the single-ion row, whose
        Lean theorem starts from the direct seed $(6,9/4)$ and proves $0.6946$.
        The mixed-field Ising seed is proved unconditionally in the Lean kernel.}
    \label{models:tab:summary}
    \begin{tabular}{@{}llcccc@{}}
        \toprule
        Model                           & Parameter           & Seed $(n_0,\beta_0)$ & $p_0\ge1-\Ps(n_0,\beta_0)$ & $q_0\ge1-\Pt(2n_0,\beta_0)$ & Certified gap $\ge$                                                                                                                                               \\
        \midrule
        Staggered field, $S=\tfrac12$   & $h_{\mathrm{st}}=1$ & $(8,21/4)$           & $6.937\times10^{-4}$       & $7.327\times10^{-4}$        & $1.1489$~\leanmark{https://sirui-lu.com/CertifiedGaps/docs/CertifiedGaps/Models/StaggeredHeisenbergPhysical.html\#CertifiedGaps.StaggeredHeisenberg.physical_gap} \\
        Dimerized, $S=\tfrac12$         & $\delta=1/2$        & $(4,6)$              & $1.133\times10^{-2}$       & $1.224\times10^{-2}$        & $0.5347$~\leanmark{https://sirui-lu.com/CertifiedGaps/docs/CertifiedGaps/Models/DimerizedHeisenbergPhysical.html\#CertifiedGaps.DimerizedHeisenberg.physical_gap} \\
        Dimerized, $S=1$                & $\delta=9/10$       & $(2,7/2)$            & $4.972\times10^{-2}$       & $3.206\times10^{-2}$        & $0.5288$~\leanmark{https://sirui-lu.com/CertifiedGaps/docs/CertifiedGaps/Models/SpinOneDimerized.html\#CertifiedGaps.SpinOneDimerized.gap}                        \\
        Single-ion anisotropy, $S=1$    & $D=4$               & $(12,29/10)$         & $3.633\times10^{-3}$       & $3.141\times10^{-2}$        & $0.8994$~\leanmark{https://sirui-lu.com/CertifiedGaps/docs/CertifiedGaps/Models/SpinOneLargeD.html\#CertifiedGaps.SpinOneLargeD.gap}                              \\
        Heisenberg ladder, $S=\tfrac12$ & $J_\perp=4$         & $(4,2)$              & $3.509\times10^{-2}$       & $2.463\times10^{-2}$        & $1.1024$~\leanmark{https://sirui-lu.com/CertifiedGaps/docs/CertifiedGaps/Models/HeisenbergLadderPhysical.html\#CertifiedGaps.HeisenbergLadder.physical_gap}       \\
        Ising ladder                    & $g=3$               & $(4,3/2)$            & $1.125\times10^{-1}$       & $5.162\times10^{-2}$        & $0.6368$~\leanmark{https://sirui-lu.com/CertifiedGaps/docs/CertifiedGaps/Models/TwoLegIsing.html\#CertifiedGaps.TwoLegIsing.gap}                                  \\
        Mixed-field Ising               & $(g,h)=(1/4,1/20)$  & $(6,97/4)$           & $9.064\times10^{-6}$       & $3.141\times10^{-5}$        & $0.3922$~\leanmark{https://sirui-lu.com/CertifiedGaps/docs/Leaderboard/Records/IsingLongitudinal.html\#Leaderboard.Records.IsingLongitudinal.g_1_4_h_1_20}        \\
        Heisenberg, $S=1$               & isotropic           & $(60,833/32)$        & $3.054\times10^{-2}$       & $4.796\times10^{-2}$        & $0.0729$~\leanmark{https://sirui-lu.com/CertifiedGaps/docs/Leaderboard/Records/SpinOneHeisenberg.html\#Leaderboard.Records.SpinOneHeisenberg.temperature_lifting} \\
        \bottomrule
    \end{tabular}
\end{table*}

Finally, to verify the $J$ doubling steps~\eqref{sm:eq:update} from
$(p_0,q_0)$ and the resulting gap bound $\Delta_L>\tfrac{\log r}{\beta_0}$ in
Lean, whose kernel decides rational inequalities but has no certified
logarithm, we express the
interval bounds~\eqref{cert:eq:interval-gap} and the quadratic tail
of Sec.~\ref{sm:sec:gap} in terms of a rational base factor $r>1$ and
tail threshold $\varepsilon\le1/100$:
\begin{equation}
    r^{2^j}z_j<1-z_j\quad(0\le j<J),\qquad
    p_J,q_J\le\varepsilon\le\frac1{100},\qquad
    5\varepsilon r^{2^J}\le1,
    \label{models:eq:chain-cert}
\end{equation}
where $z_j=1-(1-p_j)^2(1-q_j)$ along the upward-rounded rational
sequence $(p_j,q_j)$. The first inequality is equivalent to
$\tfrac{\log[(1-z_j)/z_j]}{2^j\beta_0}>\tfrac{\log r}{\beta_0}$ on $[2^jn_0,2^{j+1}n_0]$,
and the remaining two initialize the quadratic tail
inequality~\eqref{sm:eq:scalar} for all $j\ge J$. The Lean kernel checks these rational
inequalities directly, and the general theorem%
~\leanmark{https://sirui-lu.com/CertifiedGaps/docs/CertifiedGaps/Ring/Gap.html\#CertifiedGaps.RPRing.gapGT_of_chainCertificate}
concludes that any reflection-positive ring satisfying
$1-\Ps(n_0,\beta_0)\le p_0$ and $1-\Pt(2n_0,\beta_0)\le q_0$ has a unique
ground state and $\Delta_L>\tfrac{\log r}{\beta_0}$ for every even $L\ge n_0$.
All eight seeds of Table~\ref{models:tab:summary} (labeled by model) lie
inside the admissible region of the asymmetric recurrence and below the
symmetric threshold $\varepsilon_*\simeq0.1273$:
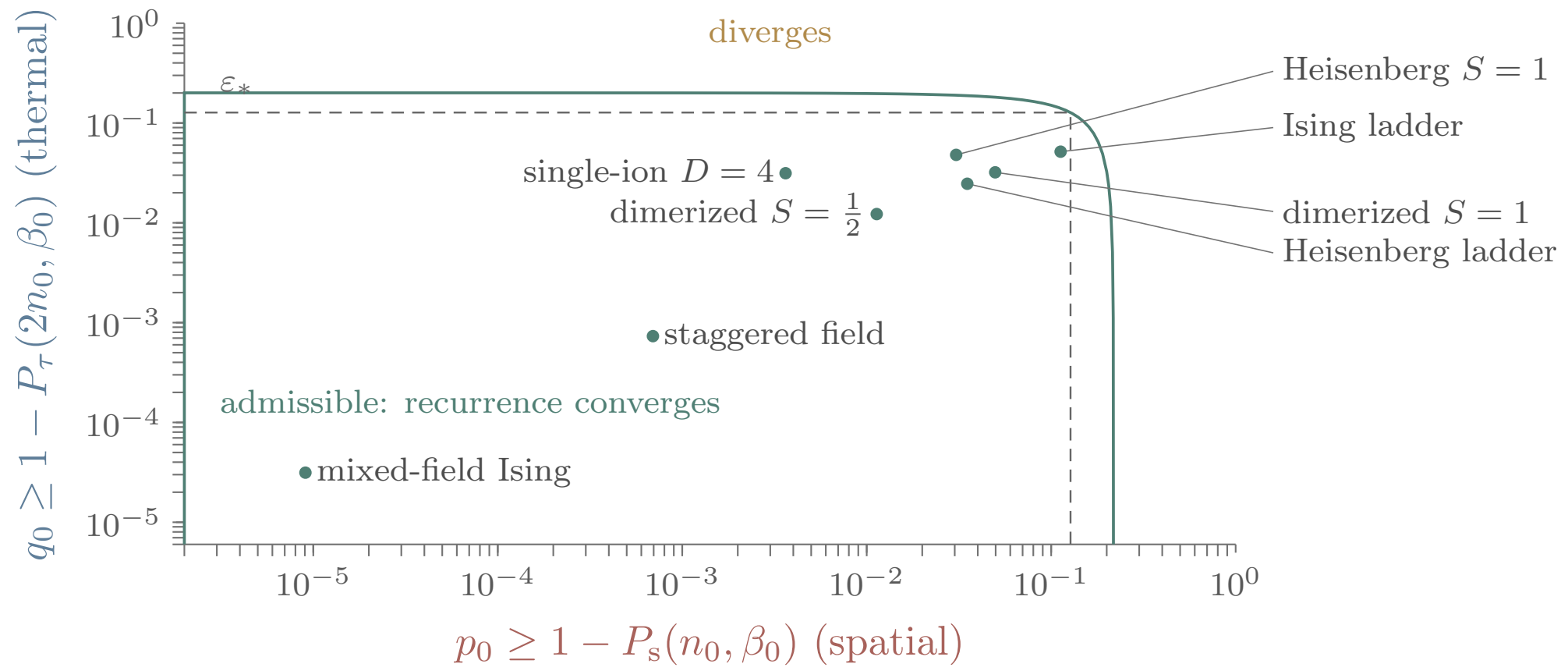
\begin{proofdiagram}
    \seedscatter
\end{proofdiagram}
The boundary is the update~\eqref{sm:eq:update} iterated in floating
point until $(p_j,q_j)$ enters $[0,1/100]^2$; it admits $p_0$ up to
$0.217$ when $q_0\to0$ and $q_0$ up to $0.200$ when $p_0\to0$, so the
recurrence tolerates $1-\Ps$ and $1-\Pt$ almost independently.

\subsection{Unconditional seed proof for the mixed-field Ising chain}
\label{models:sec:ising}

Section~\ref{models:sec:spectra} supplies the seed bounds $(p_0,q_0)$ to
the Lean theorem as hypotheses. To eliminate that hypothesis and prove an
unconditional gap theorem in Lean, the kernel must bound
$Z_{n_0}(\beta_0)$, $Z_{2n_0}(\beta_0)$, and $Z_{2n_0}(2\beta_0)$ from
above and below directly. A full $2^{12}\times2^{12}$ eigensolve inside
a proof-assistant kernel is prohibitive, but at the mixed-field Ising
point $(g,h)=(1/4,1/20)$,
\begin{equation}
    H_L=-\sum_{j=1}^L\sigma_j^z\sigma_{j+1}^z
    -\frac14\sum_{j=1}^L\sigma_j^x
    -\frac1{20}\sum_{j=1}^L\sigma_j^z,
    \label{models:eq:ising}
\end{equation}
we do not need the full spectrum. On a ring of length $L\in\{6,12\}$,
the longitudinal field splits the two classical ferromagnetic states
$\ket{\uparrow\cdots\uparrow}$ and $\ket{\downarrow\cdots\downarrow}$ by
$2hL=L/10$, whereas creating a pair of domain walls costs energy $4$.
Because the transverse field $-g\sum_{j=1}^L\sigma_j^x$ has operator
norm $gL=L/4$, which at $L\le12$ exceeds the ferromagnetic splitting
$2hL$ but remains below the domain-wall cost $4$, bounding the
transverse field by its norm separates all $2^L-2$ domain-wall states
from the two ferromagnetic levels, while two integer trial vectors
locate the two lowest eigenvalues to within $\delta=10^{-10}$:
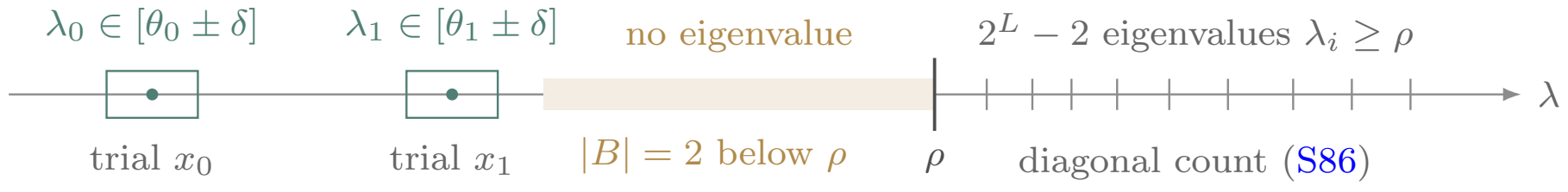
\begin{proofdiagram}
    \isingsandwich
\end{proofdiagram}

\emph{Two low-lying eigenvalues from integer trial residuals.}
For any Hermitian matrix $H$ with eigenvalues $\lambda_i$, any nonzero
trial vector $x$, and any real shift $\theta$, expanding
$\|(H-\theta\smI)x\|^2$ in the eigenbasis shows that at least one
eigenvalue $\lambda_i$ satisfies
$(\lambda_i-\theta)^2\|x\|^2\le\|(H-\theta\smI)x\|^2$. Because $H_L$ has
rational entries in the $\sigma^z$ basis, this residual inequality
reduces to an exact integer dot-product comparison. On each ring
$L\in\{6,12\}$, we supply two integer trial vectors
$x_0,x_1\in\mathbb Z^{2^L}$ scaled by $2^{52}$ and rational candidate
energies $\theta_0<\theta_1$. The Lean kernel checks in exact integer
arithmetic that%
~\leanmark{https://sirui-lu.com/CertifiedGaps/docs/SeedKernel/Ising/Seed.html\#SeedKernel.Ising.exists_near_of_check}
\begin{equation}
    \|(H_L-\theta_a\smI)x_a\|^2\le\delta^2\|x_a\|^2
    \qquad(a=0,1,\quad\delta=10^{-10}),
    \label{models:eq:residual}
\end{equation}
which places two distinct eigenvalues of $H_L$ in the disjoint intervals
$[\theta_0-\delta,\theta_0+\delta]$ and
$[\theta_1-\delta,\theta_1+\delta]$ ($\theta_0\simeq-6.3918$ and
$\theta_1\simeq-5.7966$ at $L=6$; $\theta_0\simeq-12.7835$ and
$\theta_1\simeq-11.5932$ at $L=12$).

\emph{Lower bound on the remaining $2^L-2$ eigenvalues by configuration counting.}
To verify that the two eigenvalues in $[\theta_0\pm\delta]$ and
$[\theta_1\pm\delta]$ are the two lowest levels of $H_L$ and to bound
all remaining $2^L-2$ eigenvalues from below, we split $H_L$ in the
$\sigma^z$ basis $\ket{\sigma}$ ($\sigma\in\{\pm1\}^L$) into the
diagonal Ising energy $E_{\mathrm{cl}}(\sigma)$ and the off-diagonal
transverse field $-g\sum_{j=1}^L\sigma_j^x$. Since each $\sigma_j^x$ has
operator norm one, the transverse field contributes at most $gL\|x\|^2$
to $\langle x,H_Lx\rangle$, giving the quadratic-form lower bound
\begin{equation}
    \langle x,H_Lx\rangle
    \ge\sum_{\sigma\in\{\pm1\}^L}\bigl(E_{\mathrm{cl}}(\sigma)-gL\bigr)|x(\sigma)|^2.
    \label{models:eq:gershgorin}
\end{equation}
For any threshold $\rho$, let
$B=\{\sigma:E_{\mathrm{cl}}(\sigma)-gL<\rho\}$. By the min--max
principle, at most $|B|$ eigenvalues of $H_L$ lie below $\rho$, because
Eq.~\eqref{models:eq:gershgorin} gives $\langle x,H_Lx\rangle\ge\rho\|x\|^2$
on the codimension-$|B|$ subspace of vectors vanishing on $B$. On a
ring, the two uniform configurations \spinrow{1,1,1,1,1,1} and
\spinrow{-1,-1,-1,-1,-1,-1} have classical energies $-L\mp hL$; every
other configuration, such as \spinrow{1,1,-1,-1,1,1}, has at least two
domain walls ($\sum_j\sigma_j\sigma_{j+1}\le L-4$) and at least one spin
against the field ($\sum_j\sigma_j\le L-2$), so its shifted classical
energy is at least
\begin{equation}
    \rho=-(L-4)-h(L-2)-gL,
    \label{models:eq:ising-rho}
\end{equation}
which equals $-3.7>\theta_1+\delta$ at $L=6$ and $-11.5>\theta_1+\delta$
at $L=12$. The kernel checks $|B|=2$ at this $\rho$ over all $2^L$
configurations, so, with two eigenvalues already located below
$\theta_1+\delta<\rho$, the remaining $2^L-2$ satisfy $\lambda_i\ge\rho$.

\emph{Partition-function bounds and unconditional gap theorem.}
Combining the two eigenvalue enclosures with the lower bound
$\lambda_i\ge\rho$ ($i\ge2$) sandwiches $Z_L(\beta)$ from both sides:
dropping all excited states bounds $Z_L(\beta)$ from below by the ground
state alone, $Z_L(\beta)\ge e^{-\beta(\theta_0+\delta)}$, while bounding
the two lowest eigenvalues from below by $\theta_0-\delta$ and
$\theta_1-\delta$ and all remaining $2^L-2$ eigenvalues by $\rho$ gives
the upper bound%
~\leanmark{https://sirui-lu.com/CertifiedGaps/docs/SeedKernel/Ising/Seed.html\#SeedKernel.Ising.partition_upper}
\begin{equation}
    Z_L(\beta)\le e^{-\beta(\theta_0-\delta)}
    \left[1+e^{-\beta(\theta_1-\theta_0)}
        +2^Le^{-\beta(\rho-\theta_0+\delta)}\right].
    \label{models:eq:ising-z}
\end{equation}
The leading factor $e^{-\beta\theta_0}$ cancels between numerator and
denominator in both $\Ps(6,\beta_0)=Z_{12}(\beta_0)/Z_6(\beta_0)^2$ and
$\Pt(12,\beta_0)=Z_{12}(2\beta_0)/Z_{12}(\beta_0)^2$ because
$2\theta_0(6)\simeq\theta_0(12)$ to within $10^{-7}$. Using rational
bounds on $e$ in Lean, evaluating Eq.~\eqref{models:eq:ising-z} at
$\beta_0=97/4$ proves the starting bounds at $(n_0,\beta_0)=(6,97/4)$%
~\leanmark{https://sirui-lu.com/CertifiedGaps/docs/SeedKernel/Ising/Seed.html\#SeedKernel.Ising.first_point_seed},
\begin{equation}
    p_0=\frac{9064}{10^9}\ge1-\Ps(6,97/4),\qquad
    q_0=\frac{3141}{10^8}\ge1-\Pt(12,97/4).
    \label{models:eq:ising-seed}
\end{equation}
At $j=0$, these bounds give $z_0=1-(1-p_0)^2(1-q_0)\simeq4.954\times10^{-5}$,
so $(1-z_0)/z_0>20{,}184$; substituting Eq.~\eqref{models:eq:ising-seed}
into the rational doubling inequalities~\eqref{models:eq:chain-cert}
over $J=14$ steps with base factor $r=13531935/10^3<(1-z_0)/z_0$ proves
a unique ground state and%
~\leanmark{https://sirui-lu.com/CertifiedGaps/docs/Leaderboard/Records/IsingLongitudinal.html\#Leaderboard.Records.IsingLongitudinal.g_1_4_h_1_20}
\begin{equation}
    \boxed{
        \Delta_L>\frac{\log(13531.935)}{97/4}>0.3922
        \qquad\text{for every even }L\ge6
    }
    \label{models:eq:ising-gap}
\end{equation}
unconditionally in Lean.

\section{Lean~4 formalization}
\label{sm:sec:blueprint}
\label{blueprint:sec:lean}

A computer-assisted proof combines an analytic reduction with a finite
numerical calculation. Formalizing the periodic and open-chain spin-1
proofs of Refs.~\cite{OpenAI2026Periodic,OpenAI2026Boundaryfield}
directly in Lean~4~\cite{Moura2021Lean} faces three obstacles:
\begin{enumerate}
    \item \emph{Infinite-dimensional path space.} The continuous-time
          spatial transfer kernel of Ref.~\cite{OpenAI2026Periodic} acts on
          $L^2$ of Poisson bond histories (Sec.~\ref{obc:caps}); formalizing its
          trace identities directly would require path-space measure and operator
          machinery not yet in Mathlib.
    \item \emph{Model specificity.} Writing the analytic steps around the
          isotropic spin-1 Heisenberg Hamiltonian ties the formalization to one
          model, so none of the operator inequalities can be reused for other
          spin chains or ladders.
    \item \emph{Finite numerical inputs.} The seed bounds rely on external
          computations—block diagonalizations, short-ring trace recursions, and
          matrix-product trial contractions—whose floating-point or C/Python
          implementations cannot be trusted inside a proof assistant.
\end{enumerate}

The diagrammatic reformulation in the main text and
Secs.~\ref{sm:sec:trotter}--\ref{sm:sec:models} (and
Sec.~\ref{sm:sec:obc} for open chains) resolves all three
obstacles before formalization begins. First, discretizing imaginary
time into $M$ brickwork slices replaces the path-space kernel by the
finite-dimensional Gram matrix $\smT=\smF^{\mathsf T}\smF\succeq0$
[Eq.~\eqref{sm:eq:positive}]. Instead of constructing a limiting
continuous-time transfer operator in Lean, we prove the three closed
spatial power-sum inequalities~\eqref{sm:eq:closed} at finite $M$ and
pass them to the exact partition functions $Z_L(\beta)$ via the
fixed-$(L,\beta)$ matrix product limit of Sec.~\ref{sm:sec:limit}%
~\leanmark{https://sirui-lu.com/CertifiedGaps/docs/CertifiedGaps/Ring/Gap.html\#CertifiedGaps.RPRing.spatialMomentBounds}.
Likewise, to state ground-state uniqueness and
$\Delta_L=E_1(L)-E_0(L)>c$ without sorting the eigenvalues of $H_L$, we
require a single ground-state index $m_0$ separated from every other
level $m\ne m_0$ by $E_L(m)-E_L(m_0)>c$%
~\leanmark{https://sirui-lu.com/CertifiedGaps/docs/CertifiedGaps/PurityCriterion.html\#CertifiedGaps.GapGT.lt_sub_of_monotone}.
Second, isolating the bond reflection-positivity identity
$-h=\sum_\alpha C_\alpha\otimes FC_\alpha F$ with $FoF=o$
[Eq.~\eqref{models:eq:rp}] makes the periodic scale-doubling theorem
apply uniformly to any reflection-positive unit-cell ring%
~\leanmark{https://sirui-lu.com/CertifiedGaps/docs/CertifiedGaps/Ring/Gap.html\#CertifiedGaps.RPRing.gapGT_of_chainCertificate}.

\begin{table}[ht]
    \caption{Structure of the Lean~4 formalization~\cite{Lu2026CertifiedGaps}
        and its interactive proof blueprint~\cite{Lu2026Blueprint}. Each row
        lists a chapter of the blueprint, the corresponding manuscript
        sections, the theorem proved in Lean~4, and the finite numerical
        hypotheses (if any) on which that theorem depends.}
    \label{blueprint:tab:map}
    \footnotesize
    \begin{tabular}{@{}llll@{}}
        \toprule
        Blueprint chapter & Manuscript                                                           & Theorem proved in Lean~4 & Numerical hypotheses \\
        \midrule
        2.\ Two-purity criterion
                          & Main text; Sec.~\ref{sm:sec:gap}
                          & Uniform gap from purities $\ge1-\varepsilon_0$                                                                         %
        ~\leanmark{https://sirui-lu.com/CertifiedGaps/docs/CertifiedGaps/PurityCriterion.html\#CertifiedGaps.gapGT_of_seeds}
                          & None                                                                                                                   \\
        3.\ Asymmetric doubling \& chains
                          & Secs.~\ref{sm:sec:asymmetric}, \ref{models:sec:spectra}
                          & Uniform gap from rational doubling chain                                                                               %
        ~\leanmark{https://sirui-lu.com/CertifiedGaps/docs/CertifiedGaps/ChainCheck.html\#CertifiedGaps.gapGT_of_chainCertificate}
                          & None (kernel-checked)                                                                                                  \\
        4.\ Reflection-positive rings
                          & Secs.~\ref{sm:sec:transfer}--\ref{sm:sec:limit}, \ref{models:sec:rp}
                          & Reflection-positive ring gap theorem                                                                                   %
        ~\leanmark{https://sirui-lu.com/CertifiedGaps/docs/CertifiedGaps/Ring/Gap.html\#CertifiedGaps.RPRing.gapGT_of_chainCertificate}
                          & None                                                                                                                   \\
        5.\ Trace bounds \& sector tables
                          & Sec.~\ref{sm:sec:seeds}
                          & Seed purities from short-ring moments                                                                                  %
        ~\leanmark{https://sirui-lu.com/CertifiedGaps/docs/CertifiedGaps/Seed/Ring.html\#CertifiedGaps.RPRing.seeds_of_certificate}
                          & None                                                                                                                   \\
        6.\ Spin-1 Heisenberg ring
                          & Main text; Sec.~\ref{sm:sec:seeds}
                          & Spin-1 ring gap $\Delta_L>0.07290$ ($L\ge36$)                                                                          %
        ~\leanmark{https://sirui-lu.com/CertifiedGaps/docs/Leaderboard/Records/SpinOneHeisenberg.html\#Leaderboard.Records.SpinOneHeisenberg.temperature_lifting}
                          & Short-ring traces ($4\le n\le12$)                                                                                      \\
        7.\ Open chains with end fields
                          & Sec.~\ref{sm:sec:obc}
                          & Open spin-1 gap ($n\ge1921$)                                                                                           %
        ~\leanmark{https://sirui-lu.com/CertifiedGaps/docs/CertifiedGaps/Heisenberg/OpenChain.html\#CertifiedGaps.SpinOneHeisenberg.spin_one_boundary_field_gap}
                          & Four seeds at $\beta_0=49$                                                                                             \\
        8.\ Spin chains \& ladders
                          & Sec.~\ref{sm:sec:models}; Table~\ref{models:tab:summary}
                          & Uniform gaps for Table~\ref{tab:models} models                                                                         %
        ~\leanmark{https://sirui-lu.com/CertifiedGaps/docs/CertifiedGaps/Models/StaggeredHeisenberg.html\#CertifiedGaps.StaggeredHeisenberg.gap}
                          & Seed pair $(p_0,q_0)$                                                                                                  \\
        9.\ Kernel-checked seeds
                          & Secs.~\ref{sm:sec:seeds}, \ref{models:sec:ising}
                          & Unconditional Ising gap; $S=1$ trials                                                                                  %
        ~\leanmark{https://sirui-lu.com/CertifiedGaps/docs/SeedKernel/Ising/Seed.html\#SeedKernel.Ising.ising_gap}
                          & None (Ising); $n\ge5$ traces                                                                                           \\
        \bottomrule
    \end{tabular}
\end{table}

Third, separating the universal doubling recurrence from the finite seed
allows the formalization to treat the numerical inputs at two levels
(Table~\ref{blueprint:tab:map}):
\begin{enumerate}
    \item \emph{Seed-hypothesis theorems (Chapters~6--8).} Because the
          asymmetric update~\eqref{sm:eq:update} is a rational map and the
          interval condition~\eqref{models:eq:chain-cert} replaces logarithms by
          rational power comparisons $r^{2^j}z_j<1-z_j$, the step from any
          rational seed pair $(p_0,q_0)$ to a uniform gap $\Delta_L>(\log r)/\beta_0$
          reduces to finitely many rational inequalities checked in the Lean
          kernel. Taking $(p_0,q_0)$ (or the short-ring trace table of
          Table~\ref{cert:tab:traces}) as an explicit hypothesis instantiates
          machine-checked gap theorems for all eight periodic models in
          Table~\ref{tab:models} and for the open spin-1 chain with endpoint
          fields (Theorem~\ref{obc:theorem}), while allowing the seed enclosures
          to be certified externally by verified block diagonalizations
          (Sec.~\ref{models:sec:spectra}) or exact integer trace recursions
          (Sec.~\ref{sm:sec:seeds}).
    \item \emph{Kernel-checked seed arithmetic (Chapter~9).} Where the
          finite seed calculation can be reduced to exact integer identities, the
          Lean kernel verifies the seed calculation itself. For the mixed-field
          Ising chain at $(g,h)=(1/4,1/20)$ (Sec.~\ref{models:sec:ising}), the
          kernel checks the integer trial-vector residuals~\eqref{models:eq:residual}
          and the classical configuration count~\eqref{models:eq:ising-rho},
          proving the uniform gap~\eqref{models:eq:ising-gap} unconditionally
          without external numerical hypotheses%
          ~\leanmark{https://sirui-lu.com/CertifiedGaps/docs/SeedKernel/Ising/Seed.html\#SeedKernel.Ising.ising_gap}.
          For the periodic and open spin-1 chains, the kernel checks the exact
          integer contractions~\eqref{cert:eq:mps-contractions} of the $72$- and
          $120$-site periodic matrix-product trial states, the four open-chain
          trial states on $34$, $78$, $142$, and $240$ sites, and the shortest
          ring trace recurrences ($n=4$ periodic and nine of fifteen boundary
          traces), reducing the remaining hypotheses to the short-ring integer
          trace totals for $5\le n\le12$%
          ~\leanmark{https://sirui-lu.com/CertifiedGaps/docs/SeedKernel/Heisenberg/Integers.html\#CertifiedGaps.SpinOneHeisenberg.gap_of_hornerTotalsLarge}
          and the remaining six boundary traces%
          ~\leanmark{https://sirui-lu.com/CertifiedGaps/docs/SeedKernel/Heisenberg/Integers.html\#CertifiedGaps.SpinOneHeisenberg.spin_one_boundary_field_gap_of_hornerTotalsLarge}.
\end{enumerate}

The human-readable proof blueprint~\cite{Lu2026Blueprint}
(\url{https://sirui-lu.com/CertifiedGaps/blueprint/}) links every
definition, lemma, and theorem in the dependency graph to its Lean~4
declaration, and the companion repository~\cite{Lu2026CertifiedGaps}
bundles the manuscript, standalone numerical certificate checkers, and
formal proofs following the Agentic Publication Protocol~\cite{Lu2026APP}.

\section{Open spin-1 chains with boundary fields}
\label{sm:sec:obc}

\subsection{Open-chain gap theorem and boundary remainder}
\label{obc:setup}

We extend the diagrammatic doubling argument from periodic rings
to open chains with endpoint magnetic fields. Consider the spin-1
Heisenberg chain with equal endpoint fields (\fieldsame),
\begin{equation}
    H_n^{\same}
    =\sum_{i=1}^{n-1}\bm S_i\cdot\bm S_{i+1}
    -h(S_1^z+S_n^z),\qquad h=\frac35,
    \label{obc:model}
\end{equation}
where $\sum_\alpha(S^\alpha)^2=2I$ and the exchange coupling is one.
We give a diagrammatic proof of the following theorem from
Ref.~\cite{OpenAI2026Boundaryfield}, using its certified finite inputs.

\begin{theorem}[Boundary-field gap~\cite{OpenAI2026Boundaryfield}%
        ~\leanmark{https://sirui-lu.com/CertifiedGaps/docs/CertifiedGaps/Heisenberg/OpenChain.html\#CertifiedGaps.SpinOneHeisenberg.spin_one_boundary_field_gap}]
    \label{obc:theorem}
    For every odd $n\ge1921$, $H_n^{\same}$ has a unique ground state and
    \[
        \Delta_n>\frac{\log 10}{392}\simeq0.005874.
    \]
\end{theorem}
\noindent The two constants come from three doublings of the certified
seed $(M_2,\beta_2)=(240,49)$; scale $5$ is the first at which every
tracked bound is below $10^{-2}$:
\[
    M_5=2^3\cdot240=1920,\qquad \beta_5=2^3\cdot49=392,\qquad
    n\ge M_5+1=1921.
\]

Without the endpoint fields, the effective spin-$1/2$ modes at the
two ends produce four low-lying states whose splitting is exponentially
small in the chain length~\cite{Kennedy1990Exact}. Applying equal fields on an odd chain lifts this
boundary degeneracy~\cite{OpenAI2026Boundaryfield}.

Our strategy is to bound the thermal purity of the open chain at a
scale $\beta$ proportional to the length, just as in the periodic
argument. Order the Gibbs weights of $H_n^{\same}$ as $w_0\ge w_1\ge\cdots$.
The thermal purity $\Pt^{\same}(n,\beta)=\sum_m w_m^2$ then bounds the
leading weight and the gap:
\begin{equation}
    w_0\ge\Pt^{\same}(n,\beta),\qquad
    e^{-\beta\Delta_n}
    \le\frac{1-\Pt^{\same}(n,\beta)}{\Pt^{\same}(n,\beta)},\qquad
    \Delta_n=E_1(n)-E_0(n).
    \label{obc:target}
\end{equation}
A thermal purity above $1/2$ forces a unique ground state and bounds the
gap from below. What changes in the open-chain setting is how we
propagate the thermal purity across doubled lengths and inverse
temperatures.

To see where the new difficulty enters, consider the spatial transfer
operator at a fixed temperature. Suppose that $X_\beta$ has a simple
positive leading eigenvalue $\lambda$ (\specbars[mpored]{.9,.06,.04}), and
write the normalized transfer operator as
\begin{equation}
    T:=X_\beta/\lambda=\Pi+R,\qquad
    \Pi=\ket{\psi}\bra{\psi},\qquad
    \Pi R=R\Pi=0,\quad \|\psi\|=1.
    \label{obc:split}
\end{equation}
In a periodic ring, the trace weights the leading projector $\Pi$ by one.
In an open chain, the chain ends are contracted with boundary vectors,
so the leading contribution is weighted instead by how much of $v$ lies
along $\psi$ (\overlapglyph):
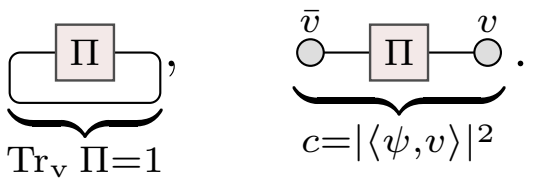
\begin{equation}
    \underbrace{\tnchain[ring]{\Pi}}_{\trv\Pi=1},
    \qquad
    \underbrace{\tnchain[capl=\bar v,capr=v]{\Pi}}_
    {c=|\langle\psi,v\rangle|^2}.
    \label{obc:leading-diagram}
\end{equation}
Because $c$ is not fixed by normalization and could a priori vanish,
squaring in space no longer closes on its own. With $T$ self-adjoint, the
same boundary vector at both ends, and an even length $m$, the remainder of the
normalized boundary matrix element is a squared norm, hence nonnegative:
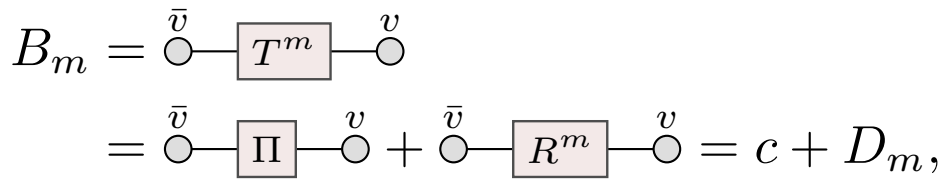
\begin{align}
    B_m
        &=\tnchain[capl=\bar v,capr=v]{T^m}\notag\\
        &=\tnchain[capl=\bar v,capr=v]{\Pi}
    +\tnchain[capl=\bar v,capr=v]{R^m}
    =c+D_m,\notag\\
    D_m &=\|R^{m/2}v\|^2\ge0.
    \label{obc:boundary-split}
\end{align}
To show that the leading eigenvector dominates $B_m$, we bound the
remainder's share of $B_m$ rather than the remainder itself,
\begin{equation}
    e_m:=\frac{D_m}{B_m}=1-\frac{c}{B_m},
    \label{obc:defect}
\end{equation}
the grey fraction of \stackbar{.14} at the certified seed value
$e_{142}\le.14$ of Eq.~\eqref{obc:seeds}. Increasing the chain length
suppresses $D_m$ by powers of $\|R\|$ while leaving $c$ unchanged.

Doubling the inverse temperature changes both $v$ and $\psi$, so $c$
does not bound the new overlap $c'$; Step~1 (Sec.~\ref{obc:move-one})
removes $c'$ by comparing two lengths at the new temperature.

The recurrence therefore tracks three contractions at every scale $j$,
each with its small quantity and its certified seed value:
\begin{center}
    \renewcommand{\arraystretch}{1.5}
    \begin{tabular}{@{}clllc@{}}
        \toprule
            & contraction                        & tracked quantity                & seed ($j=2$) & controlled in                                  \\
        \midrule
        (R) & \obcring\ ring purity              & $\varepsilon_j\ge1-\Ps,\,1-\Pt$ & $.052$       & Sec.~\ref{obc:ring}                            \\
        (E) & \obcends\ boundary overlap         & $e_{k_j}(j)=D/B$                & $.14$        & Sec.~\ref{obc:move-one}                        \\
        (T) & \obcthermal\ opposite-field purity & $1-\Pt^{\opp}(M_j,\beta_j)$     & $.21$        & Secs.~\ref{obc:move-two}--\ref{obc:move-three} \\
        \bottomrule
    \end{tabular}
\end{center}
Section~\ref{obc:caps} constructs the boundary vectors, Sec.~\ref{obc:initialization}
runs three explicit steps from $j=2$ to $j=5$ and closes the induction,
and Sec.~\ref{obc:comparison} passes to equal fields and odd lengths.

\subsection{Continuous-time transfer kernel and boundary vectors}
\label{obc:caps}

We first represent the open-chain partition functions as transfer-operator
matrix elements using the continuous-time bond expansion of
Ref.~\cite{OpenAI2026Boundaryfield}, which sums over the times of bond
insertions rather than placing them on a finite Trotter grid.
As in the periodic proof (Sec.~\ref{cert:sec:transfer}), we shift the
Hamiltonian by a constant $\kappa$ per site. The shift is chosen so that $-\kappa$
lies just above the energy per site of the $72$-site matrix-product
trial state of Ref.~\cite{OpenAI2026Periodic} ($E_0(72)/72<-\kappa$,
Sec.~\ref{cert:sec:inputs}):
\begin{equation}
    \kappa=\frac{700741}{500000}=1.401482.
    \label{obc:kappa}
\end{equation}
Throughout this section, $Z_n$ and all open-chain partition functions
include the shift $\kappa nI$. It places the leading transfer eigenvalue
near one at the seed scales and cancels from every purity and spectral
gap.

In the Cartesian basis, $G_\alpha=iS^\alpha$ is real antisymmetric and
\[
    -\bm S_i\cdot\bm S_{i+1}
    =\sum_{\alpha=x,y,z}G_{\alpha,i}G_{\alpha,i+1}.
\]
Expanding $e^{-\beta H}$ on each bond produces a bond history $a$, an
ordered list of events $(\tau,\alpha)$ with $0<\tau<\beta$ and
$\alpha\in\{x,y,z\}$. The bond measure $\mu_\beta$ sums the labels and
integrates the ordered times, with unit mass on the empty history:
\[
    \mu_\beta(\text{all histories})
    =\sum_{k\ge0}\underbrace{3^k}_{\text{labels}}\,
    \underbrace{\frac{\beta^k}{k!}}_{0<\tau_1<\cdots<\tau_k<\beta}
    =e^{3\beta}.
\]
Between the left bond history $a$ and the right bond history $b$ of a
site, the shifted local tensor is
\begin{equation}
    W_\beta^{ss'}(a,b)
    =e^{-\kappa\beta}
    \bra{s}\mathcal T_\tau
    \prod_{(\tau,\alpha)\in a\cup b}G_\alpha\ket{s'},
    \label{obc:local}
\end{equation}
where $\mathcal T_\tau$ orders the events of both bonds by time, later
ones to the left:
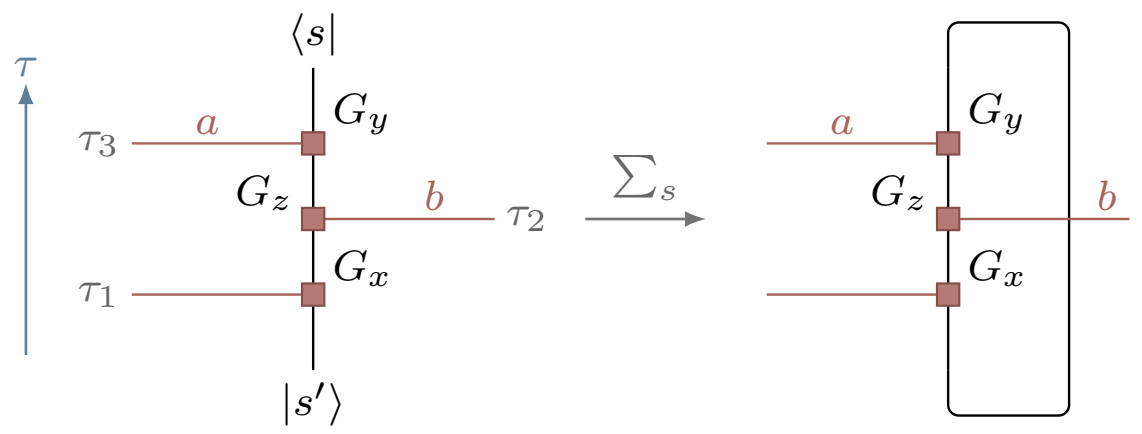
\begin{proofdiagram}
    \worldlinekernel
\end{proofdiagram}
Tracing the physical indices defines the transfer kernel $X_\beta(a,b)$
on $L^2(\mu_\beta)$. It is real, and symmetric because the product
follows the times, not the sides.

To represent a boundary field along $z$, we contract the endpoint bond
history with the boundary vector
\begin{equation}
    v_\beta(a)=
    \begin{cases}
        (ih)^{|a|}, & \text{all event labels are }z, \\
        0,          & \text{otherwise},
    \end{cases}
    \qquad \|v_\beta\|^2=e^{\beta h^2}.
    \label{obc:vector}
\end{equation}
Its value on the empty history is one, and its norm follows by summing
$(\beta h^2)^k/k!$ over the number $k$ of $z$ events.

Each event on the boundary bond inserts $G_z=iS^z$ at the end site at
its time. It enters the time-ordered product like an interior bond
event, and summing counts and integrating times exponentiates the
endpoint field. Conjugation reverses the sign of $ih$, so contracting
with $v_\beta$ at both ends gives the auxiliary opposite-field chain
\fieldopp\ (left). The $\pi$ rotation $P=\operatorname{diag}(1,-1,-1)$
about $x$ acts on the right boundary vector as $U_P$; it negates every
$z$ event there and restores equal fields \fieldsame\ (right):
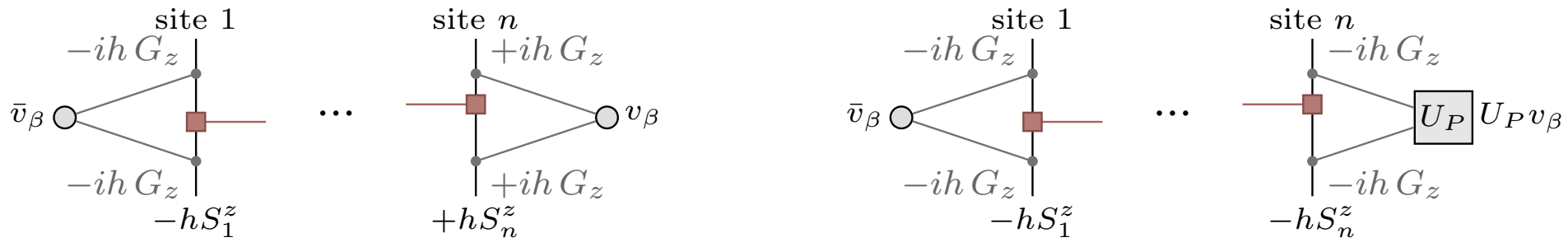
\begin{equation}
    \capfields{0}\qquad\qquad\capfields{1}
    \label{obc:capfields}
\end{equation}
\begin{equation}
    H_n^{\opp}=H_n^{\same}+2hS_n^z.
    \label{obc:opposite}
\end{equation}
The ring bounds of Sec.~\ref{obc:ring} concern the spectrum of $X_\beta$,
so both open-chain partition functions must become matrix elements of
it. In diagram form,
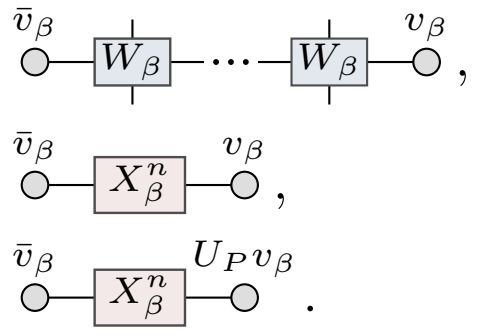
\begin{align}
    e^{-\beta(H_n^{\opp}+\kappa nI)} &=\obcW,\notag\\
    Y_n(\beta)
    :=\Tr e^{-\beta(H_n^{\opp}+\kappa nI)}
                                     &=\obccap{X_\beta^n}{v_\beta},\notag\\
    Y_n^{\same}(\beta)
    :=\Tr e^{-\beta(H_n^{\same}+\kappa nI)}
                                     &=\obccap{X_\beta^n}{U_Pv_\beta}.
    \label{obc:capping}
\end{align}
The two end bond histories meet the boundary vectors, and the physical
trace at every site turns the row of tensors into $X_\beta^n$; the two
chains differ only by $U_P$ at the right end.

Spin rotations act on bond histories by real orthogonal operators that
commute with $X_\beta$, so $[U_P,X_\beta]=0$. Contracting the
time-ordered expansion site by site is exact by Fubini's theorem, since
$\|G_\alpha\|=1$ bounds the absolutely summed series by
$3^ne^{3\beta(n-1)+2\beta h}$ before the energy shift.

The recurrence runs on the auxiliary opposite-field chain at even
length $m$, not on the odd equal-field chain of the theorem. With the
same boundary vector at both ends, the remainder of $Y_m(\beta)$ is the
squared norm $D_m\ge0$ of Eq.~\eqref{obc:boundary-split}, on which the
doubling closes.

\subsection{Bulk ring purities and transfer-operator bounds}
\label{obc:ring}

The open-chain recurrence of Steps~1--3 needs three facts about the bulk
transfer operator at every scale $j$. They are a simple positive leading
eigenvalue that $U_P$ fixes, a decay rate for the remainder across half
a chain, and the ratio $r_j$ of leading eigenvalues at consecutive
temperatures. None is certified directly at large $j$, but the periodic
argument proves ring purity bounds at every scale from one certified
seed, and this subsection extracts the three facts from those purities.
The open chain thus borrows its bulk control from the ring; only the
boundary needs new inputs.

The recurrence runs over the dyadic scales
\begin{equation}
    \beta_j=49\,2^{j-2},\qquad n_j=120\,2^{j-2},\qquad
    M_j=2n_j,\qquad j\ge2,
    \label{obc:scales}
\end{equation}
starting at the certified point $j=2$ of Ref.~\cite{OpenAI2026Boundaryfield}.
The ring length controls the bulk and the open length carries the
thermal purity. The boundary ratio $e_k$ is tracked at the certified length
$142$ first and at the previous open length thereafter,
\begin{equation}
    k_2=142,\qquad k_{j+1}=M_j,
    \label{obc:k}
\end{equation}
so $k_j=M_j/2$ for $j\ge3$:
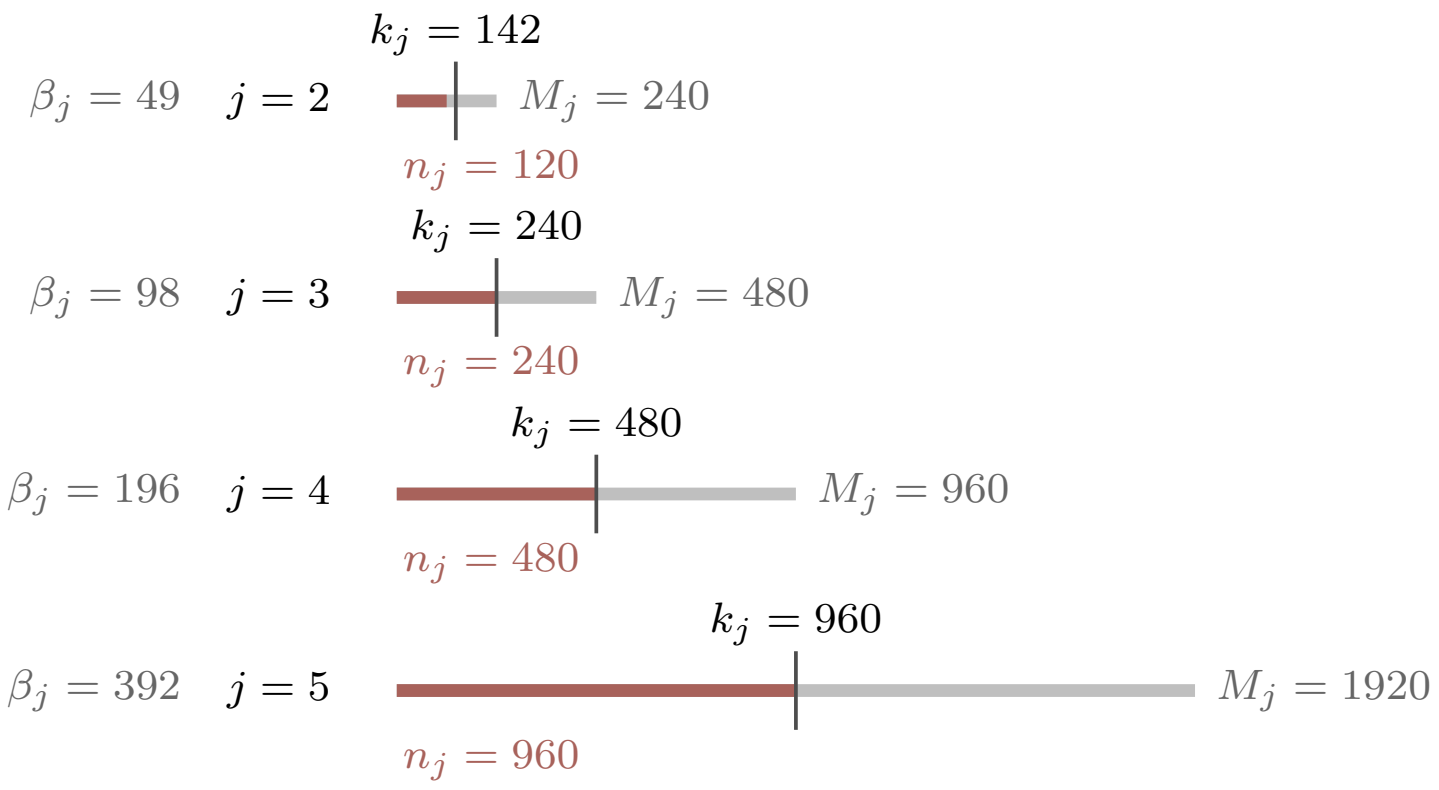
\begin{proofdiagram}
    \obcscales
\end{proofdiagram}
Ref.~\cite{OpenAI2026Boundaryfield} certifies four inputs at $j=2$%
~\leanmark{https://sirui-lu.com/CertifiedGaps/docs/CertifiedGaps/Heisenberg/BoundaryInitTable.html\#CertifiedGaps.SpinOneHeisenberg.BoundaryInit.boundaryDefect_le_of_table}:
\begin{equation}
    \renewcommand{\arraystretch}{1.4}
    \begin{tabular}{@{}llcl@{}}
        \toprule
        input                      & controls         & weights                        & bound                       \\
        \midrule
        $\Ps(120,49)\ge.948$       & bulk spectrum    & \specbars[mpored]{.95,.03,.02} & $\varepsilon_2=1-.948=.052$ \\
        $\Pt(240,49)\ge.948$       & bulk spectrum    & \specbars{.95,.03,.02}         & $\varepsilon_2$             \\
        $e_{142}(2)\le.14$         & boundary overlap & \stackbar{.14}                 & $E_2=.14$                   \\
        $\Pt^{\opp}(240,49)\ge.79$ & thermal          & \specbars{.79,.13,.08}         & $\eta_2=1-.79=.21$          \\
        \bottomrule
    \end{tabular}
    \label{obc:seeds}
\end{equation}
Here $\Ps(n,\beta)=Z_{2n}(\beta)/Z_n(\beta)^2$ and
$\Pt(n,\beta)=Z_n(2\beta)/Z_n(\beta)^2$ are ring purities, $e_m(j)$ is
Eq.~\eqref{obc:defect} at $\beta_j$, and
$\Pt^{\opp}(m,\beta)=Y_m(2\beta)/Y_m(\beta)^2$.

The periodic argument squares $1-\Ps$ and $1-\Pt$ with constant $5$ only
below the threshold $1/100$, which the seed $\varepsilon_2=.052$
exceeds; repeating the four-step update with constant $6$ gives
\begin{equation}
    \Ps(n_j,\beta_j),\ \Pt(M_j,\beta_j)\ge1-\varepsilon_j,\qquad
    \varepsilon_2=.052,\qquad
    \varepsilon_{j+1}=6\varepsilon_j^2.
    \label{obc:ring-purities}
\end{equation}
One scale of the periodic argument, with $\varepsilon=\varepsilon_j$,
updates the spatial and thermal purities in four moves:
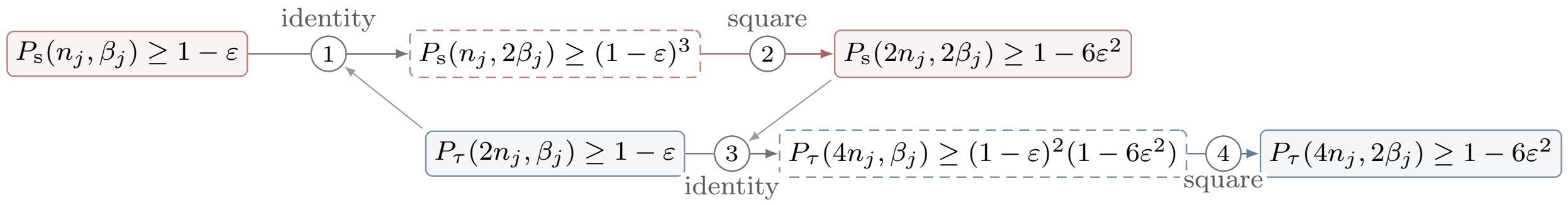
\begin{proofdiagram}
    \ringsteps
\end{proofdiagram}
Squaring and renormalizing a positive
trace-one operator of purity $P$ concentrates its weights
($\specbars[black!55]{.5,.3,.2}\mapsto\specbars[black!55]{.66,.24,.1}$),
and its new purity $P'$ satisfies
\begin{equation}
    1-P'\le f(1-P),\qquad
    f(z)=\frac{z^2}{2(1-z)^2}.
    \label{obc:square}
\end{equation}
For $0\le\varepsilon\le.052$, Move~2 gives
\begin{equation}
    f(1-(1-\varepsilon)^3)
    =\frac{\varepsilon^2}{2}
    \left[\sum_{s=1}^3(1-\varepsilon)^{-s}\right]^2
    \le6\varepsilon^2\le\varepsilon,
    \label{obc:six}
\end{equation}
the bracket being at most $3.35$ at $\varepsilon=.052$, so the prefactor is
at most $5.6<6$. Since
$(1-\varepsilon)^2(1-6\varepsilon^2)\ge(1-\varepsilon)^3$, Move~4
gives the same bound $6\varepsilon^2$, and induction on $j\ge2$ proves
Eq.~\eqref{obc:ring-purities}.

Write $\lambda_j$ for the leading eigenvalue of $X_{\beta_j}$ and
$T_j=X_{\beta_j}/\lambda_j=\Pi_j+R_j$ as in Eq.~\eqref{obc:split}. The
three bulk facts, each with the step that consumes it, are%
~\leanmark{https://sirui-lu.com/CertifiedGaps/docs/CertifiedGaps/Open/Criterion.html\#CertifiedGaps.OpenMomentWitness.mu_pow_ge}
\begin{align}
     &\text{(i)}\quad \lambda_j>0\text{ simple},\quad U_P\Pi_j=\Pi_j
     &                                                                           &\text{overlap $c$ in Step~1 and Sec.~\ref{obc:comparison}},\notag\\
     &\text{(ii)}\quad \|R_j\|^{n_j}\le d(\varepsilon_j)<1,\quad
    d(\varepsilon)=\frac{\varepsilon}{2-3\varepsilon}
     &                                                                           &\text{suppression factor $q$ in Step~1},\notag\\
     &\text{(iii)}\quad r_j^{M_j}\ge(1-\varepsilon_j)(1-\varepsilon_{j+1}),\quad
    r_j=\frac{\lambda_{j+1}}{\lambda_j^2}\le1
     &                                                                           &\text{bulk factor in Step~2 and Sec.~\ref{obc:comparison}}.
    \label{obc:ring-bounds}
\end{align}

No transfer eigenvalue is computed, so (i) and (ii) must follow from
the ring purity alone. A purity at least $1-\varepsilon>1/2$ forces one
weight $p_0$ above $1/2$ (\dominantglyph), and its tail bounds every
subleading ratio. The purity is at most $p_0$,
so the tail $z=1-p_0$ is below $1/2$. The purity is also at most
$p_0^2+z^2=1-2z(1-z)$, so $2z(1-z)\le\varepsilon$ and
\[
    z\le\varepsilon,\qquad
    z\le\frac{\varepsilon}{2(1-\varepsilon)},\qquad
    \frac{z}{1-z}\le\frac{\varepsilon}{2(1-\varepsilon)-\varepsilon}=d(\varepsilon).
\]
Applying this estimate to the normalized weights
$p_i=|\lambda_i(\beta_j)|^{n_j}/Z_{n_j}(\beta_j)$ yields the subleading tail bound
\begin{equation}
    \sum_{i\ge1}
    \left|\frac{\lambda_i(\beta_j)}{\lambda_j}\right|^{n_j}
    \le d(\varepsilon_j).
    \label{obc:full-tail}
\end{equation}
Since $d(\varepsilon_j)<1$, three consequences follow.
\begin{itemize}
    \item $\lambda_j$ is simple, and $\|R_j\|^{n_j}\le d(\varepsilon_j)<1$
          is property (ii).
    \item $\lambda_j>0$. A negative leading eigenvalue would make
          $\trv X_{\beta_j}^N<0$ for all large odd $N$, because $X_{\beta_j}$ is
          Hilbert--Schmidt and the rest of its spectrum has smaller radius;
          but $Z_N(\beta_j)>0$.
    \item $U_P\Pi_j=\Pi_j$. The cyclic axis permutation and the three $\pi$
          rotations commute with the real kernel $X_{\beta_j}$, so each acts by a
          sign on the leading eigenline. The permutation has order three, so its
          sign is $+1$. It conjugates the three rotations into one another, so
          their signs agree; two of them multiply to the third, so the common
          sign is $+1$.
\end{itemize}

Property (iii) bounds the bulk factor that Step~2 loses when the
temperature doubles. The two transfer operators do not commute, so we
compare both leading eigenvalues with one ring partition function.
First, $\Pt(N,\beta_j)\le1$ and even $N$th roots as $N\to\infty$
give $\lambda_{j+1}\le\lambda_j^2$, so $r_j\le1$. At length
$M_j=n_{j+1}$,
\begin{align}
    \lambda_{j+1}^{M_j}
     &\ge(1-\varepsilon_{j+1})Z_{M_j}(\beta_{j+1})
     &                                             &\text{leading weight at }\beta_{j+1},\notag\\
     &\ge(1-\varepsilon_{j+1})(1-\varepsilon_j)
    Z_{M_j}(\beta_j)^2
     &                                             &\text{ring thermal purity at }\beta_j,\notag\\
     &\ge(1-\varepsilon_{j+1})(1-\varepsilon_j)
    \lambda_j^{2M_j}
     &                                             &\text{even }M_j\text{: trace}\ge\text{leading term},
    \label{obc:bulk}
\end{align}
and dividing by $\lambda_j^{2M_j}$ proves (iii). The bulk is now
controlled at every scale; the rest of this section concerns the
boundary only.

\subsection{One doubling step for the opposite-field chain}
\label{obc:steps}

Normalizing by $\lambda_j^m$ removes the extensive part of the
partition function, so the recurrence tracks only the overlap and the
remainder. At scale $j$, write $v_j=v_{\beta_j}$ and, for even $m\ge2$,
define the normalized opposite-field partition function, its remainder,
and the boundary remainder ratio by
\begin{align}
    c_j    &=|\langle\psi_j,v_j\rangle|^2,\notag\\
    B_m(j) &=\frac{Y_m(\beta_j)}{\lambda_j^m}
    =c_j+D_m(j)>0,\notag\\
    D_m(j) &=\folded{R_j^{m/2}}=\|R_j^{m/2}v_j\|^2,\qquad
    e_m(j)=\frac{D_m(j)}{B_m(j)}.
    \label{obc:normalized}
\end{align}
Folding only $R_j^{k/2}$, $k\le m$ even, onto each boundary vector
leaves $R_j^{m-k}$ in the middle, and the remainder decays
geometrically with the extra length (\decayglyph[mpored]):

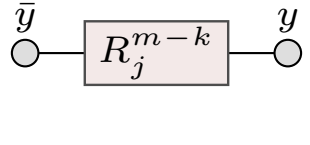
\begin{equation}
    D_m(j)
    =\tnchain[capl={\bar y},capr=y]{R_j^{m-k}}
    \le\|R_j\|^{m-k}D_k(j),
    \qquad y=R_j^{k/2}v_j,\quad D_m(j)\ge0.
    \label{obc:decay}
\end{equation}
In particular $c_j\le B_m(j)\le B_k(j)$; Steps~1 and~2 use this
monotonicity in the length.

Steps~1 and~2 also need the normalized contraction at the doubled
temperature in terms of quantities at $\beta_j$. Two rows of
Eq.~\eqref{obc:capping} stacked in imaginary time do not fuse site by
site, but the full operators multiply:
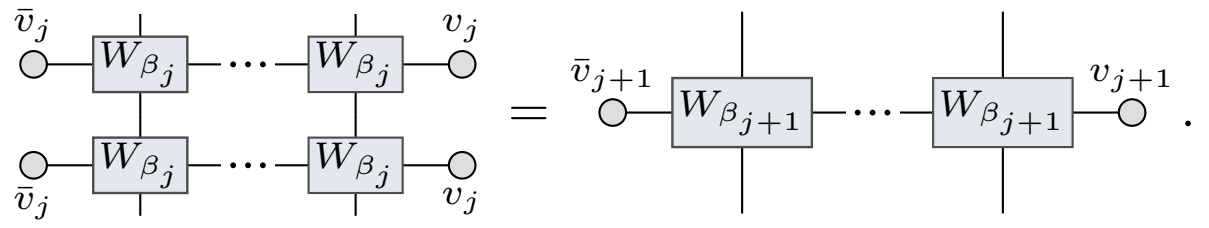
\begin{equation}
    \obcstack=\obcstackdouble.
    \label{obc:thermal-diagram}
\end{equation}
Its trace, normalized by $Y_m(\beta_j)^2$, is the thermal purity
$\Pt^{\opp}(m,\beta_j)=\Tr\rho_m^{\opp}(\beta_j)^2\in(0,1]$ of the
Gibbs state $\rho_m^{\opp}(\beta)=e^{-\beta(H_m^{\opp}+\kappa mI)}/Y_m(\beta)$.
Substituting $Y_m(\beta_j)=\lambda_j^mB_m(j)$ and
$\lambda_{j+1}=r_j\lambda_j^2$ relates the normalized contractions at
the two temperatures:
\begin{equation}
    B_m(j+1)
    =\Pt^{\opp}(m,\beta_j)\,r_j^{-m}B_m(j)^2.
    \label{obc:bridge}
\end{equation}
The thermal purity is the only factor not yet tracked; it is entry (T)
of the recurrence.

Given the boundary remainder ratio at length $M_j/2$ and the thermal purity at
length $M_j$, both at $\beta_j$, one step to scale $j+1$ moves through
the (length, inverse temperature) plane in a fixed order; the key lists
the three bounds that Steps~1--3 below derive:
\begin{proofdiagram}
    \obcstairs
\end{proofdiagram}
The proved pair sits at $k_{j+1}=M_j$ and $M_{j+1}=2M_j$, so the step
repeats.

\subsubsection{Step 1: Boundary remainder at the doubled inverse temperature}
\label{obc:move-one}

Step~1 must bound the boundary remainder ratio at the doubled
temperature, which Step~2 then uses. The old overlap says nothing about
the new one $c'$. But $c'$ does not depend on the length, so it cancels
from the difference of two lengths at the new temperature, and
Eq.~\eqref{obc:bridge} returns their ratio to scale $j$. Fix even
lengths $m>k\ge2$, assume $e_k(j)<1$, and write primes for scale $j+1$.
Choose $q<1$ with $\|R_{j+1}\|^{m-k}\le q$, so that
Eq.~\eqref{obc:decay} gives $D_m'\le qD_k'$. Subtracting the two
lengths cancels $c'$:
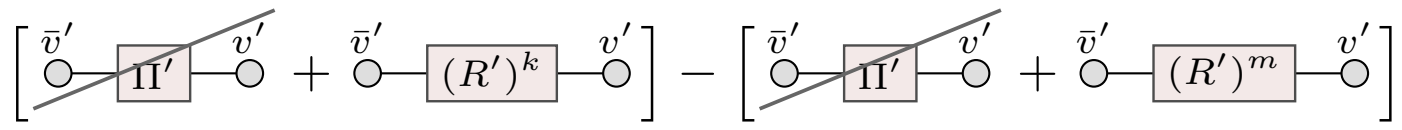
\begin{align}
    B_k'-B_m'
     &=\Bigl[\cancelled{$\tnchain[capl={\bar v'},capr={v'}]{\Pi'}$}
           +\tnchain[capl={\bar v'},capr={v'}]{(R')^k}\Bigr]
    -\Bigl[\cancelled{$\tnchain[capl={\bar v'},capr={v'}]{\Pi'}$}
         +\tnchain[capl={\bar v'},capr={v'}]{(R')^m}\Bigr]
    =D_k'-D_m'.
    \label{obc:cancellation-diagram}
\end{align}
Combining $D_k'=D_m'+B_k'-B_m'$ with $D_m'\le qD_k'$ gives
\begin{equation}
    (1-q)D_m'\le q(B_k'-B_m'),\qquad
    e_m'\le\frac{q}{1-q}\left(\frac{B_k'}{B_m'}-1\right).
    \label{obc:new-defect}
\end{equation}

Equation~\eqref{obc:bridge} at both lengths expresses the
new-temperature ratio in terms of quantities at scale $j$:
\begin{align}
    \frac{B_k'}{B_m'}
     &=\underbrace{\frac{\Pt^{\opp}(k,\beta_j)}{\Pt^{\opp}(m,\beta_j)}}_{\le1/\Pt^{\opp}(m,\beta_j)}
    \underbrace{r_j^{m-k}}_{\le1}
    \underbrace{\left[\frac{B_k(j)}{B_m(j)}\right]^2}_{\le[1-e_k(j)]^{-2}}\notag\\
     &\le\frac{1}{\Pt^{\opp}(m,\beta_j)[1-e_k(j)]^2},
    \label{obc:ratio}
\end{align}
the last since $B_m(j)\ge c_j=[1-e_k(j)]B_k(j)>0$. Once $e_m'<1$
follows, so does $c'=(1-e_m')B_m'>0$.

Equation~\eqref{obc:new-defect} is a product of two small factors, the
decay $q/(1-q)$ over the length difference and the excess of
$B_k'/B_m'$ above one, which is why the boundary remainder ratio squares.
For the induction, both factors must be $O(x)$ once every tracked bound
is at most $x$. Take
$m=M_j$ and $k=M_j/2$ with $\varepsilon_j\le x\le.01$. Property (ii) at
scale $j+1$ gives $\|R_{j+1}\|^{M_j}\le d(\varepsilon_{j+1})\le d(6x^2)$,
so across the half-chain $m-k=M_j/2$ we may take
\begin{equation}
    q=\sqrt{d(\varepsilon_{j+1})}
    \le\sqrt{\frac{6x^2}{2-18x^2}}\le2x.
    \label{obc:half-chain}
\end{equation}
Separately, when $e_k(j)\le x$ and
$\Pt^{\opp}(m,\beta_j)\ge1-x$, Eq.~\eqref{obc:ratio} bounds the excess above one by
\begin{equation}
    \frac{B_k'}{B_m'}-1\le(1-x)^{-3}-1=O(x).
    \label{obc:ratio-excess}
\end{equation}
Multiplying Eqs.~\eqref{obc:half-chain} and~\eqref{obc:ratio-excess}
in Eq.~\eqref{obc:new-defect} bounds the new boundary remainder ratio $e_m'$ by
$O(x^2)$.

\subsubsection{Step 2: Thermal purity at the doubled length}
\label{obc:move-two}

Step~2 must bound the thermal purity at length $2M$ and the old
temperature $\beta_j$, which Step~3 squares. Doubling an open chain is
not a squaring, since the boundary vectors appear once, so no purity
inequality applies directly. Instead, for even $M$, dividing
Eq.~\eqref{obc:bridge} at length $2M$ by that at length $M$ gives three
factors, each already controlled:
\begin{align}
    \frac{\Pt^{\opp}(2M,\beta_j)}{\Pt^{\opp}(M,\beta_j)}
     &=\underbrace{r_j^M}_{\text{(iii)}}\;
    \underbrace{\frac{B_{2M}(j+1)}{B_M(j+1)}}_{\ge1-e_M(j+1)\ \text{(Step 1)}}\;
    \underbrace{\left[\frac{B_M(j)}{B_{2M}(j)}\right]^2}_{\ge1\ \eqref{obc:decay}}\notag\\
     &\ge r_j^M[1-e_M(j+1)].
    \label{obc:length}
\end{align}
The doubling thus costs only the bulk factor of (iii) and the Step~1
ratio.

\subsubsection{Step 3: Thermal squaring at the doubled length}
\label{obc:move-three}

Step~3 returns to the dyadic grid at $(2M_j,\beta_{j+1})$, where the
next scale starts. The open-chain Gibbs state is positive with unit
trace, so the periodic squaring inequality applies to it unchanged; the
boundary plays no role. Squaring the Gibbs state $\rho_{2M}^{\opp}(\beta_j)$,
Eq.~\eqref{obc:thermal-diagram}, and applying the squaring
inequality~\eqref{obc:square} doubles the inverse temperature at length
$2M$:
\begin{equation}
    1-\Pt^{\opp}(2M,\beta_{j+1})
    \le f(1-\Pt^{\opp}(2M,\beta_j)).
    \label{obc:thermal-square}
\end{equation}
With Steps~1 and~2 this closes one scale of the recurrence.

\subsection{Initial rational steps and the small-remainder induction}
\label{obc:initialization}

The three steps of Sec.~\ref{obc:steps} now become a recurrence across
$j\ge2$ on two numbers per scale, iterated from the seeds, with the
comparison lengths $k_j$ of Eq.~\eqref{obc:k}.
Let $E_j$ and $\eta_j$ denote upper bounds on $e_{k_j}(j)$ and
$1-\Pt^{\opp}(M_j,\beta_j)$. Step~1 needs the decay over the length
difference, while (ii) bounds it over the ring length; a fractional
power converts one to the other. One scale $j\to j+1$ updates them by%
~\leanmark{https://sirui-lu.com/CertifiedGaps/docs/CertifiedGaps/Open/Criterion.html\#CertifiedGaps.OpenMomentWitness.open_step}
\begin{align}
    F(q,e,t)   &=\frac{q}{1-q}
    \left[\frac{1}{(1-e)^2t}-1\right],
               &                                       &\text{Step 1, Eqs.~\eqref{obc:new-defect}--\eqref{obc:ratio}},\notag\\
    q_j        &=d(\varepsilon_{j+1})^{(M_j-k_j)/M_j},
               &                                       &\text{(ii) at scale $j+1$, where $n_{j+1}=M_j$},\notag\\
    E_{j+1}    &\ge F(q_j,E_j,1-\eta_j),
               &                                       &\text{Step 1},\notag\\
    A_j        &=(1-\eta_j)(1-\varepsilon_j)
    (1-\varepsilon_{j+1})(1-E_{j+1}),
               &                                       &\text{Step 2, Eq.~\eqref{obc:length}: }\Pt^{\opp}(2M_j,\beta_j)\ge A_j,\notag\\
    \eta_{j+1} &\ge f(1-A_j),
               &                                       &\text{Step 3, Eq.~\eqref{obc:thermal-square}},
    \label{obc:updates}
\end{align}
valid whenever all four bounds lie in $[0,1)$; $F$ increases in $q$
and $e$ and decreases in $t$, so any upper bound on $q_j$ may be
substituted.

The seed bounds $E_2=.14$ and $\eta_2=.21$ are too large for the
analytical induction below, which needs $x\le.01$, so we first run
Eq.~\eqref{obc:updates} for three explicit rational steps:
\begin{center}
    \begin{minipage}[c]{.42\textwidth}
        \centering
        \begin{tabular}{@{}rrrcr@{}}
            \toprule
            $j$ & $E_j$ & $\eta_j$ & $(M_j-k_j)/M_j$ & $q_j\le$ \\
            \midrule
            2   & .14   & .21      & $49/120$        & .142     \\
            3   & .12   & .155     & $1/2$           & .0282    \\
            4   & .016  & .026     & $1/2$           & .0028    \\
            5   & .001  & .001     & ---             & ---      \\
            \bottomrule
        \end{tabular}
    \end{minipage}\hfill
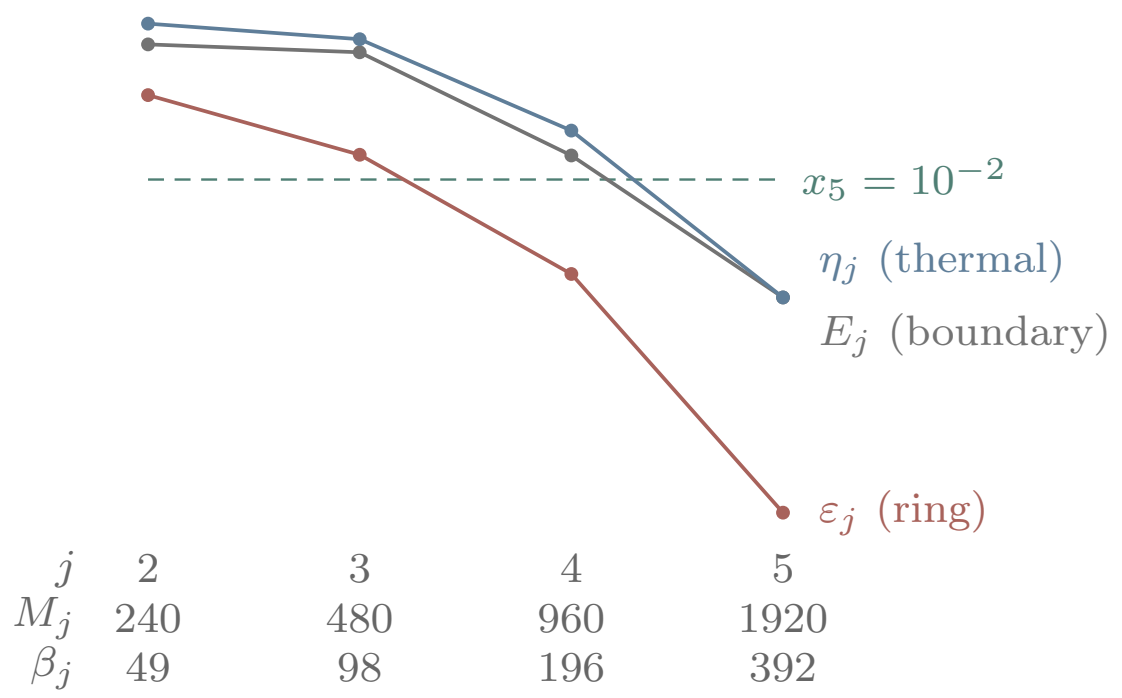
    \begin{minipage}[c]{.55\textwidth}
        \centering
        \obcladder
    \end{minipage}
\end{center}
Following Ref.~\cite{OpenAI2026Boundaryfield}, each entry is rounded
upward to a simple rational. The fractional powers then reduce to
integer power comparisons that a proof assistant checks directly,
$d(\varepsilon_3)^{49}<.142^{120}$ for $q_2$. Each row follows from
the previous one; at $j=2$,
\begin{align*}
    E_3    & :\ \frac{q_2}{1-q_2}\simeq.166,\quad
    \frac{1}{(1-E_2)^2(1-\eta_2)}-1\simeq.712,\quad
    .166\cdot.712\simeq.118\le.12,                                      \\
    \eta_3 & :\ A_2=(1-\eta_2)(1-\varepsilon_2)(1-\varepsilon_3)(1-E_3)
    \simeq.79\cdot.948\cdot.984\cdot.88=.648,\quad
    f(.352)\simeq.147\le.155.
\end{align*}
The first step gains little because $q_2$ is still large; once $q_j$
is small, the bounds square, as the plot above shows. With
$\varepsilon_5=6\varepsilon_4^2<.000015$, the $j=5$ row brings all
three bounds below $.01$, the base case of the small-remainder
induction:
\begin{equation}
    \varepsilon_j,\ e_{k_j}(j),\
    1-\Pt^{\opp}(M_j,\beta_j)\le x_j,\quad j\ge5;\qquad
    x_5=.01,\quad x_{j+1}=10x_j^2,\quad
    x_j=10^{-1-2^{j-5}}.
    \label{obc:induction}
\end{equation}

The induction step of Eq.~\eqref{obc:induction} must return all three
bounds from $x$ to $10x^2$ in one scale. Each step is quadratic in $x$,
but Step~2 consumes the Step~1 output and the ring bound $\varepsilon_j$ enters both
$q$ and $A$, so the constants are tracked jointly. Assume that the
three bounds hold at scale $j\ge5$ and write $x=x_j\le.01$. One step
from scale $j$ to scale $j+1$ ($M_{j+1}=2M_j$) updates the boundary and
thermal bounds in turn:
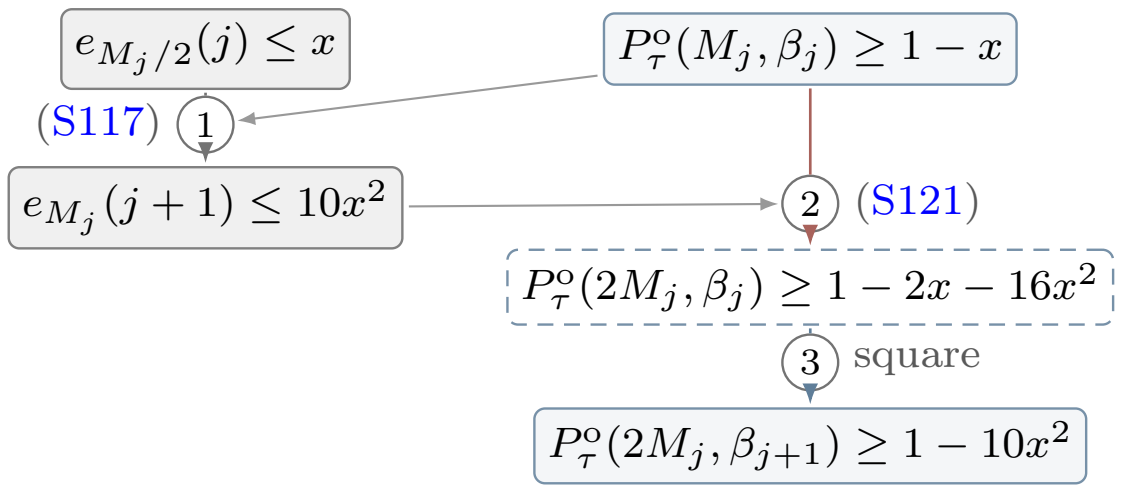
\begin{proofdiagram}
    \begin{prooflanes}\def\lanedy{1.05}
        \lanebound{b}{e0}{0}{0}{$e_{M_j/2}(j)\le x$}
        \lanebound{t}{t0}{1}{0}{$\Pt^{\opp}(M_j,\beta_j)\ge1-x$}
        \lanebound{b}{e1}{0}{1}{$e_{M_j}(j+1)\le10x^2$}
        \lanebound{to}{t1}{1}{1.5}{$\Pt^{\opp}(2M_j,\beta_j)\ge1-2x-16x^2$}
        \lanebound{t}{t2}{1}{2.5}{$\Pt^{\opp}(2M_j,\beta_{j+1})\ge1-10x^2$}
        \lanemove{e0}{e1}{1}{\eqref{obc:new-defect}}
        \laneuse{t0}{1}
        \lanemove[mpored]{t0}{t1}{2}{\eqref{obc:length}}
        \laneuse{e1}{2}
        \lanemove[mpoblue]{t1}{t2}{3}{square}
    \end{prooflanes}
\end{proofdiagram}
For Step~1, Eqs.~\eqref{obc:half-chain} and~\eqref{obc:ratio-excess}
give $q_j\le2x$ and
\begin{equation}
    F(2x,x,1-x)
    \le\frac{6x^2}{(1-2x)(1-x)^3}<7x^2,
    \label{obc:boundary-step}
\end{equation}
so $E_{j+1}=10x^2$ bounds $e_{M_j}(j+1)$. For Step~2, the four factors
of $A_j$ are at least $1-x$, $1-x$, $1-6x^2$, and $1-10x^2$. Hence
$A_j\ge(1-x)^2(1-16x^2)\ge1-2x-16x^2$, with $16=6+10$ and the positive
cross terms dropped. For Step~3, since $0<x\le.01$, thermal squaring
yields
\begin{equation}
    f(2x+16x^2)<3x^2,
    \label{obc:thermal-step}
\end{equation}
so $\eta_{j+1}=10x^2$ bounds $1-\Pt^{\opp}(2M_j,\beta_{j+1})$, and
$\varepsilon_{j+1}\le6x^2\le10x^2$ closes the induction at scale $j+1$;
the common prefactor $10$ covers the three constants $7$, $3$, and $6$.

\subsection{Equal endpoint fields, odd lengths, and the uniform gap}
\label{obc:comparison}

Equation~\eqref{obc:induction} controls even opposite-field chains at
every $j\ge5$; the theorem needs the thermal purity of odd equal-field
chains, which Eq.~\eqref{obc:target} turns into the gap. Three things
change: the right boundary vector carries the twist $U_P$, the odd
length makes the remainder neither a square nor necessarily positive,
and only even lengths are certified. Define the normalized equal-field
partition function $Q_j(n)=Y_n^{\same}(\beta_j)/\lambda_j^n$. The twist
is harmless for the leading term, because $U_P$ commutes with the
transfer operator and fixes $\psi_j$ by (i), so the overlap $c_j$
survives:
\begin{equation}
    \tnchain[last=sym]{\Pi_j,U_P}=\tnchain{\Pi_j}.
    \label{obc:twist}
\end{equation}
For the remainder, pull half of an even length $M\le n$ onto each
boundary vector, commuting $U_P$ through $R_j^{M/2}$ ($[U_P,R_j]=0$ because
$[U_P,X_\beta]=0$ and $U_P\Pi_j=\Pi_j$); the middle operator then has norm
at most one ($\|R_j\|\le1$, $U_P$ orthogonal):
\begin{align}
    Q_j(n)-c_j
     &=\tnchain[capl={\bar v_j},capr={v_j}]
       {R_j^{M/2},R_j^{n-M}U_P,R_j^{M/2}}\notag\\
     &=\tnchain[capl={\bar y},capr={y}]{R_j^{n-M}U_P},\qquad
    y=R_j^{M/2}v_j.
    \label{obc:signed}
\end{align}
Cauchy--Schwarz bounds the open contraction by the squared norm of its
caps, the even-length opposite-field remainder:
\begin{equation}
    \schwarzfold=D_M(j).
    \label{obc:absolute}
\end{equation}
The odd equal-field chain is thus controlled by the even opposite-field
remainder at any even $M\le n$, and the recurrence certifies $M=M_j$.

Fix a scale $j\ge5$ and an odd length $n$ with $M_j\le n\le2M_j$. The
equal-field purity is a ratio of $Q$ at the two temperatures, as in
Eq.~\eqref{obc:bridge}, so $Q$ needs an upper bound at $\beta_j$ and a
lower bound at $\beta_{j+1}$. Both follow from Eq.~\eqref{obc:absolute}
at the certified even length $M=M_j=k_{j+1}$, where the recurrence
bounds the remainder at both temperatures:
\begin{align}
    Q_j(n)     &\le c_j+D_M(j)=B_M(j),\notag\\
    Q_{j+1}(n) &\ge c_{j+1}-D_M(j+1)\notag\\
               &=[1-2e_M(j+1)]B_M(j+1)\notag\\
               &\ge(1-2x_{j+1})B_M(j+1)>0.
    \label{obc:comparison-bounds}
\end{align}
Equation~\eqref{obc:bridge} at length $M$ and
Eq.~\eqref{obc:comparison-bounds} then bound the equal-field thermal
purity factor by factor; the bulk factor is the only term that still
depends on $n$:
\begin{align}
    \Pt^{\same}(n,\beta_j)
     &=r_j^n\frac{Q_{j+1}(n)}{Q_j(n)^2}\notag\\
     &\ge\underbrace{(1-2x_{j+1})}_{\eqref{obc:comparison-bounds}}\;
    \underbrace{\Pt^{\opp}(M,\beta_j)}_{\ge1-x_j}\;
    \underbrace{r_j^{n-M}}_{\ge r_j^{M},\ n-M\le M}\notag\\
     &\ge(1-2x_{j+1})(1-x_j)
    \underbrace{(1-\varepsilon_j)(1-\varepsilon_{j+1})}_{\text{(iii)}}\notag\\
     &\ge1-2x_j-26x_j^2\notag\\
     &\ge1-4x_j,
    \label{obc:final-purity}
\end{align}
where $26=20+6$ collects $2x_{j+1}=20x_j^2$ and
$\varepsilon_{j+1}\le6x_j^2$, and $26x_j^2\le2x_j$ for $x_j\le1/13$.
Every odd $n$ in $[M_j,2M_j]$ thus inherits the even-length bound at
scale $j$ up to a factor $4$, which the proof below converts into the gap.

\begin{proof}[Proof of Theorem~\ref{obc:theorem}]
    Each odd $n$ gets its own scale $j$, so the gap bound must be uniform
    in $j$. Both $-\log(1-\Pt^{\same}(n,\beta_j))$ and the inverse temperature
    $\beta_j$ double from one scale to the next, so their ratio in
    Eq.~\eqref{obc:target} is scale independent.
    The intervals $[M_j,2M_j]$, $j\ge5$, cover every odd $n\ge M_5+1=1921$,
    each at its own inverse temperature:
    \begin{proofdiagram}
        \oddcoverage
    \end{proofdiagram}
    Given odd $n$, choose $j\ge5$ with $M_j\le n\le2M_j$.
    Since $x_j\le x_5=.01$, Eq.~\eqref{obc:final-purity} gives
    $w_0\ge\Pt^{\same}(n,\beta_j)\ge1-4x_j>1/2$, so $H_n^{\same}$ has a unique ground state. At this
    length $n$ and finite inverse temperature $\beta_j$,
    Eq.~\eqref{obc:target} gives
    \begin{align}
        e^{-\beta_j\Delta_n}
         &\le\frac{4x_j}{1-4x_j}
         &                               &x_j=10^{-1}\,10^{-2^{j-5}},\quad 1-4x_j\ge.96,\notag\\
         &\le\frac{5}{12}\,10^{-2^{j-5}}
        <10^{-2^{j-5}},
         &                               &\tfrac{.4}{.96}=\tfrac{5}{12}<1.
        \label{obc:gap-final}
    \end{align}
    Taking logarithms and dividing by $\beta_j=\beta_5\,2^{j-5}$ with
    $\beta_5=49\cdot2^3=392$ proves $\Delta_n>\tfrac{\log10}{392}$ for every $j$.
\end{proof}

In Lean~4, we formalize both the general open-chain theorem from a
rational certificate%
~\leanmark{https://sirui-lu.com/CertifiedGaps/docs/CertifiedGaps/Open/Certificate.html\#CertifiedGaps.OpenCertificate.gapGT_of_openCertificate}
and the derivation of the four seeds~\eqref{obc:seeds} from the short
boundary trace table%
~\leanmark{https://sirui-lu.com/CertifiedGaps/docs/CertifiedGaps/Heisenberg/BoundaryInitTable.html\#CertifiedGaps.SpinOneHeisenberg.BoundaryInit.boundaryDefect_le_of_table},
and verify the open-chain integer matrix-product trial contractions on
$m=34,78,142,240$ sites and the nine shortest boundary trace totals
directly in the Lean kernel, leaving only the remaining six short-chain
trace totals as hypotheses%
~\leanmark{https://sirui-lu.com/CertifiedGaps/docs/SeedKernel/Heisenberg/Integers.html\#CertifiedGaps.SpinOneHeisenberg.spin_one_boundary_field_gap_of_hornerTotalsLarge}.
Rerunning the same three rational steps of Eq.~\eqref{obc:updates} from
the four seeds~\eqref{obc:seeds} with four-digit rounding and the ring
constant $5.583$ in place of $6$, and entering the quadratic tail at the
scale-$5$ defect $2.241\times10^{-4}$ instead of $10^{-2}$, sharpens the
uniform bound to $\Delta_n>\log(10^6/2241)/392\simeq0.01556$ for every
odd $n\ge1921$%
~\leanmark{https://sirui-lu.com/CertifiedGaps/docs/CertifiedGaps/Heisenberg/OpenChain.html\#CertifiedGaps.SpinOneHeisenberg.gapGT_of_boundaryFieldInputs_sharp}.

\end{document}